\documentclass[trackchanges,twocolumn]{aastex7}

\usepackage{comment}
\usepackage{gensymb}
\usepackage[english]{babel}
\usepackage[autostyle]{csquotes}
\usepackage{amsmath}
\usepackage{textcomp}
\usepackage{xcolor}
\usepackage{graphicx} 
\usepackage{multirow}
\usepackage{subcaption}
\usepackage{dirtytalk}
\usepackage{subcaption}

\begin{document}

\title{Precise Radii of Visual Binaries with Dynamical Masses: $\xi$ Bootis $\&$ 70 Ophiuchi}

\author[orcid=0009-0005-8004-2351,gname=Becky,sname=Flores]{Becky E. Flores}
\affiliation{Georgia State University, Department of Physics and Astronomy, Atlanta, GA 30303, USA}
\email[show]{bflores5@gsu.edu}  

\author[orcid=0000-0001-5313-7498,gname=Russel, sname=White]{Russel J. White} 
\affiliation{Georgia State University, Department of Physics and Astronomy, Atlanta, GA 30303, USA}
\email{rwhite31@gsu.edu}

\author[orcid=0000-0001-5415-9189, gname=Gail,sname=Schaefer]{Gail Schaefer}
\affiliation{The CHARA Array of Georgia State University, Mt. Wilson, CA 91203, USA}
\email{gshaefer@gsu.edu} 

\author[0000-0002-0078-5288]{Mark R. Giovinazzi}
\affiliation{Amherst College, Department of Physics and Astronomy, Amherst, MA 01002, USA}
\email{mgiovinazzi@amherst.edu}

\author[0009-0000-4050-4172]{Simone Lilavois}
\affiliation{Amherst College, Department of Physics and Astronomy, Amherst, MA 01002, USA}
\email{slilavois28@amherst.edu}

\author[0009-0007-5640-0061]{Ryan Purviance}
\affiliation{Amherst College, Department of Physics and Astronomy, Amherst, MA 01002, USA}
\email{rpurviance@amherst.edu}

\author[0009-0001-4688-6949]{Henry Dennen}
\affiliation{Amherst College, Department of Physics and Astronomy, Amherst, MA 01002, USA}
\email{hdennen26@amherst.edu}

\author[0000-0001-8170-7072]{Daniella Bardalez Gagliuffi}
\affiliation{Amherst College, Department of Physics and Astronomy, Amherst, MA 01002, USA}
\email{dbardalezgagliuffi@amherst.edu}

\author[orcid=0000-0001-5737-1687,sname='Harada']{Caleb K.\ Harada}
\affiliation{Institute for Astronomy, University of Hawai`i, 2680 Woodlawn Drive, Honolulu, HI 96822, USA}
\affiliation{Department of Earth Sciences, University of Hawai`i at Man\=oa, 1680 East-West Road, Honolulu, HI 96822, USA}
\affiliation{Department of Astronomy, 501 Campbell Hall \#3411, University of California, Berkeley, CA 94720, USA}

\email{ckharada@hawaii.edu}

\author[0000-0003-4824-0938]{Brian D. Mason}
\affiliation{US Naval Observatory, 3450 Massachusetts Avenue, NW, Washington, DC, 20392, USA}
\email{brian.d.mason.civ@us.navy.mil}

\author[0000-0001-7233-7508,sname=Matson,gname=Rachel]{Rachel A. Matson}
\affiliation{US Naval Observatory, 3450 Massachusetts Avenue, NW, Washington, DC, 20392, USA}
\email{rachel.a.matson2.civ@us.navy.mil}

\author[0000-0001-8365-5982]{Isabelle Baraffe}
\affiliation{ University of Exeter, Department of Physics and Astronomy, Stocker Road, Exeter EX4 4QL, UK}
\affiliation{Ecole Normale Sup\'erieure de Lyon, CRAL (UMR CNRS 5574), Universit\'e de Lyon, 69007, Lyon, France}
\email{I.Baraffe@exeter.ac.uk}

\author[0009-0001-4619-7369]{Chi Thanh Nguyen}
\affiliation{University of Trieste, Department of Physics, Trieste, Friuli Venezia Giulia, IT   }
\email{chi.nguyen@inaf.it}

\author[0000-0002-2012-7215]{Greg Feiden}
\affiliation{ University of North Georgia, Department of Physics and Astronomy, Dahlonega, GA 30597, USA }
\email{gregory.feiden@ung.edu}

\author[orcid= 0000-0002-3380-3307]{John D. Monnier}
\affiliation{University of Michigan, Department of Astronomy, Ann Arbor, MI 48109, USA}
\email{monnier@umich.edu}

\author[orcid=0000-0001-6017-8773 ]{Stefan Kraus}
\affiliation{University of Exeter, Department of Physics \& Astronomy, Stocker Road, Exeter, EX4 4QL, UK}
\email{S.Kraus@exeter.ac.uk}

\author[orcid=0000-0002-2208-6541 ]{ Narsireddy Anugu }
\affiliation{The CHARA Array of Georgia State University, Mt. Wilson, CA 91203, USA}
\email{nanugu@gsu.edu}

\author[orcid=0000-0001-9764-2357 ]{Claire L. Davies}
\affiliation{University of Exeter, Department of Physics \& Astronomy, Stocker Road, Exeter, EX4 4QL, UK}
\email{C.Davies3@exeter.ac.uk}

\author[orcid=0000-0002-3003-3183 ]{Tyler Gardner}
\affiliation{University of Colorado Boulder, Cooperative Institute for Research in Environmental Sciences, 216 UCB, Boulder, CO, 80309, USA}
\email{Tyler.Gardner@colorado.edu}

\author[orcid=0009-0006-0225-4444]{Mayra Gutierrez}
\affiliation{University of Michigan, Department of Astronomy, Ann Arbor, MI 48109, USA}
\email{mgutie60@ucsc.edu}

\author[orcid=0000-0002-1788-9366 ]{Noura Ibrahim}
\affiliation{University of Michigan, Department of Climate and Space Sciences and Engineering, Ann Arbor, MI 48109, USA}
\email{inoura@umich.edu}

\author[orcid=0000-0001-9745-5834 ]{ Cyprien Lanthermann }
\affiliation{The CHARA Array of Georgia State University, Mt. Wilson, CA 91203, USA}
\email{ clanthermann@gsu.edu}

\author[]{ Norman Vargas }
\affiliation{The CHARA Array of Georgia State University, Mt. Wilson, CA 91203, USA}
\email{nvargas@gsu.edu}

\author[]{ Olli Majoinen }
\affiliation{The CHARA Array of Georgia State University, Mt. Wilson, CA 91203, USA}
\email{omojoinen@gsu.edu}

\begin{abstract}

A new orbital solution for $\xi$ Bootis (GJ 566 AB, P=152 yrs) and the recently revised orbital solution for 70 Ophiuchi (GJ 702 AB, P=88 yrs) yield component masses precise to 0.4\%. Their relatively wide separations ($> 4.0\arcsec$) allow the components to be studied independently, making them ideal for testing the predictions of stellar evolutionary models. We use the six-telescope near-infrared MIRC-X and MYSTIC beam combiners at the CHARA Array to measure the angular diameters of all 4 stars. Combined with precise Gaia DR3 parallax values and bolometric fluxes, these interferometric measurements translate to stellar radii that are precise to $<2.4\%$ and effective temperatures precise to $<1.7\%$. We compare the measured radii and temperatures to four stellar evolutionary models (Lyon, Dartmouth, PARSEC, MIST). Over the mass range $0.67 - 0.91$ M$_\odot$, all models reproduce the observed properties within a few standard deviations, with no systematic offsets.
The pairs are coeval and consistent with independent age estimates, further validating model predictions.
With physical separations greater than 20 au, these pairs show no evidence of radius inflation, despite being spectroscopically classified as chromospherically active, and in the case of $\xi$ Bootis, probable members of the adolescent age ($\sim 0.4$ Gyr) Ursa Majoris Moving Group.
In contrast, a set of 21 short-period ($P\lesssim$ 10 days) eclipsing binary stars with similar dynamical masses and metallicities are larger and often cooler than these same model predictions. Within this mass range, radius inflation appears to be associated with short-period binary star systems.

\end{abstract}


\section{Introduction}\label{sec:intro}

Precisely measured stellar properties of nearby stars, such as mass, radius, luminosity, age and metallicity, are essential for testing stellar evolution theory and for predicting the properties of stars throughout the Galaxy. Many of the nearest stars are potential targets for the proposed Habitable Worlds Observatory \citep[HWO,][]{mamajek_provisional_2024, harada_setting_2024, harada_sporeshwo_2025}, where accurate stellar characterization will be critical for detecting and interpreting Earth‑like exoplanets.
The majority of stars with direct mass and radius measurements are those in short-period ($<100\,\mathrm{days}$) eclipsing binary (EB) systems \citep{andersen_EBs1991, torres_accurate_2010}. For these, a well‑sampled light curve that resolves both eclipses, together with radial‑velocity measurements detecting both components, yields stellar masses and radii measured to a few‑percent precision.
Precision improves significantly when utilizing space-based photometry
\citep[e.g.,][]{mahadevan_survey2019, Han_magnetic_inflation2019, miller_fundamental2020, sebastian_EBLMIX2023, cakirli_low_mass_companions2025, spejcher_moderate_spin-orbit2026}.
Being coeval and of similar composition, the components of EBs have served as long-standing benchmarks for testing stellar evolution and establishing mass–radii relations \citep[e.g.,][]{hillenbrand_assessment2004, fernandes_revisiting_2021}. 

While observational constraints from these systems have improved evolutionary theory, problems remain. It has been known for decades that stars in the lower part of the main sequence (spectral types K and M) are often larger than their predicted sizes by as much as 10\% 
\citep{popper_orbits1997, torres_absolute_2002, kraus_massradius_2011ApJ...728...48K, birkby_discovery_2012, dittman_discovery2017, morrell_exploring_2019, 
torres_eclipsing_2021, cruz_detached2022}.
The stars are also cooler than expected, although the magnitude of this has been difficult to quantify 
\citep{torres_stringent2021, morales_fundamental2022}
and may be correlated with metallicity \citep[e.g.,][]{swayne_EBLMXI2024, pan_analysis_2025}.
This phenomenon known as “radius inflation” is thought to stem from a pernicious bias of EBs – they are often short-period systems that are tidally locked, rapidly rotating, and chromospherically active 
\citep{garrido_radius_2019, pass_active_2023, pan_starspot_2024}. One explanation for the radius inflation problem is that rapid rotation is generating strong magnetic fields that inhibit convective energy transport, causing the star to expand 
\citep{lopez_morales_correlation2007, feiden_chaboyer_magnetic2013, feiden_magnetic_2016, kochukhov_shylyak_magnetic2019}.
 
\cite{clausen_0711_1999} and \cite{torres_eclipsing_2006} show that evolutionary models can reproduce this effect, in an ad hoc fashion, by reducing the mixing length parameter used to model convection.
Similar reconciliations were achieved by \cite{chabrier_evolution_2007} and \cite{somers_spots_2020} who considered the reduced convective efficiency and star spots caused by large magnetic fields.
The discrepancies of model predictions with observed radii are not exclusive to K and M stars. Some active stars with masses near that of the Sun also display enlarged radii and cooler temperatures than predicted \citep[e.g.,][]{clausen_absolute_2009, vos_absolute_2012, southworth_rediscussion_2022}.

Assessing the magnitude and cause of the radius and temperature discrepancies for EBs is difficult because the stellar properties depend, to a mass-dependent extent, on the age and chemical composition of the systems \citep[e.g.,][]{jennings_revising_2023}. But these are generally poorly determined for EBs because the blended spectra compromise metallicity measurements, and the induced rapid rotation makes age tracers like gyrochronology ineffective \citep{barnes_ages_2007}. Additionally, reddening caused by interstellar extinction \citep{prsa_influence_2005, jones_physics_2020}, the effects of star spots, and the assumed limb-darkening profile can compromise the accuracy of EB star measurements \citep{tofflemire_lowmass_2023}. Despite these challenges, there is tentative evidence that radius inflation is more severe for metal-poor stars \citep{boetticher_eblm_2019}, although the trend is blurred by an underlying correlation with orbital period \citep{swayne_eblm_2024}; the shortest period systems are inflated the most.

The development of Optical Long Baseline Interferometry (OLBI) has enabled a new way to test evolutionary model predictions. OLBI has yielded precise limb-darkened angular diameter measurements for hundreds of stars at a precision level of $5\%$ or better \citep{segransan_first_2003, vanbelle_directly_2009, ligi_radii2016, vonbraun_extrasolar_2017, karovicova_fundamental_2020, karovicova_fundamental_2021, karovicova_fundamental_2022, baines_33_2023, baines_vintage_2025}. The radius is determined directly when combined with distance, and in combination with the bolometric flux, the stellar effective temperature $(T_\mathrm{EFF})$ can be determined by inverting an observational form of Stefan’s Law \citep{stefan_uber_1879, boltzmann_ableitung_1884}.  
An advantage of this approach is that most stars observed with OLBI are single (for observing practicalities) and thus are not biased by tidally induced rapid rotation and activity. Collectively, OLBI has confirmed the general discrepancies with model predictions, though to a lesser extent. Stars cooler than 5000 K appear to be larger than model predictions by 3-5\% \citep{boyajian_stellar_2012b, rabus_discontinuity_2019}. One caveat in these comparisons is that the masses of these stars are unknown; the comparisons of radius and temperature are done relative to luminosity, which may mitigate the offset. Additionally, more recent interferometric measurements have identified some discrepancies with first-generation measurements. \cite{white_interferometric_2018} measured the angular diameters of stars, at optical wavelengths, to be as much as 15\% smaller than measurements done at infrared wavelengths \citep[see also][]{casagrande_stellar_2014, tayar_guide_2022}. While the exact cause of the differences is unknown, the trend of increasing discrepancies toward smaller sizes suggests an issue with measuring angular diameters near the resolution limit when the visibility curve is sparsely sampled. The use of “bad” point-source calibration stars (see Section \ref{sec:angular-diameters}) is also a possibility.

In this paper, we use OLBI at near-infrared wavelengths to measure precise radii of the components of the visual binaries GJ 566 AB ($\xi$ Bootis) and GJ 702 AB (70 Ophiuchi), potential HWO targets \citep[][]{mamajek_provisional_2024, harada_setting_2024, harada_sporeshwo_2025}. These two systems have precisely determined dynamical masses, allowing for direct comparisons with different stellar evolution models. The star systems are introduced in Section \ref{sec:gj566-gj702}. Interferometric observations are described in Section \ref{sec:interferometry}. Measured angular diameters are presented in Section \ref{sec:angular-diameters} and translated into radii and effective temperatures in Section \ref{sec:Stellar-prop}.  In Section \ref{sec:model-comp}, we present comparisons with stellar evolution models and in Section \ref{sec:EBcomps} we present comparisons with EBs. Our findings are summarized in Section \ref{sec:Summary}.


\section{The Visual Binary Stars $\xi$ Bootis and 70 Ophiuchi}\label{sec:gj566-gj702}

GJ 566 AB ($\xi$ Bootis, HD 131156, HR 5544) is a visual binary consisting of a G8V and a K4V at a distance of 6.7 pc \citep[see Table \ref{tab:stellar-properties};][]{gliese_nearby1995, gray_contributions_2003, gaiacollaboration_gaia_2023}. The pair is well-resolved in most observations, with a semi-major axis of 4.9\arcsec.  Fits to the available astrometric data show the orbit is well-constrained, with a period of 152 yrs and total dynamical mass of $\sim 1.6$ M$_\odot$ \citep[e.g.,][]{mason_WDS2001}. 

Here we present a fully solved orbit derived from a comprehensive dataset that includes 623 relative astrometry measurements spanning 199 years from the Washington Double Star catalog \citep[WDS;][]{mason_WDS2001}, 554 radial velocity measurements spanning more than 30 years from the Hamilton \citep{vogt_lick_1987}, HIRES \citep{vogt_hires_1994}, and Automated Planet Finder \citep[APF][]{radovan_automated_2014} spectrographs \citep[published in][]{rosenthal_california_2021}, and absolute astrometric constraints from Hipparcos and Gaia \citep{leeuwen_validation_2007, gaiacollaboration_gaia_2023}. To determine the system's three-dimensional architecture and robust, model-independent masses, we employ the orbit-fitting software \texttt{orvara} \citep{brandt_orvara_2021} that relies on parallel-tempering Markov Chain Monte Carlo sampling.

\begin{figure*}
    \centering
    \includegraphics{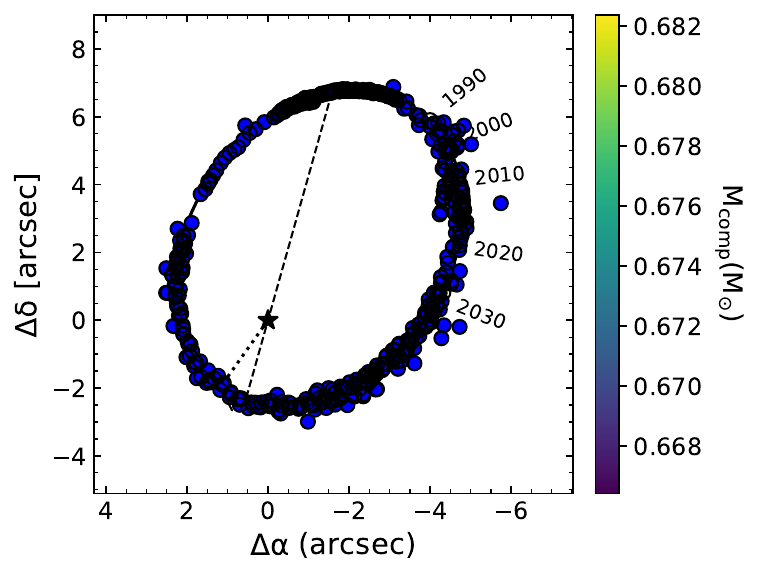}
\caption{Projected relative astrometric orbit of GJ 566 B around A. Filled circles represent the relative astrometric measurements colored by companion mass. Several decades are labeled for reference. The dashed line indicates the line of nodes, and the star marks the barycenter.}
    \label{fig:566-orbit}
\end{figure*}

For the GJ 566 system, this technique yields a refined orbital period of $P=152.292\pm0.049~\mathrm{yr}$ and dynamical masses of $M_\mathrm{A}=0.906\pm0.004~\mathrm{M_\odot}$ for GJ 566 A and $M_\mathrm{B}=0.674\pm0.003~\mathrm{M_\odot}$ for GJ 566 B (Giovinazzi et al. 2026, in prep.). This corresponds to mass precisions $\lesssim0.4\%$ for both components. Similar precisions ($<1\%$) using this method for other nearby systems have been achieved \citep[e.g.,][]{giovinazzi_neid_2025, an_orbits_2025}.
The best fit astrometric orbit is illustrated in Figure \ref{fig:566-orbit} and the best fit orbital elements are presented in Table \ref{tab:orbital-properties}.

Both GJ 566 A and B exhibit chromospheric emission \citep{olah_magnetic2016}, consistent with their modest rotation periods of $\sim 6.4$ days and $\sim 11.9$ days, respectively \citep[see][for a discussion of previous measurements]{strassmeier_doppler2023}. Both have projected rotational velocities ($v\mathrm{sin}i$) below 
$6.0 \mathrm{\,km\,s^{-1}}$, based on upper limit measurements presented in \citet{luck_abundances_2017}.
\citet{mamajek_improved_2008} use the rotation periods to estimate a gyrochronology age of 389 Myr, an average age for the pair.  Based on signatures of adolescence \citep{gruner_wide_binaries2023}, including elevated lithium abundances \citep{strassmeier_lithium2022} and kinematics, the pair are thought to be members of the Ursa Majoris Moving Group \citep[UMMG;][]{king_UMa2003, dopcke_UMa2019}, with an age of $414\pm 23$ Myr \citep{jones_UMa2015}. Consistent with this, the BANYAN $\Sigma$ Bayesian algorithm assigns a UMMG membership probability of 93.9\% 
\citep{gagne_montreal2026}.
Metallicity measurements via high dispersion spectra scatter around solar metallicity \citep{strassmeier_lithium2022}, which is consistent with the solar value ([Fe/H] $= 0.01\pm0.07$ dex) spectroscopically measured for the UMMG by \citet{dopcke_UMa2019}.

GJ 702 AB (70 Oph, HD 165341) is a visual binary consisting of a K0V and a K5V at a distance of 5.1 pc \citep[see Table \ref{tab:stellar-properties};][]{gliese_nearby1995, gray_contributions_2003,gaiacollaboration_gaia_2023}. The pair is also well-resolved in most observations, with a semi-major axis of 4.6\arcsec. With a period of 88 yr, the pair has completed nearly 3 orbital revolutions since first observed by Herschel in 1779 \citep{izmailov_orbits_2019}. 
Using the astrometric orbit and radial velocity measurements of both components, \citet{li_century_2026} determine dynamical masses of $M_A=0.8827\pm0.0044$ M$_\odot$ and $M_B=0.7319\pm0.0031 $ M$_\odot$. Their results are consistent with, but more precise than previous orbital analyses \citep{eggenberger_analysis_2008}; the dynamical masses have precisions of 0.4\%. Best-fit orbital properties are summarized in Table \ref{tab:orbital-properties}.

Both GJ 702 A and GJ 702 B exhibit chromospheric activity, based on Ca II H\&K emission and variability 
\citep{baliunas_chromospheric1995}. Neither has a projected rotational velocity above the measurement limit of 
$4.3 \mathrm{\,km\,s^{-1}}$ \citep{luck_abundances_2017, morris_stellar_2019}. 
GJ 702 A has a measured rotation period of $19.9 \pm 0.5$ days 
\citep{colliercameron_2009, noyes_1984}. However, \citet{colliercameron_2009} speculate that the true rotation period is
twice this value, $39.8 \pm 1.0$ days since Ca II H and K light curves sometimes exhibit double-peaked modulation \citep{baliunas_chromospheric1995}. No rotation period for GJ 702 B has been reported. We caution that
the value of 34 days listed in \citet{gruner_wide_binaries2023} is estimated from rotation-activity relations. 
\citet{colliercameron_2009} provides a gyrochonology age of $2.0 \pm 0.1$ Gyr for GJ 702 A using the faster rotation period, or an age of $7.0 \pm 0.3$ Gyr adopting twice this period. The latter agrees best with the asteroseismic age of $6.0 \pm 1.0$ Gyr reported in \citet{eggenberger_analysis_2008}. Metallicity measurements determined from high dispersion spectra are consistent with the near solar value \citep[Fe/H $ = 0.04 \pm 0.05$ dex;][]{eggenberger_analysis_2008}.


\begin{deluxetable}{cccccccccccc}

\tablecaption{Component Properties \label{tab:stellar-properties}}

\tablehead{\colhead{GJ} & \colhead{RA} & \colhead{DEC} & \colhead{Spectral }& \colhead{Plx} & \colhead{$v$sin$i$} & \colhead{V} & \colhead{H} & \colhead{K}  \\ 
\colhead{Name} & \colhead{(hh mm ss)} & \colhead{(dd mm ss)} &\colhead{Type} & \colhead{(mas)} & \colhead{$\mathrm{(km\,s^{-1})}$} & \colhead{(mag)} & \colhead{(mag)} & \colhead{(mag)} } 
\startdata
566 A & 14 51 23.3 & +19 06 01.6 & G8V & $148.070 \pm 0.132$ & 6.0 & 4.8 & 2.3  & 2.0  \\
566 B & 14 51 23.0 & +19 06 06.9 & K4V & $148.179 \pm 0.055$ & 5.1 & 7.0 & 4.2* & 4.1*  \\
702 A & 18 05 27.2 & +02 30 00.5 & K0V & $195.567 \pm 0.196$ & 4.3 & 4.2 & 2.4* & 2.3**  \\
702 B & 18 05 27.5 & +02 29 56.2 & K5V & $195.856 \pm 0.354$ & 2.1 & 6.1 & 3.6* & 3.1**  \\
\enddata

\tablerefs{RA \& DEC (J2000), Spectral Type: \cite{gliese_nearby1995, gray_contributions_2003}, Plx: \cite{gaiacollaboration_gaia_2023},   $v$sin$i$: \cite{luck_abundances_2017, morris_stellar_2019}, Vmag: \cite{hog_tycho2_2000}, Hmag \& Kmag: \cite{cutri_2mass_2003}. Values with an * are estimated using the relations from \cite{pecaut_intrinsic_2013} and values with ** are from \cite{boyajian_stellar_2012b} }

\end{deluxetable}

\begin{deluxetable}{ccc}
\tablecaption{Orbital Properties \label{tab:orbital-properties}}
\tablehead{\colhead{Element} & \colhead{GJ 566} & \colhead{GJ 702} } 
\startdata
WDS                  & 14514+1906 AB        & 18055+0230 AB       \\
P (yr)               & ${152.292}_{-0.049}^{+0.049}$  & ${88.126}_{-0.010}^{+0.010}$ \\
a ('')               & ${4.9181}_{-0.0027}^{+0.0027}$  & ${4.54332}_{-0.00088}^{+0.00089}$ \\
i ($\degree$)        & ${140.611}_{-0.079}^{+0.079}$  & ${121.144}_{-0.028}^{+0.028}$ \\
$\Omega$ $(\degree)$ & ${166.79}_{-0.11}^{+0.11}$  & ${121.666}_{-0.032}^{+0.032}$ \\
$T_0$ (JD)           & ${2474014}_{-27}^{+27}$     & ${2477904.6}_{-8.4}^{+8.4}$ \\
e                    & ${0.51037}_{-0.00047}^{+0.00048}$   & ${0.50015}_{-0.00016}^{+0.00017}$ \\
$\omega$ $(\degree)$ & ${22.29}_{-0.17}^{+0.17}$  & ${193.154}_{-0.058}^{+0.058}$ \\
Mass ratio           & ${0.7447}_{-0.0053}^{+0.0054}$   & ${0.8292}_{-0.0057}^{+0.0058}$
\enddata
\tablecomments{Orbital elements for GJ 566 are from Section \ref{sec:gj566-gj702} and for GJ 702 are from 
\cite{li_century_2026}.}
\end{deluxetable}

\section{Interferometric Observations} \label{sec:interferometry}
 Georgia State University's Center for High Angular Resolution Astronomy (CHARA) Array is an infrared/optical interferometer located on Mt. Wilson, California. The CHARA Array is made up of six 1-meter telescopes that are spread out in a Y-configuration with distances between pairs of telescopes (referred to as ‘baselines’) ranging from 34-331 meters \citep{mcalister_first_2005, tenbrummelaar_first_2005}. For this study, we use the six-telescope MIRC-X and MYSTIC beam combiners \citep{anugu_mircx_2020, setterholm_mystic_2022} that operate simultaneously in the \emph{H-} (1.67 $\mu m$) and \emph{K-}band (2.13 $\mu m$), respectively, yielding formal spatial resolution limits as small as 0.52 mas in \emph{H-} and 0.66 mas in \emph{K-}band. Observations are obtained in one of two spectral modes, Grism 190 (G190) or Prism 50 (P50) for MIRC-X, and Grism 278 (G278) or Prism 49 (P49) for MYSTIC. The higher spectral modes (Grism) are best for bright targets ($H \lesssim 5.5$ mag) to avoid saturation. 

With projected separations of $4.6\arcsec- 4.9\arcsec$ and visual magnitude differences of $\sim$2 mags, observations of the secondary stars were sometimes challenging. Nominal CHARA observations require the tiptilt and adaptive optics systems that operate at visual wavelengths to lock onto the target.  At separations less than $\sim$6\arcsec, the tiptilt and adaptive optics of all six telescopes can struggle to stay locked on the secondary without switching to the brighter primary. This was especially problematic for GJ 566 B, since it has the largest magnitude difference and is the faintest star in the sample. This leads to data from different baselines being unusable and more scatter in the visibility data overall for GJ 566 B and GJ 702 B. Despite this, combining data from multiple nights provides broad visibility coverage, mitigating systematic errors. In total, we have three nights of observations for each of the four components taken between 2022-2025, with at least three bracketed observations each; a bracket here consists of a calibrator star (see Section \ref{subsec:calibrators}) and a science star observed consecutively. The data acquisition involves 5 minutes of interferometric data and 12 minutes of calibration sequence data on every star observed (target or calibrator). The calibration sequence data, often called 'shutters' in reduction scripts, 
consist of four sequences while off-fringe - the first records flux from each beam individually (beams), the second records flux with all beams open (foregrounds), the third records flux of the sky while away from the target (sky), and the fourth records background flux with all of the beams closed (backgrounds). 

An observing log listing the filter, spectral mode, number of brackets, and calibrators used for each night is in Table \ref{tab:obs-log}. 

\begin{deluxetable}{ccccccc}
\tablecaption{MIRC-X $\&$ MYSTIC Observing Log \label{tab:obs-log}}
\tablehead{\colhead{GJ} & \colhead{UT Date} & \colhead{Wavelength} & \colhead{Spectral Mode} & \colhead{Brackets} &  \colhead{Cal HD} } 
\startdata
566 A & 2022-05-14 & H      & G190     & 1 &  129974, 131528\\
      & 2022-06-01 & H$\&$K & G190/P49 & 1/1 &  125352, 137510\\
      & 2024-03-11 & H$\&$K & G190/P49 & 2/2 & 126141, 136923\\
566 B & 2023-06-05 & H$\&$K & G190/P49 & 1/1 & 127739, 136923 \\
      & 2024-03-11 & H$\&$K & G190/P49 & 1/1 & 126141, 136923\\
      & 2025-06-04 & H$\&$K & P50/P49 & 1/1 & 126141, 136923\\
\hline
702 A & 2024-05-30 & H$\&$K & G190/P49 & 1/1 & 167450 \\
      & 2024-07-18 & H$\&$K & P50/P49 & 1/1 & 161662 \\
      & 2024-07-19 & H$\&$K & G190/G278 & 1/1 &  161662, 170920  \\
702 B & 2023-06-05 & H$\&$K & G190/P49 & 1/1 & 160158, 170920 \\
      & 2024-07-18 & H$\&$K & P50/P49 & 1/1 & 161662 \\
      & 2024-07-19 & H\&K & G190/G278 & 1/1 & 161662, 170920 \\
\enddata
\end{deluxetable}

\subsection{Calibrator Star Selection} \label{subsec:calibrators}
Calibrator stars are unresolved point sources that are observed to calibrate visibility measurements that may be distorted by instrumental and atmospheric effects. In selecting calibrator stars, we avoid rapidly rotating stars ($v$sin$i>100$ $\mathrm{\,km\,s^{-1}}$) as they may be oblate, exhibit gravity darkening, and have an orientation-dependent
size.  We avoid candidate binaries by excluding stars with a Gaia DR3 re-normalized weight error (RUWE) value greater than 1.4 and stars with an image parameter determination fraction of multi-peak (IPDfmp) value less than 2 \citep{LL:LL-124, gaiacollaboration_gaia_2023}. Stars that exhibit proper motion anomalies in \cite{kervella_stellar_2019} are also excluded.

Calibrators are selected from the JMMC Stellar Diameter Catalog, version 2 \citep{bourges_jmmc_2014}. We choose calibrators that are within 10 degrees of the target, within 3 magnitudes of the target, and most of which are smaller than the resolution limit in \emph{H-}band ($\sim 0.52$ mas). Four calibrator stars are larger than this resolution limit and, therefore, are marginally resolved. However, the angular diameters of calibrators are accounted for in the calibration process, and since these 4 calibrators were only used for the 
3 stars larger than 1 mas (GJ 566 A, GJ 702 A, and GJ 702 B), the effects of their slightly larger size are expected to be minimal. The properties of the calibrator stars are listed in Table \ref{tab:cals}.

We strove to take observations in bracketed sequences with at least 2 calibrator stars, to create a cal1-target-cal2 bracket for each observation. This allows the calibrators to be cross-checked for consistency and to identify unknown binaries or a non-uniform surface brightness. The one exception to this strategy was the use of only 1 calibrator for two of the three nights when observing GJ 702 A and B. In these cases, we rely on closure phase data to validate the calibrator. A non-symmetric source will have closure phases that vary between 0 and 180$\degree$, and sinusoidal variation usually indicates a companion star.
All calibrators pass this closure phase test.  Additionally, the calibrator from UT 2024-07-18, HD 161662, was observed on a different night with the calibrator star HD 170920, allowing its size to be cross-checked. In summary, every target was observed on 3 different nights using multiple calibrators, and at least one calibrator satisfied all criteria.


\begin{deluxetable}{cccccccccc}

\tablecaption{Calibrators \label{tab:cals}}

\tablehead{\colhead{Calibrator} & \colhead{Spectral } & \colhead{V} & \colhead{H} & \colhead{K} & \colhead{$\theta_\mathrm{UD,H}$} & \colhead{$\theta_\mathrm{UD,K}$} & \colhead{RA} & \colhead{DEC} & \colhead{Target(s)} \\ 
\colhead{HD} &\colhead{Type} & \colhead{(mag)} & \colhead{(mag)} & \colhead{(mag)} & \colhead{(mas)} & \colhead{(mas)} & \colhead{(hh mm ss)} & \colhead{(dd mm ss)} & \colhead{GJ} } 

\startdata
129974 &  G5 & 6.91 & 4.67 & 4.51 & 0.563 $\pm$ 0.013 & 0.565  $\pm$ 0.013 & 14 45 22 & +10 35 56 & 566 A \\
131528 & K0 & 6.94 & 4.64 & 4.65 & 0.554 $\pm$ 0.052 & 0.556 $\pm$ 0.052 & 14 53 31 & +19 53 18 & 566 A \\
125352 & K0 & 7.03 & 4.62 & 4.39 & 0.639 $\pm$ 0.052 & 0.642 $\pm$ 0.052 & 14 18 21 & +21 18 15 & 566 A \\
137510 &  G0IV-V & 6.26 & 4.94 & 4.84 & 0.450 $\pm$ 0.011 & 0.452 $\pm$ 0.011 & 15 25 53 & +19 28 51 & 566 A \\
126141 &  F5V & 6.22 & 5.28 & 5.23 & 0.360  $\pm$ 0.008 & 0.361 $\pm$ 0.008 & 14 23 07 & +25 20 17 & 566 A \\
136923 &  G9V & 7.16 & 5.37 & 5.29 & 0.409 $\pm$ 0.009 & 0.411 $\pm$ 0.009 & 15 22 47 & +18 55 08 & 566 A, 566 B \\
127739 &  F2IV & 5.91 & 5.02 & 4.99 & 0.397 $\pm$ 0.010 & 0.398 $\pm$ 0.010 & 14 32 33 & +22 15 36 & 566 B \\
167450 &  K0III & 7.26 & 4.92 & 4.71 & 0.541 $\pm$ 0.012 & 0.543 $\pm$ 0.012 & 18 15 34 & -05 28 31 & 702 A \\
161662 &  K0III & 7.646 & 5.09 & 4.87 & 0.512 $\pm$ 0.012 & 0.514  $\pm$ 0.012 & 17 46 54 & -01 43 10 & 702 A, 702 B \\
170920 &  A5IV/V & 5.93 & 5.46 & 5.40 & 0.298  $\pm$ 0.008 & 0.299  $\pm$ 0.008 & 18 31 57 & -01 00 11 & 702 A, 702 B \\
160158 &  G5 & 8.33 & 5.50 & 5.36 & 0.399  $\pm$ 0.010 & 0.401  $\pm$ 0.010 & 17 38 08 & +06 12 50 & 702 B \\
\enddata

\tablecomments{Uniform disk diameters are from the JMMC Stellar Diameter Catalog \citep{bourges_jmmc_2014}. }

\end{deluxetable}

Observations are reduced using the standard MIRC-X Python reduction pipeline (version 1.4.0)\footnote{https://gitlab.chara.gsu.edu/lebouquj/mircx\_pipeline} with the number of coherent coadds set to 5, signal-to-noise threshold to 3, and integration time set to 30 seconds. The MYSTIC observations were reduced using the same reduction pipeline and parameters, but in this case sky frames are included for \emph{K-}band data. All data are calibrated with the MIRC IDL pipelines \citep{monnier_imaging_2007}, which yield calibrated squared visibilities that include recommended calibration errors of $5\%$ to each MIRC-X and MYSTIC visibility.

\section{Angular Diameters} \label{sec:angular-diameters}

In the simplest form, a star can be modeled as a uniform disk, but limb-darkening is needed to account for the true intensity distribution of a star. We solve for uniform disk $(\theta_\mathrm{UD})$ and limb-darkened disk $(\theta_\mathrm{LD})$ diameters by fitting the calibrated visibilities, $V^2$, to a visibility model. This model is described as a first-order Bessel function, $J_1(x)$, scaled by the wavelength dependent linear limb-darkening coefficient, $\mu_\lambda$:

\begin{equation} \label{eq:vis-eq}
\begin{split}
    V^2 = V^2_{\lambda,0}\biggl[ \Bigr(\frac{1-\mu_\lambda}{2}+\frac{\mu_\lambda}{3} \Bigl)^{-1} \\ \Bigr( (1 - \mu_\lambda) \cdot \frac{J_1(x)}{x} + \mu \sqrt{\frac{\pi}{2}} \frac{J_{3/2}(x)}{x^{3/2}} \Bigl) \biggr].
\end{split}
\end{equation}

 Here, \(  x=\frac{\pi B \theta}{\lambda}\), where B is the baseline, $\lambda$ is the wavelength of the observations, and $\theta$ is the angular diameter  \citep{hanburybrown_effects_1974}. A visibility model nominally converges towards 
 $V^2=1$ at short baselines. In practice, however, visibility calibration is typically uncertain at the few-percent level, introducing a multiplicative scaling error in the measured visibilities. This effect is most apparent at the shortest baselines, where the data systematically lie above or below $V^2=1$ by a comparable fractional amount.
 When the visibility curve is well sampled, a scale factor can be determined and applied to mitigate systematic errors in angular diameter measurements.
 \citep{caballero_detailed_2022, evans_orbit_2024, nunez_nearinfrared_2017}. 
  As a specific example, observations of GJ 566 B in 2025, obtained under marginal seeing conditions, likely included incoherent light from its 4.6\arcsec\ companion, GJ 566 A (see Section \ref{sec:interferometry}), resulting in a scale factor less than unity. We therefore determine a scale factor for each observation bracket, allowing us to account for variations between individual data sets and observing nights. This approach is particularly important for close binary systems, where small calibration offsets can bias the inferred visibilities and propagate directly into errors in the derived stellar parameters.

The uniform disk diameter depends on wavelength, being smaller at visible wavelengths by nearly by $4\%$ compared to infrared wavelengths; the difference between \emph{H} and \emph{K} is only about $0.4\%$. 
In our analysis, we first solve for the uniform disk diameter in both \emph{H-} and \emph{K-} bands separately by setting $\mu_\lambda=0$. These values are reported in Table \ref{tab:diameters}. We then solve for a combined limb-darkened diameter using limb-darkening coefficients at the respective wavelengths, tuned to the stellar properties. 
We adopt linear limb-darkening coefficients derived from PHOENIX atmosphere models using the least-squares method in 
\cite{ claret_gravity_2011}. We adopt coefficients calculated with solar metallicity (see Section \ref{sec:gj566-gj702}), a surface gravity (log\,$g$) value of $+4.5$ dex, and, initially, effective temperature from \cite{harada_setting_2024}. These temperatures are rounded to the nearest grid value 
\cite[200 K increments; ][]{claret_gravity_2011}. 
Since angular diameters combined with bolometric fluxes enable direct measures of effective temperatures (see Section \ref{sec:Stellar-prop}), we iterate our choice of limb-darkening coefficient based on the resulting temperature. The procedure converges after two or three iterations. 

The $V^2$ values are then modeled as a uniform disk and as a limb-darkened disk using an MPFIT prescription that minimizes the $\chi^2$ \citep{markwardt_nonlinear_2009}, including the scale factor noted above. 
As recommended in the \textit{MIRC-X/MYSTIC Data Reduction User Manual}\footnote{See Section 10 of the MIRC-X/MYSTIC Data Reduction User Manual: https://chara.gsu.edu/instrumentation/mirc}, we applied the appropriate corrections to the wavelength scale depending on the year the data was taken. 

\begin{figure*}[h]
     \centering
     \begin{subfigure}{\columnwidth}
         \centering
         \includegraphics[width=\linewidth]{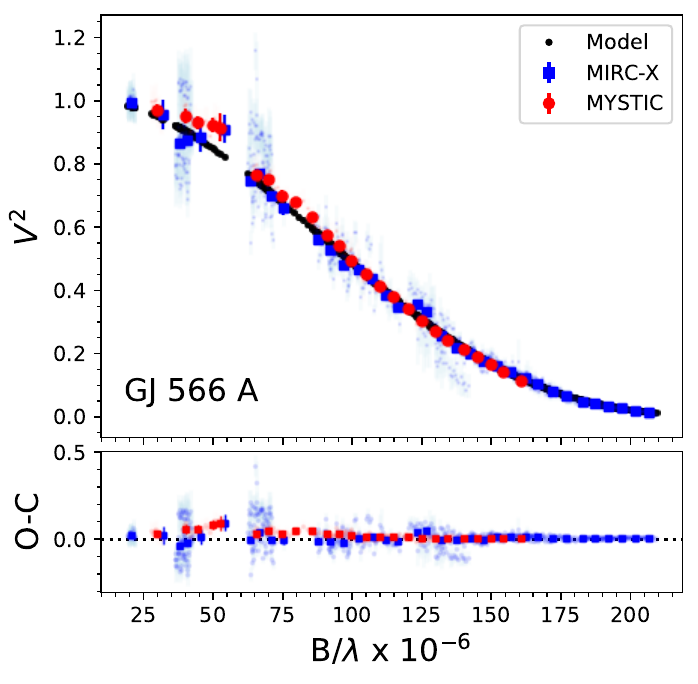}
         \caption{ }
         \label{fig:gj566a-vis}
     \end{subfigure}
     \begin{subfigure}{\columnwidth}
         \centering
         \includegraphics[width=\linewidth]{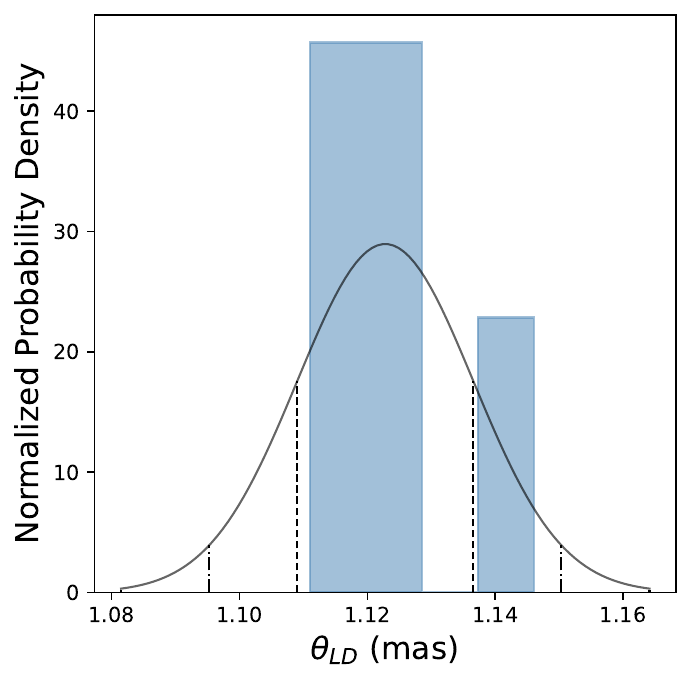}
         \caption{ }
         \label{fig:gj566a-hist}
     \end{subfigure}
     \caption{
     (a) MIRC-X (small light blue) and MYSTIC (small light red) squared visibility data vs. spatial frequency for GJ 566 A. The visibility data are binned with equal spatial-frequency spacing, and the inverse-variance-weighted average is shown as large blue squares and large red circles. The best-fit model (to the unbinned data) is shown as black circles, and the residuals (O-C) are plotted below the best fit. (b) The histogram of all individual MIRC-X and MYSTIC limb-darkened diameter measurements is fitted with a Gaussian distribution. Limits at 1$\sigma$ and 2$\sigma$ are denoted by dashed and dashed-dotted lines, respectively.
     }
     \label{fig:gj566a-vis-hist}
\end{figure*}

\begin{figure*}
     \centering
     \begin{subfigure}{\columnwidth}
         \centering
         \includegraphics[width=\linewidth]{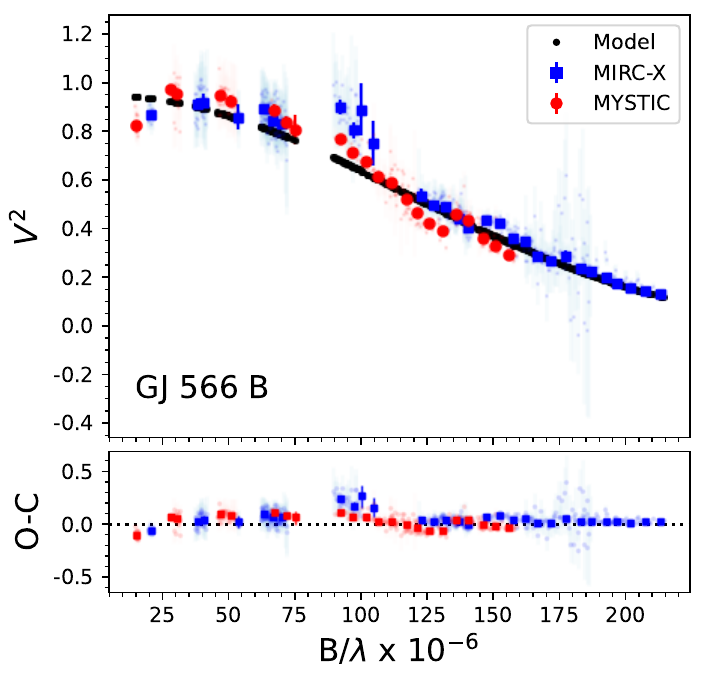}
         \caption{ }
         \label{fig:gj566b-vis}
     \end{subfigure}
     \begin{subfigure}{\columnwidth}
         \centering
         \includegraphics[width=\linewidth]{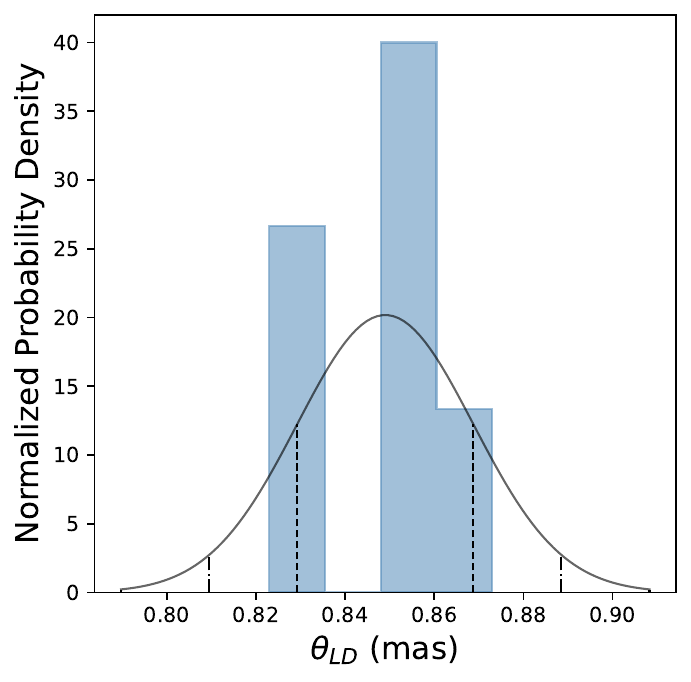}
         \caption{}
         \label{fig:gj566b-hist}
     \end{subfigure}
     \caption{(a)  MIRC-X (small light blue) and MYSTIC (small light red) squared visibility data vs. spatial frequency for GJ 566 B. The visibility data are binned with equal spatial-frequency spacing, and the inverse-variance-weighted average is shown as large blue squares and large red circles. The best-fit model (to the unbinned data) is shown as black circles, and the residuals (O-C) are plotted below the best fit. (b)  The histogram of all individual MIRC-X and MYSTIC limb-darkened diameter measurements is fitted with a Gaussian distribution. Limits at 1$\sigma$ and 2$\sigma$ are denoted by dashed and dashed-dotted lines, respectively. }
     \label{fig:gj566b-vis-hist}
\end{figure*}

\begin{figure*}
     \centering
     \begin{subfigure}{\columnwidth}
         \centering
         \includegraphics[width=\linewidth]{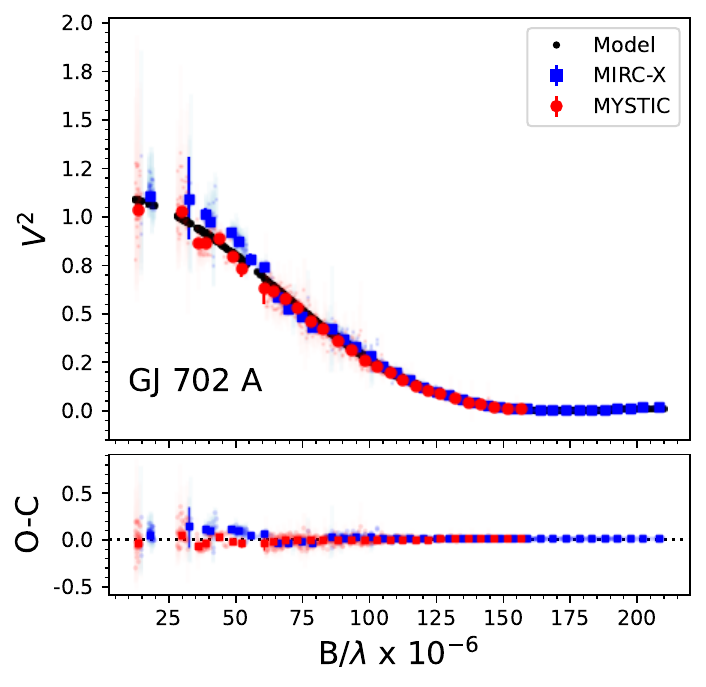}
         \caption{  }
        \label{fig:gj702a-vis}

     \end{subfigure}
     \begin{subfigure}{\columnwidth}
         \centering
         \includegraphics[width=\linewidth]{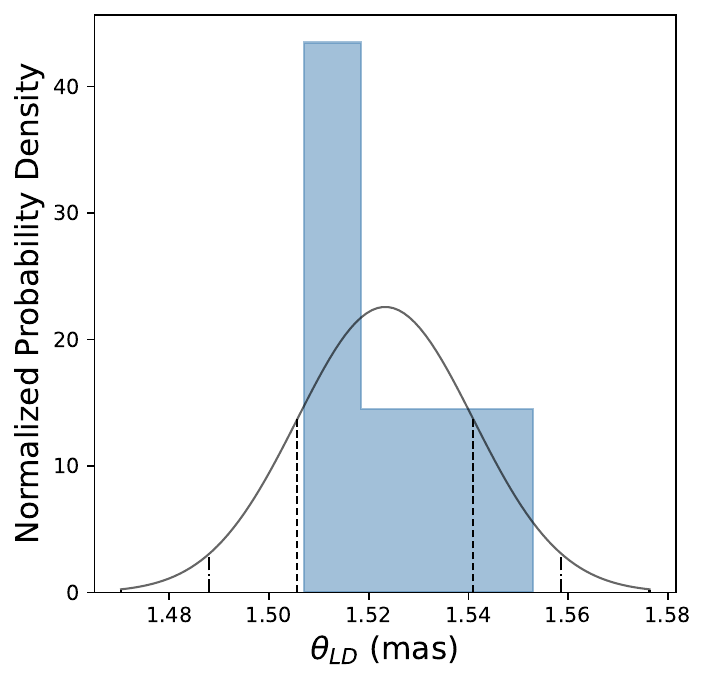}
         \caption{}
         \label{fig:gj702a-hist}
     \end{subfigure}
     \caption{ MIRC-X (small light blue) and MYSTIC (small light red) squared visibility data vs. spatial frequency for GJ 702 A. The visibility data are binned with equal spatial-frequency spacing, and the inverse-variance-weighted average is shown as large blue squares and large red circles. The best-fit model (to the unbinned data) is shown as black circles, and the residuals (O-C) are plotted below the best fit.(b) The histogram of all individual MIRC-X and MYSTIC limb-darkened diameter measurements is fitted with a Gaussian distribution. Limits at 1$\sigma$ and 2$\sigma$ are denoted by dashed and dashed-dotted lines, respectively.  }
     \label{fig:gj702a-vis-hist}
\end{figure*}

\begin{figure*}
     \centering
     \begin{subfigure}{\columnwidth}
         \centering
         \includegraphics[width=\linewidth]{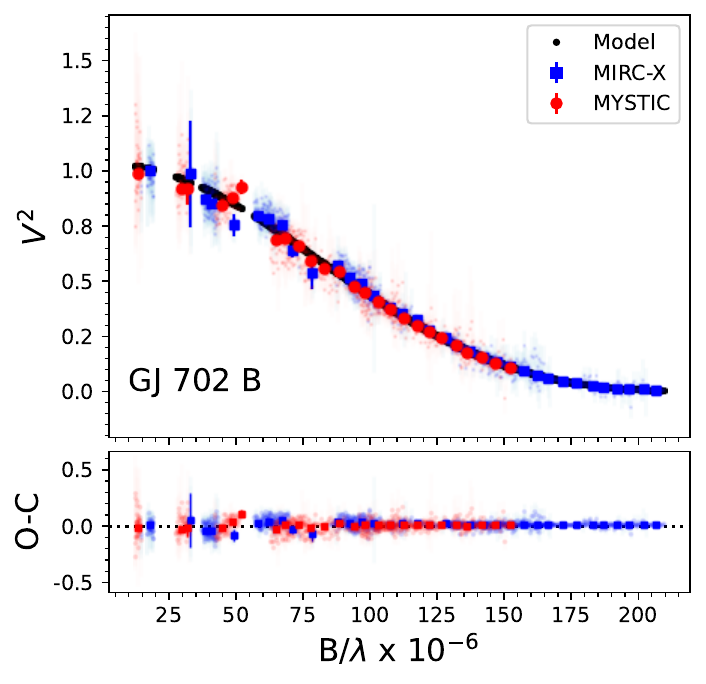}
         \caption{ }
         \label{fig:gj702b-vis}
     \end{subfigure}
     \begin{subfigure}{\columnwidth}
         \centering
         \includegraphics[width=\linewidth]{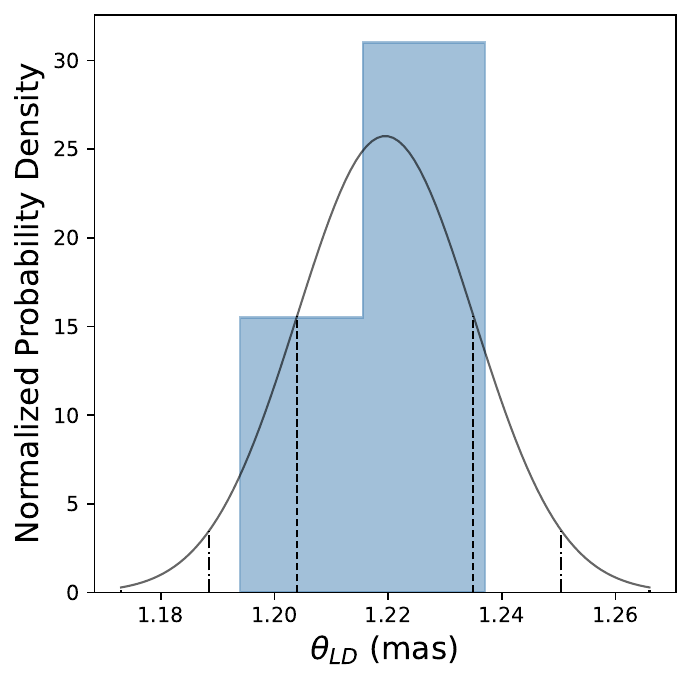}
         \caption{}
         \label{fig:gj702b-hist}
     \end{subfigure}
     \caption{ MIRC-X (small light blue) and MYSTIC (small light red) squared visibility data vs. spatial frequency for GJ 702 B. The visibility data are binned with equal spatial-frequency spacing, and the inverse-variance-weighted average is shown as large blue squares and large red circles. The best-fit model (to the unbinned data) is shown as black circles, and the residuals (O-C) are plotted below the best fit. (b) The histogram of all individual MIRC-X and MYSTIC limb-darkened diameter measurements is fitted with a Gaussian distribution. Limits at 1$\sigma$ and 2$\sigma$ are denoted by dashed and dashed-dotted lines, respectively.}
     \label{fig:gj702b-vis-hist}
\end{figure*}

The MIRC-X and MYSTIC $V^2$ data, along with the best-fit limb-darkened angular diameter model, are illustrated in Figures \ref{fig:gj566a-vis}-\ref{fig:gj702b-vis}. The Figures also show the data averaged in bins of equal spatial frequency with inverse-variance weighting. This binned data is for illustration only, and is not used in the best fit analysis.

The dominant uncertainty in our angular diameter measurements stems from unknown systematic errors that occur using two different instruments and from night-to-night variations. To account for this, we measure separate uniform and limb-darkened diameters for MIRC-X and MYSTIC for each night, yielding 6-7 independent uniform disk and limb-darkened measurements for each star. Each measurement also has its own scale factor. Figures \ref{fig:gj566a-hist}$-$\ref{fig:gj702b-hist} show the distributions of limb-darkened diameter measurements; the distributions of uniform disk measurements and scale factors are similar. A Gaussian function is fit to each distribution to determine 1$\sigma$ systematic uncertainties. We add the statistical uncertainties provided by the MPFIT routine and these systematic uncertainties in quadrature to determine the final uncertainties for the uniform and limb-darkened diameters. 

In Table \ref{tab:diameters} we report uniform disk diameters measured from each instrument, $\theta_\mathrm{UD}$, the limb-darkening coefficients, $\mu_\lambda$, the combined MIRC-X and MYSTIC diameters, $\theta_\mathrm{LD, HK}$, the average of scale factors applied, $V^2_\mathrm{\lambda,0}$ and the dispersion of scale factors, calculated as the standard deviation of factors used for that star. The last columns provide the reduced $\chi^2$ values of the best fit $\theta_\mathrm{LD}$ and the percent uncertainties in the angular diameters. The percent uncertainties range from 1.2-1.3\% for 3 of the 4 stars, with the fourth star (GJ 566 B), the faintest of the sample, having a percent uncertainty of 2.4\% (Table \ref{tab:diameters}).


\begin{deluxetable*}{cccccccccc}

\tablecaption{Angular Diameters\label{tab:diameters}}

\tablehead{\colhead{GJ} & \colhead{N. of } & \colhead{$\theta_\mathrm{UD,H} \pm \sigma$} & \colhead{$\theta_\mathrm{UD,K} \pm \sigma$} & \colhead{$\mu_\mathrm{H,K}$} & \colhead{$\theta_\mathrm{LD,HK} \pm \sigma$} & \colhead{Mean} & \colhead{Disp.} & \colhead{Reduced} & \colhead{$\theta_\mathrm{LD}$} \\ 
\colhead{Name} & \colhead{brackets} & \colhead{(mas)} & \colhead{(mas)} & \colhead{} & \colhead{(mas)} & \colhead{$V^2_\mathrm{\lambda,0}$} & \colhead{$V^2_\mathrm{\lambda,0}$} & \colhead{ $\chi^2$} & \colhead{$\%$ err} } 
\startdata
566 A & 5 & $1.083\pm0.012$ & $1.092\pm0.008$ & 0.3402, 0.2860 & $1.121\pm0.014$ & 1.01 & 0.04 & 2.24 & 1.3 \\
566 B & 6 & $0.814\pm0.023$ & $0.861\pm0.019$ & 0.4233, 0.3608 & $0.849\pm0.020$ & 0.95 & 0.08 & 2.29 & 2.4 \\
702 A & 6 & $1.472\pm0.008$ & $1.465\pm0.025$ & 0.3525, 0.2969 & $1.523\pm0.018$ & 1.10 & 0.11 & 1.06 & 1.2 \\
702 B & 6 & $1.170\pm0.012$ & $1.175\pm0.018$ & 0.4233, 0.3608 & $1.217\pm0.016$ & 1.03 & 0.05 & 1.00 & 1.3 \\
\enddata
\end{deluxetable*}
\subsection{Comparisons with Previous Angular Diameter Measurements}\label{subsec:comp-previous-angdia}

We assemble previous angular diameter measurements of GJ 702 A, GJ 702 B, and GJ 566 A, for comparison with new measurements in Table \ref{tab:diameter-comps}; GJ 566 B has no previous angular diameter measurements. All three have been observed using CHARA's Classic two-telescope beam combiner that operates in the infrared \citep[either \emph{H-} or \emph{K-} band;][]{tenbrummelaar_classic_2013} and GJ 566 A has been observed using CHARA's PAVO two-telescope beam combiner that operates in the visible \citep[\emph{R-}band;][]{ireland_sensitive_2008}. The Classic measurements of GJ 702 A and GJ 702 B agree to within 1\% of our measurements, while GJ 566 A is larger by 6.5\%. However, the new measurement is essentially identical ($<1\sigma$ difference) to the PAVO measurement of GJ 566 A. 
The shorter wavelength, higher resolution PAVO data relies on spectrally dispersed fringe measurements that mitigate broadband calibration systematics inherent to Classic’s undispersed design \citep{ireland_sensitive_2008, white_interferometric_2018}. 
MIRC-X and MYSTIC also utilize a spectrally dispersed design and further benefit from the use of all 6 telescopes, providing up to 15 interferometric baselines and broad, well-sampled coverage of the visibility function (e.g, Figure \ref{fig:gj566a-vis}). This enables the identification and correction of systematic errors (e.g., the scaling coefficient, $V^2_\mathrm{\lambda,0}$, in equation \ref{eq:vis-eq}) that would be difficult to recognize in single baseline observations, making the measurements more robust. We suggest that the slightly larger Classic measurement of GJ 566 A is spurious, and perhaps caused by incoherent light from GJ 566 B. The otherwise excellent agreement with previous measurements provides validation that the measured diameters are accurate to $\le 1.3$\% for GJ 566 A, GJ 702 A and GJ 702 B, and 2.4\% for GJ 566 B.

\begin{deluxetable*}{cccccc}
\tablecaption{Angular Diameter Comparisons\label{tab:diameter-comps}}
\tablehead{\colhead{GJ} & \colhead{$\theta_\mathrm{LD} \pm \sigma$} & \colhead{$\theta_\mathrm{LD}$} & \colhead{Instrument} & \colhead{Reference} \\
\colhead{Name} & \colhead{(mas)} & \colhead{$\%$ diff} &\colhead{Name} & \colhead{} }
\startdata
566 A & $1.121 \pm 0.014$ &  & MIRC-X/MYSTIC & This work \\
      & $1.124 \pm 0.009$ & 0.3 & PAVO &  \cite{karovicova_fundamental_2022}\\
      & $1.196 \pm 0.014$ & 6.5 & CLASSIC  &  \cite{boyajian_stellar_2012a}\\
702 A & $1.523 \pm 0.018$ &  & MIRC-X/MYSTIC & This work \\ 
      & $1.515 \pm 0.005$ & 0.5  & CLASSIC  & \cite{boyajian_stellar_2012b}\\
702 B & $1.217 \pm 0.016$ &  & MIRC-X/MYSTIC & This work \\
      & $1.221 \pm 0.015$ & 0.3 & CLASSIC  & \cite{boyajian_stellar_2012b} \\
\enddata
\end{deluxetable*}


\begin{deluxetable*}{cccccccc}

\tablecaption{Fundamental Properties \label{tab:fund.prop}}
\tablehead{
\colhead{GJ} & \colhead{Spectral} & \colhead{Mass} & \colhead{$L_\mathrm{BOL}$} & \colhead{$F_\mathrm{BOL}$} & \colhead{R} & \colhead{log $g$} & \colhead{$T_\mathrm{EFF}$} \\
\colhead{Name} & \colhead{Type} & \colhead{($M_\odot$)} & \colhead{($L_\odot$)} & \colhead{($10^{-8}\,\mathrm{erg\,cm^{-2}\,s^{-1}}$)} & \colhead{($R_\odot$)} & \colhead{(cm s$^{-2}$)} & \colhead{(K)} 
}
\startdata
566 A & G8V & $0.9056\pm0.0039^{a}$ & $0.555\pm0.016$ & $38.883\pm1.107$ & $0.814\pm0.010$ & $4.574\pm0.011$ & $5520\pm52$ \\
566 B & K4V & $0.6744\pm0.0034^{a}$ & $0.127\pm0.006$ & $8.944\pm0.397$ & $0.616\pm0.015$ & $4.688\pm0.021$ & $4392\pm71$ \\
702 A & K0V & $0.8827\pm0.0044^{b}$ & $0.514\pm0.020$ & $62.763\pm2.390$ & $0.837\pm0.010$ & $4.542\pm0.014$ & $5339\pm60$ \\
702 B & K5V & $0.7319\pm0.0031^{b}$ & $0.157\pm0.010$ & $19.317\pm1.219$ & $0.668\pm0.009$ & $4.652\pm0.013$ & $4447\pm76$ \\
\enddata 
\tablerefs{
Spectral Types: \cite{gliese_nearby1995, gray_contributions_2003}, a: Giovinazzi et al. 2026 (in prep); b: \cite{li_century_2026},
bolometric luminosities and fluxes: \cite{harada_setting_2024}. The radii, surface gravity, and effective temperatures are this work. 
}

\end{deluxetable*}

\section{Stellar Properties} \label{sec:Stellar-prop}

We use the limb-darkened angular diameters in combination with parallax measurements, $\varpi$, from Gaia DR3 (see Table \ref{tab:stellar-properties}) to directly determine stellar radii, 

\begin{equation}\label{eq:radius}
    R = \frac{ \theta_\mathrm{LD}}{2\varpi}.
\end{equation}

The stellar radii and dynamical masses are used to calculate each star's surface gravity, 
 
 \begin{equation}
     g = \frac{GM}{R^2}.
 \end{equation}
Uncertainties in the stellar radii and surface gravities are determined via error propagation. We note that the determined surface gravities are within 4\% of the values used to obtain limb-darkening coefficients (see Section \ref{sec:angular-diameters}). Values are reported in Table \ref{tab:fund.prop}.

\cite{harada_setting_2024} calculates stellar bolometric luminosities for all 4 stars in this study using the open-source \texttt{ARIADNE} package \citep{vines_ariadne_2022}, which fits the photometric energy distributions of each component with synthetic atmosphere models. \texttt{ARIADNE} fits multiple model grids independently and then averages their posterior samples using Bayesian evidence as weights; the final best-fit luminosities and associated uncertainties are derived from this weighted posterior distribution. For completeness, we include the photometry used in these fits along with the best-fit energy distributions in Appendix \ref{app:phot}. We note that Gaia saturation corrections \citep{riello_gaia_2021} were applied to Gaia G magnitudes. The resulting luminosities are determined to better than 6.3\% precision and are listed in Table \ref{tab:fund.prop}. The \texttt{ARIADNE} posteriors for bolometric luminosity and parallax measurements are used to calculate the bolometric flux of each component as

\begin{equation} \label{eq:Fbol}
F_\mathrm{BOL} = \frac{L_\mathrm{BOL}}{4\pi d^2}.
\end{equation}

Uncertainties in the bolometric flux are from half the difference between the 84th and 16th percentiles of the bolometric flux posterior distributions and treated as $1\sigma$ errors. We note that the larger uncertainties in luminosities and bolometric fluxes arise from spatially resolved photometry being limited to optical measurements, a limitation common in bright, close binary systems. 

Finally, the angular diameters and bolometric flux values are used to determine effective temperatures,  

\begin{equation}\label{eq:teff}
    T_\mathrm{EFF} = \Bigl( \frac{4 F_\mathrm{BOL}}{\sigma  \theta_\mathrm{LD}^2} \Bigr)^{1/4},
\end{equation}

with uncertainties determined via error propagation. The determined stellar properties have precisions that are less than 2.4\% for radii, less than 0.4\% for log g, less than 6.3\% for bolometric flux, and less than 1.7\% for effective temperatures. 
Such precisions are better than typical spectroscopic estimates \citep[e.g,][]{valenti_spectroscopic_2005} and less subject to systematic uncertainties, highlighting the power of combining interferometrically determined radii, Gaia DR3 parallaxes, and bolometric fluxes. All of the determined stellar properties are listed for each component in Table \ref{tab:fund.prop}.

\section{Comparisons with Stellar Evolution Models} \label{sec:model-comp}

The precisely determined stellar masses, radii, and effective temperatures of the components of GJ 566 and GJ 702 provide a powerful tool for testing the predictions of stellar evolution models. We consider four evolutionary models for this test, including 
BHAC15 \citep{baraffe_new_2015}, Dartmouth \citep{dotter_dartmouth_2008}, MIST \citep{choi_mesa_2016}, and PARSEC v2.0 \citep{costa_mixing_2019, costa_multiple_2019, nguyen_parsec_2022}. 
We use the solar metallicity versions of these models, consistent with the metallicity values measured for these stars (Section \ref{sec:gj566-gj702}). We note that the models adopt slightly different values for solar metallicity: Z = 0.0153 (BHAC15), Z = 0.01885 (Dartmouth), and Z = 0.0142 (MIST and PARSEC). For testing purposes, the authors of the PARSEC and BHAC15 models kindly provided mass tracks sampled in 0.02 M$_\odot$ increments, between 0.6 M$_\odot$ and 1.0 M$_\sun$, which is finer than the available original model release\footnote{The finer grid of BHAC15 models is available at https://perso.ens-lyon.fr/isabelle.baraffe/BHAC15dir/.}. 
This sampling is consistent with that of the Dartmouth models and with MIST models generated with the user-selected evolutionary track interpolator.
We note that the PARSEC tracks with masses $\geq0.76\,M_\odot$ are computed with the PARSEC v2.0 code

\citep[see][]{nguyen_parsec_2022, nguyen_parsec_2025}, while tracks with masses $<0.76\,M_\odot$ are computed using the previous PARSEC code \citep[][]{chen_improving_2014}. A visual comparison of tracks from the two models at the transition mass shows them to be indistinguishable.

Figures \ref{fig:gj566--modcomp} and \ref{fig:gj702-modcomp} illustrate the measured radii and effective temperatures of GJ 566 A and B and GJ 702 A and B. These are compared to evolutionary mass tracks consistent with their dynamical masses. Since the B components in both systems have dynamical masses between the available grids, we create mass tracks by linearly interpolating between the grid; these are used for an illustrative comparison only.
The exact mass and age predicted by each evolutionary model for each star are determined via a least-squares analysis similar to that of \citet{jorgensen_lindegren2005} and \citet{ligi_radii2016}, but implemented differently. 
The measured radii and effective temperatures are compared with the predictions of each mass track in steps of 1 million years.  When normalized by the uncertainties, this provides a likelihood at each mass and age that accounts for not only the star's proximity to the mass track, but the temporal evolution along the mass track. When summed, these give probability distributions in mass and age. Along the main sequence, the probability distributions for mass are nearly Gaussian-like, and consequently, we fit Gaussian functions to these distributions to determine the predicted stellar masses and 1$\sigma$ uncertainties. The age distributions have multiple peaks because of the quantized mass tracks; for each, the largest peak stems from the mass track evolving closest to the measured radius and effective temperature, and additional peaks stem from mass tracks just above or below this in mass. In this case, a cumulative distribution function is used to find the most probable age, and $1\sigma$ uncertainties are set at the 16th and 84th percentiles of this distribution. The predicted masses and ages for all 4 models are reported in Table \ref{tab:566-mod-masses-ages} and \ref{tab:702-mod-masses-ages}. The precisely determined stellar radii and effective temperatures translate to predicted mass uncertainties that range from 1.1\% to 3.2\%, enabling detailed comparison with dynamical masses. We note that a fine grid of mass tracks (e.g., $\lesssim 0.02$ M$_\odot$) is needed to conduct this analysis without significant interpolation of the evolutionary models.

\begin{figure*}
    \centering
    \includegraphics[scale=0.7]{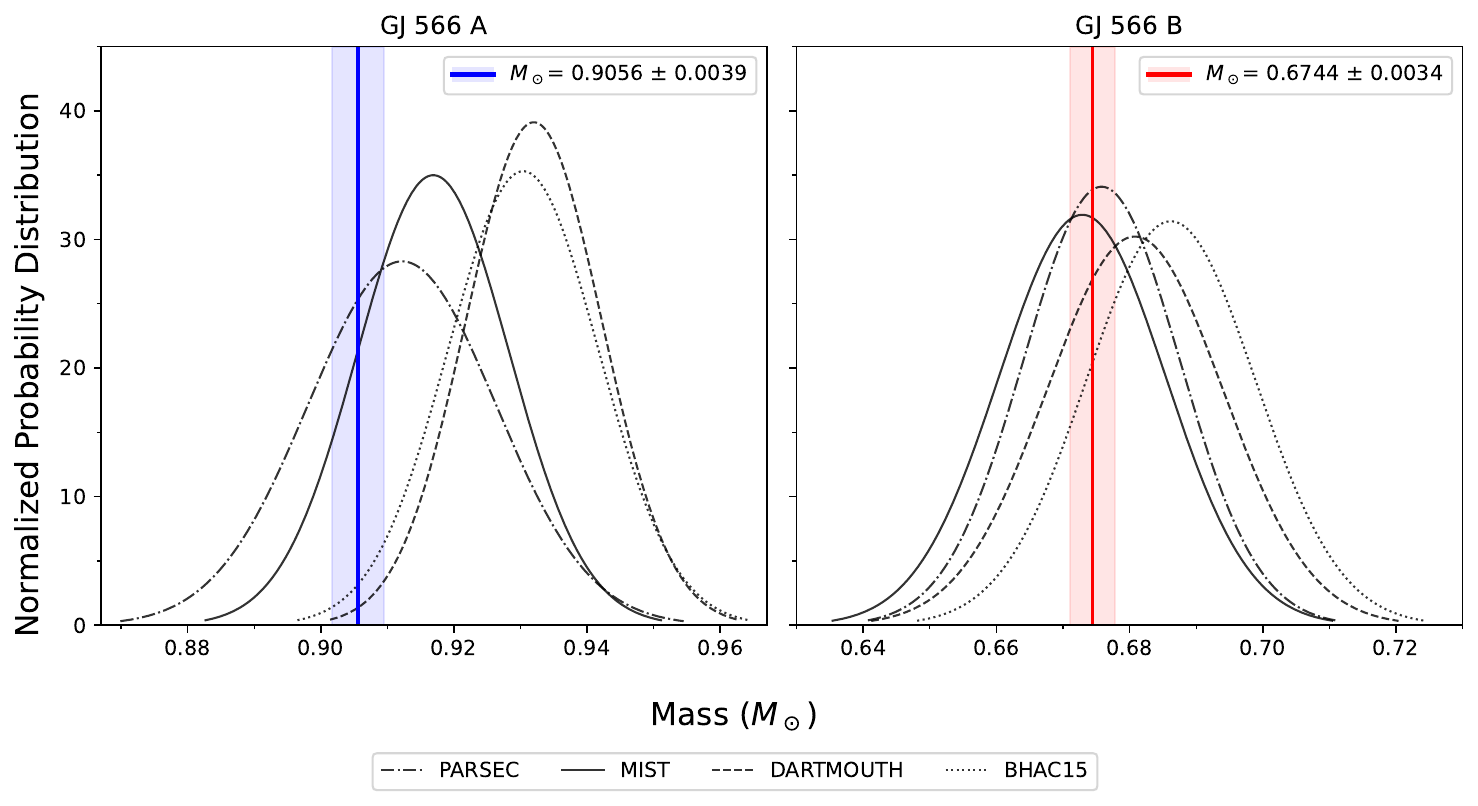}
    \caption{Mass probability distribution of four model predictions for GJ 566 A (left) and GJ 566 B (right). Each component's dynamical mass is denoted by a solid vertical line, with the shaded region representing the uncertainty.}
    \label{fig:gj566-massdist}
\end{figure*}

\begin{figure*}
    \centering
    \includegraphics[scale=0.7]{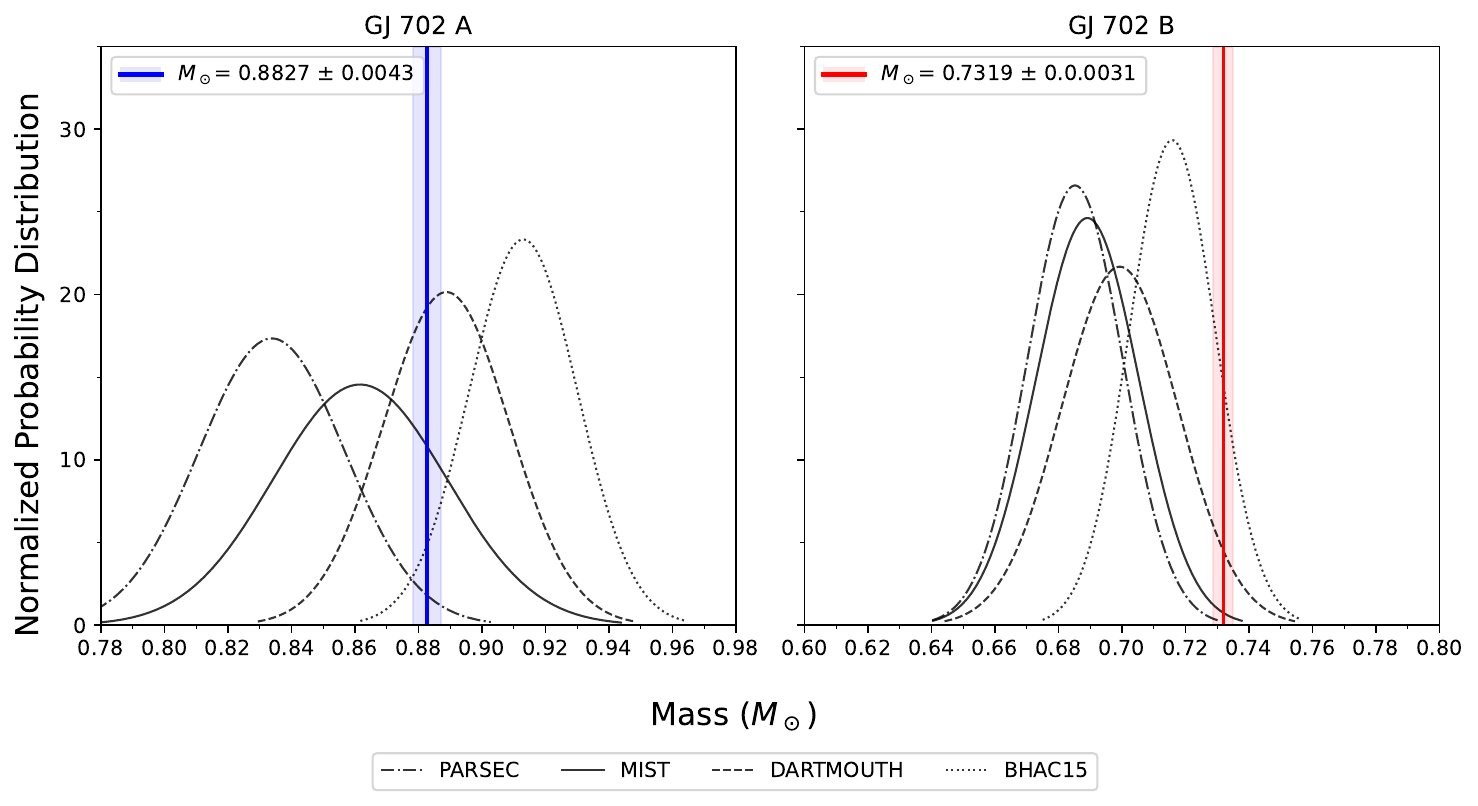}
    \caption{Mass probability distribution of four model predictions for GJ 702 A (left) and GJ 702 B (right). Each component's dynamical mass is denoted by a solid vertical line, with the shaded region representing the uncertainty.}
    \label{fig:gj702-massdist}
\end{figure*}

\begin{figure*}
    \centering
    \includegraphics[scale=0.8]{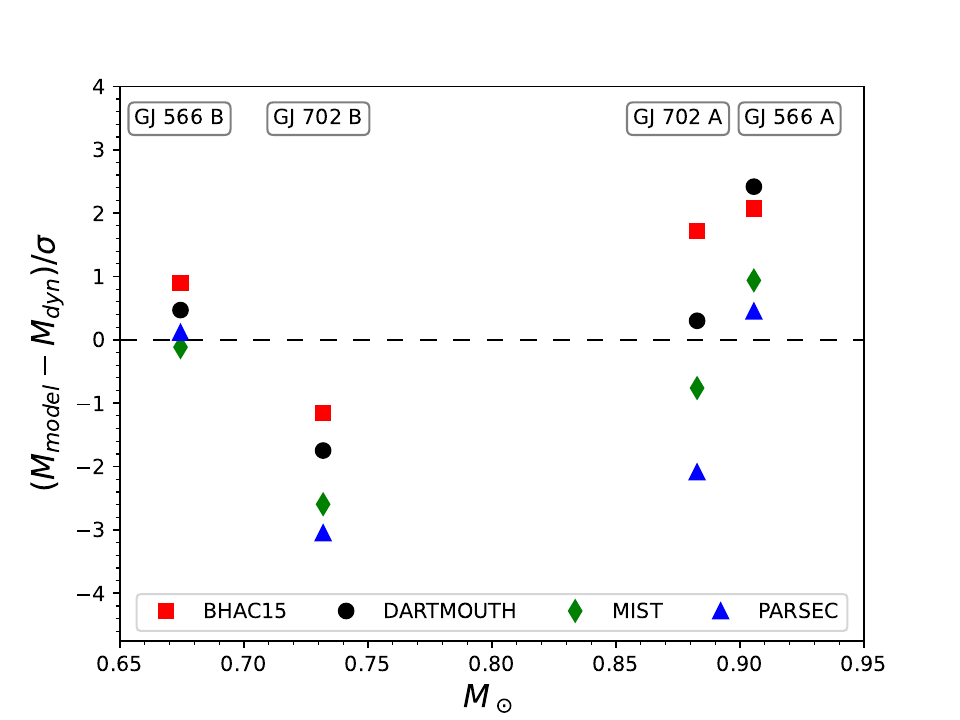}
\caption{Model predicted mass minus dynamical mass normalized by the combined uncertainty, is plotted versus 
dynamical mass for each star (labeled). Each symbol denotes a different stellar evolution model (BHAC15, Dartmouth, MIST, and PARSEC). The dashed horizontal line marks perfect agreement between model and dynamical masses. }
    \label{fig:model_mass_comp}
\end{figure*}

In Figures \ref{fig:gj566-massdist} and \ref{fig:gj702-massdist}, we illustrate the Gaussian mass probability distributions of each model for comparison with the dynamically measured values. Agreement is assessed quantitatively by calculating the difference between the model-predicted and dynamically measured masses, divided by the uncertainties added in quadrature. The result yields differences with model predictions in standard deviations, with positive values corresponding to model predictions being larger than dynamical values. These differences are tabulated in Table \ref{tab:model-residuals} and illustrated in Figure \ref{fig:model_mass_comp}.

Over the mass range of $0.6 - 1.0\,\mathrm{M_\odot}$ at solar metallicities, all 4 models predict masses that are reasonably consistent with dynamically measured values. GJ 566 A has model predictions that are larger by $0.45\sigma$ (PARSEC) to $+2.42\sigma$ (DARTMOUTH). However, its companion, GJ 566 B, exhibits the best agreement with model predictions, being within $1\sigma$ for all 4 models. For GJ 702 A, model predictions scatter about the dynamical value, ranging from $-2.09\sigma$ (PARSEC) to $+1.72\sigma$ (BCAH15), while values for GJ 702 B are all less than the dynamical value, ranging from $-1.15\sigma$ (BCAH15) to $-3.05\sigma$ (PARSEC). In summary, all model-predicted masses are within a few standard deviations ($<3.1\sigma$) of the dynamical values, and over all masses there is no systematic offset in model predictions. These results do not favor one model over another for predicting stellar mass over this mass range, at solar metallicity. 
Consequently, we suggest that the dispersion in stellar mass estimates derived from multiple evolutionary models provides a valuable indication of the global uncertainties inherent in model‑based stellar mass predictions.

The ages inferred for GJ 702 A range from 3.0 Gyr to 9.1 Gyr, with an average value of 5.7 Gyr. For GJ 702 B, the ages range from 5.3 Gyr to 8.6 Gyr, with an average of 6.6 Gyr; individual uncertainties are typically a few Gyr. The mean ages are consistent with previous independent age estimates of GJ 702 A using asteroseismology \citep[$6.0 \pm 1.0$ Gyr;][]{eggenberger_analysis_2008} and gyrochronology \citep[$7.0 \pm 0.3$ Gyr][]{colliercameron_mainsequence_2009}, assuming a period of 39.8 days (see Section \ref{sec:gj566-gj702}). The ages inferred for the GJ 566 system are younger, ranging from 0.4 Gyr to 1.3 Gyr for GJ 566 A, with an average value of 0.8 Gyr, and ranging from 1.4 Gyr to 1.6 Gyr for GJ 566 B, with an average value of 1.5 Gyr. Uncertainties are near 1 Gyr, and are larger for the lower mass, more slowly evolving star GJ 566 B. These ages are slightly larger than but consistent with the gyrochronology age of 389 Myr \citep{mamajek_improved_2008} and the age of 414 Myr for the Ursa Majoris Moving Group \citep{jones_UMa2015}, further strengthening its case as a bona fide member.
While the uncertainties in the model-predicted ages remain relatively large, their general agreement with previous age estimates and the consistent ages of the pairs, being coeval, provides further validation of the models. We provide averages of the model ages for each star in Tables \ref{tab:566-mod-masses-ages} and \ref{tab:702-mod-masses-ages}; uncertainties are set to the average uncertainty in age.

Altogether, these results demonstrate that the measured radii of these four stars are not inflated compared to model predictions. All four have been classified as chromospherically active, and the components of GJ 566 A and B have a relatively young age ($\lesssim 1$ Gyr). If chromospheric activity and/or youth is responsible for some radius inflation (see Section \ref{sec:intro}), these stars are not active enough or young enough for radius inflation to be manifested.

As an aside, we note that at 6.7 pc, GJ 566 A and B are the closest Moving Group or open cluster members with angular diameter measurements. The only Moving Group or open cluster members closer than GJ 566 are Luhman 16AB, a binary brown dwarf \citep[2.0 pc;][]{luhman_discovery2014}, and YZ CMi, an M5 flare star \citep[6.0 pc;][]{schofer_CARMENES2022}; both are members of the $\sim 500$ Myr Oceanus Moving Group \citep{gagne_oceanus2023}.

\section{Comparisons with Eclipsing Binaries}\label{sec:EBcomps}
To investigate radius inflation within this mass range more completely, we assemble
the components of EB stars with stellar masses that are identical to within 3\% of the dynamical masses of GJ 566 A (5 stars), GJ 566 B (5 stars), GJ 702 A (5 stars), and GJ 702 B (8 stars). The EB stars are assembled from \cite{southworth_debcat_2014}, \cite{morales_fundamental2022}, and \cite{torres_accurate_2010}.  We exclude stars that are identified as pre-main sequence \citep[e.g., V1174 Ori;][]{stassun_dynamical_2004}, and exclude stars with measured metallicities that are less than $-0.2$ dex or greater than $+0.2$ dex, if known.

All stars have dynamical masses that are determined to 2\% precision, radii to 4\% precision, and effective temperatures to 5\% precision. Three stars have masses that are consistent with both GJ 566 A and GJ 702 A (RT And B, HS Aur A, and HS Aur B); the sample of EB stars in total consists of 20 unique stars.
The spectral types (if known), orbital periods, dynamical masses, effective temperatures, stellar radii, and metallicities (if known) for these stars are listed in Table \ref{tab:EBs}. The spectral types with all 4 subsets vary by several subclasses, despite their similar masses.  At least 11 of these 20 stars have measured metallicities.

We assume solar metallicity for those without measured values, consistent with previous work \cite[e.g.,][]{torres_eclipsing_2006,lopez-morales_testing_2007, rozyczka_new_2010,
torres_accurate_2010,  helminiak_orbital_2011}.

\begin{figure*}
    \centering
    \includegraphics[scale=0.8]{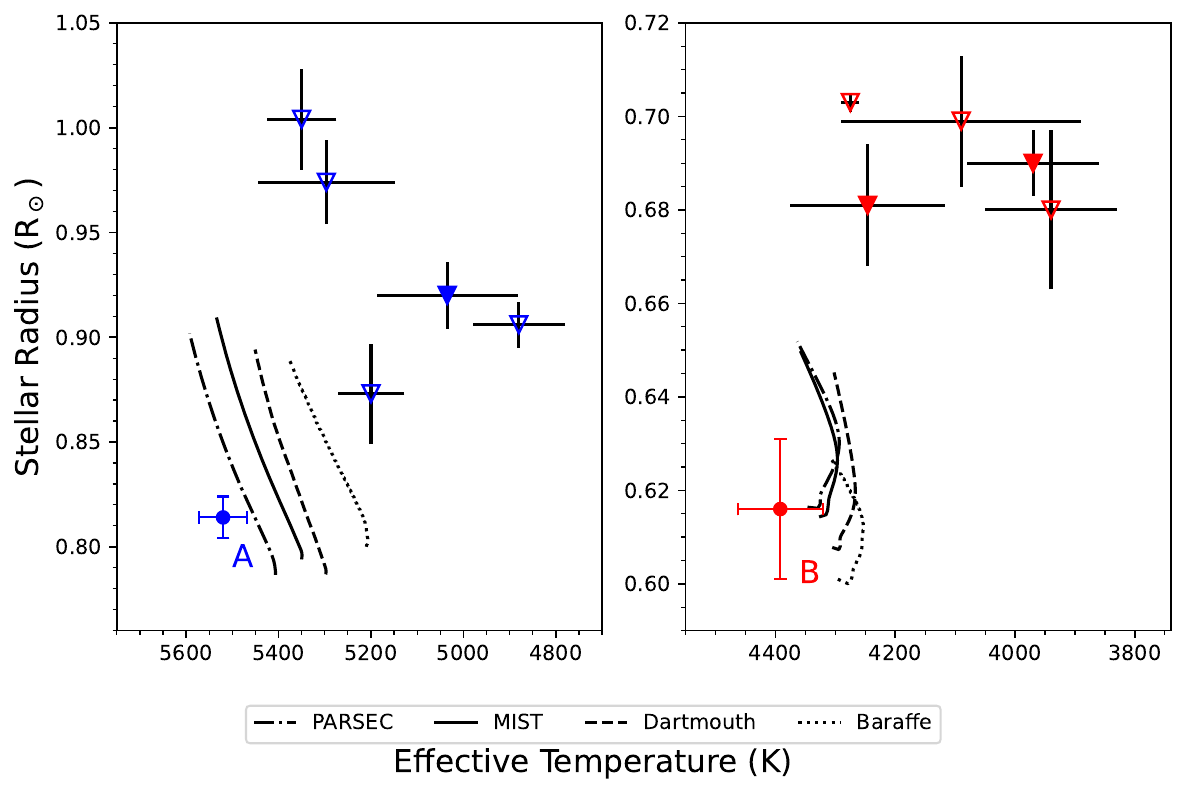}
    \caption{Mass tracks at 0.90 M$_\odot$ (left) and 0.67 M$_\odot$ (right) from four stellar evolution models are compared with the measured radii and temperatures of GJ 566 A (left; filled circle) and GJ 566 B (right; filled circle). All models are computed with solar metallicity and span from 0.1 - 7.0 Gyr. A sample of EB stars with masses identical to within 3\% and nearly solar metallicity (filled triangles) or unknown metallicity (unfilled triangles) are shown for comparison. GJ 566 A and B are generally consistent with model predictions, while most EB stars exhibit the classic signatures of radius inflation. }
    \label{fig:gj566--modcomp}
\end{figure*}

\begin{figure*}
    \centering
    \includegraphics[scale=0.8]{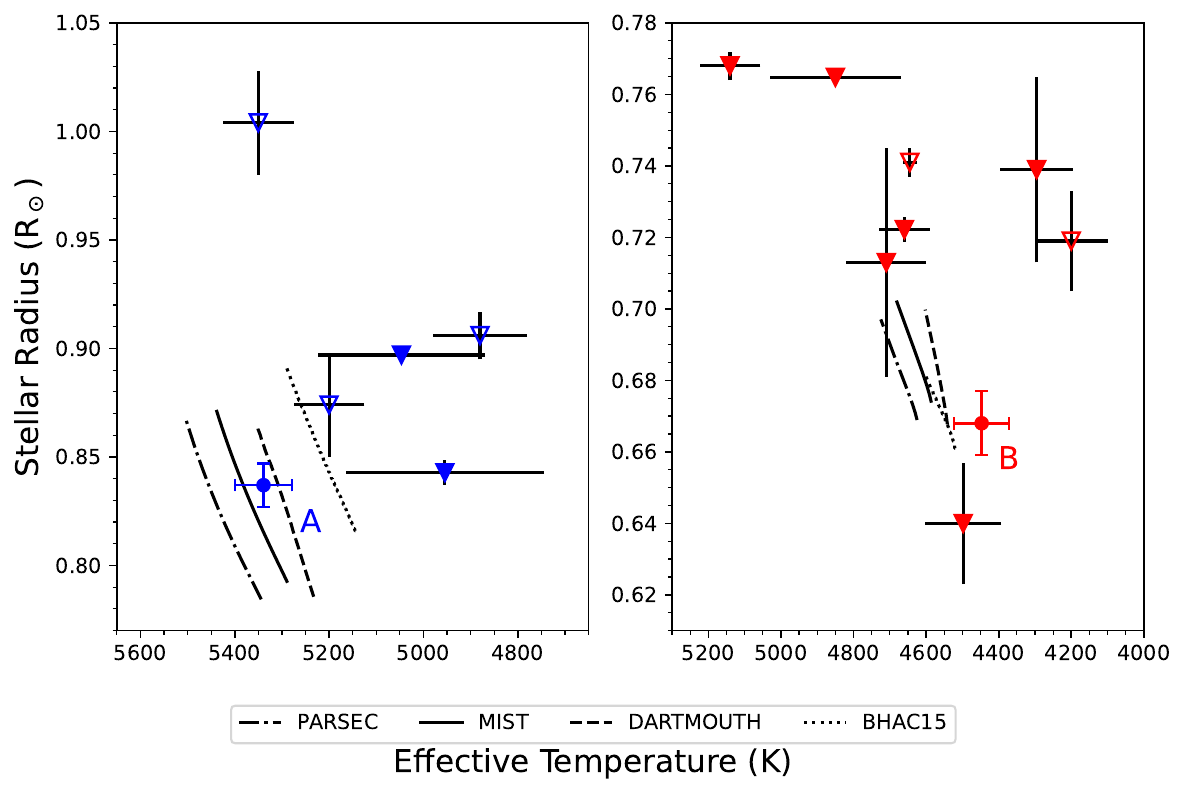}
    \caption{Mass tracks at 0.88 M$_\odot$ (left) and 0.73 M$_\odot$ (right) from four stellar evolution models are compared with the measured radii and temperatures GJ 702 A (left; filled circle) and GJ 702 B (right; filled circle). All models are computed with solar metallicity and span from 1 - 7 Gyrs. A sample of EB stars with masses identical to within 3\% and nearly solar metallicity (filled triangles) or unknown metallicity (unfilled triangles) are shown for comparison. GJ 702 A and B are generally consistent with model predictions, while most EB stars exhibit the classic signatures of radius inflation. One exception is TYC 6296-2012-1 B which is smaller than GJ 566 B. }
    \label{fig:gj702-modcomp}
\end{figure*}

The radii and effective temperatures of these EB stars are compared with those of GJ 566 A and B (Figure \ref{fig:gj566--modcomp}) and GJ 702 A and B (Figure \ref{fig:gj702-modcomp}), and the nearest mass tracks. EB stars with a known metallicity are distinguished from those with an unknown metallicity. The EB stars are systematically larger than the measured radii by several standard deviations in most cases. The EB stars are also, in most cases, colder than the effective temperatures by several standard deviations. The one exception is TYC 6296-2012-1 B \citep[M=0.733 M$_\odot$][]{torres_eclipsing_2019}, which is slightly smaller than GJ 702 B. Its relatively small size (compared with model predictions) is discussed in the discovery paper \citep{torres_eclipsing_2019}. Overall, most EB stars are larger and colder than this comparison sample and than model predictions. As a group, these stars exhibit the classic signature of radius inflation. The apparent inflation could be reconciled somewhat if the EB stars are more metal-rich than solar; an increase in metallicity to +0.25 dex will shift models to cooler temperatures by as much as $\sim$ 350 K and slightly higher luminosities. However, considering the subset of 12 stars with near-solar metallicity, the discrepancy remains.

\begin{figure*}
    \centering
    \includegraphics[scale=0.7]{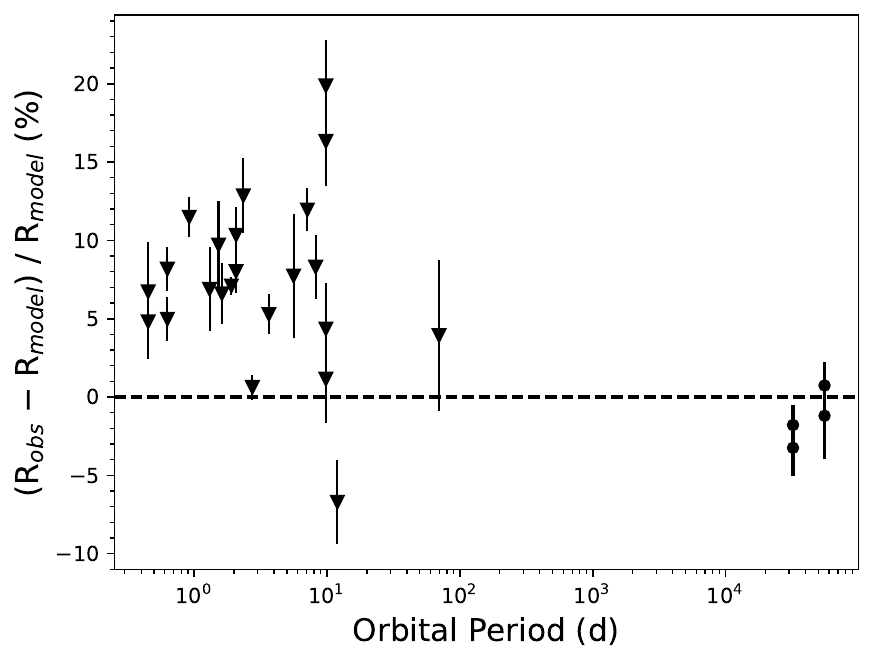}
    \caption{Percent differences between the observed radii and the average model radii for eclipsing binary stars (triangles) and the four stars in this study (circles) as a function of orbital period. The eclipsing binaries are compared to the mean radius predicted by four stellar evolution models (Baraffe, Dartmouth, MIST, and PARSEC; see Table \ref{tab:EBs}), assuming an age of 5 Gyr. The components of GJ 566 and GJ 702 are compared to the mean model radii at their own masses and ages at 1 Gyr and 6 Gyr, respectively (see Tables \ref{tab:566-mod-masses-ages} and \ref{tab:702-mod-masses-ages}). Error bars include the observational radius uncertainty and the dispersion among the model radii, added in quadrature. Most eclipsing binaries exhibit radii inflated relative to model predictions, whereas the long-period binaries GJ 566 and GJ 702 do not.}
    \label{fig:radii-period-comp}
\end{figure*}

Figure~\ref{fig:radii-period-comp} compares the fractional differences between observed stellar radii and the average radii predicted by four stellar evolution models as a function of orbital period. The eclipsing binaries are compared to the mean radius predicted by four stellar evolution models (Baraffe, Dartmouth, MIST, and PARSEC; see Table \ref{tab:EBs}), assuming an age of 5 Gyr. The components of GJ 566 and GJ 702 are compared to the mean model radii at their dynamical masses and ages at 1 Gyr and 6 Gyr (see Tables \ref{tab:566-mod-masses-ages} and \ref{tab:702-mod-masses-ages}), respectively.
Consistent with previous studies of low-mass eclipsing binaries, most of the eclipsing binaries in the sample exhibit radii larger than predicted by the models. However, within this relatively small comparison sample, there is no clear evidence that the degree of radius inflation depends on orbital period. In contrast, the components of GJ 566 and GJ 702 show no radius inflation relative to model expectations. All four stars are consistent with the model predictions to within $1-2\sigma$.

Perhaps most problematic for the EB stars is that the four sets do not have radii and temperatures that are consistent among themselves within the uncertainties, despite having masses that are nearly identical. Coupled with radius inflation that appears to be inherent to EB stars, the broad spread in values suggests caution when using EB stars to test stellar evolutionary models designed for single stars.




\begin{deluxetable*}{lcccc}
\tablecaption{Model Masses and Ages for GJ 566 \label{tab:566-mod-masses-ages}}
\tablehead{
\colhead{} &
\multicolumn{2}{c}{GJ 566 A} & 
\multicolumn{2}{c}{GJ 566 B} \\
\cline{2-5}
\colhead{Model} &
\colhead{Mass} & \colhead{Age} &
\colhead{Mass} & \colhead{Age} \\
\colhead{} &
\colhead{($M_\odot$)} & \colhead{(Gyr)} &
\colhead{($M_\odot$)} & \colhead{(Gyr)} 
}
\startdata
PARSEC    & $0.9122\pm0.0141$ & $1.34\pm1.02$ & $0.6758\pm0.0117$ & $1.36\pm1.17$ \\
MIST      & $0.9169\pm0.0114$ & $0.87\pm0.73$ & $0.6729\pm0.0125$ & $1.45\pm1.18$ \\
BCAH15    & $0.9304\pm0.0113$ & $0.43\pm0.40$ & $0.6862\pm0.0127$ & $1.61\pm1.21$ \\
DARTMOUTH & $0.9320\pm0.0102$ & $0.69\pm0.57$ & $0.6808\pm0.0132$ & $1.40\pm1.14$ \\
Average Age   & \nodata & $0.83\pm0.62$        & \nodata  & $1.46\pm0.18$ \\
\enddata
\tablecomments{The average age uncertainty is set to the average uncertainty in age.}
\end{deluxetable*}


\begin{deluxetable*}{lcccc}
\tablecaption{Model Masses and Ages for GJ 702\label{tab:702-mod-masses-ages}}
\tablehead{
\colhead{} &
\multicolumn{2}{c}{GJ 702 A} & 
\multicolumn{2}{c}{GJ 702 B} \\ 
\cline{2-5}
\colhead{Model} &
\colhead{Mass} & \colhead{Age} &
\colhead{Mass} & \colhead{Age} \\
\colhead{} &
\colhead{($M_\odot$)} & \colhead{(Gyr)} &
\colhead{($M_\odot$)} & \colhead{(Gyr)} 
}
\startdata
PARSEC    & $0.8338\pm0.0230$ & $9.08\pm2.47$ & $0.6852\pm0.0150$ & $7.76\pm3.93$ \\
MIST      & $0.8616\pm0.0274$ & $6.29\pm2.70$ & $0.6891\pm0.0162$ & $6.93\pm4.22$ \\
DARTMOUTH & $0.8888\pm0.0198$ & $2.96\pm2.34$ & $0.6993\pm0.0184$ & $5.27\pm3.23$ \\
BHAC15    & $0.9130\pm0.0171$ & $4.47\pm1.78$ & $0.7158\pm0.0136$ & $6.30\pm3.90$ \\
Average Age   & \nodata & $5.70\pm2.32$        & \nodata  & $6.57\pm3.82$ \\
\enddata
\tablecomments{The average age uncertainty is set to the average uncertainty in age.}
\end{deluxetable*}
\begin{deluxetable*}{ccccc}

\tablecaption{[Model Mass $-$ Dynamical Mass] in Standard Deviation Units \label{tab:model-residuals}}
\tablehead{\colhead{Model}  & \colhead{GJ 566 A} & \colhead{GJ 566 B} & \colhead{GJ 702 A} & \colhead{GJ 702 B} }
\startdata
BCAH15	    &2.075	&0.898	&1.716	&-1.154\\
DARTMOUTH	&2.418	&0.470	&0.301	&-1.747\\
MIST	    &0.938	&-0.116	&-0.760	&-2.595\\
PARSEC	    &0.451	&0.115	&-2.088	&-3.049\\
\enddata
\end{deluxetable*}


\begin{deluxetable*}{lccccccl}

\tablecaption{Eclipsing Binary Stars with Similar Dynamical Masses \label{tab:EBs}}
\tablehead{\colhead{Component} &
\colhead{Spectral} & 
\colhead{Period} & 
\colhead{Mass} & 
\colhead{Effective Temperature} &
\colhead{Radius} & 
\colhead{[M/H]} & 
\colhead{Ref} \\ 
\colhead{Name} & 
\colhead{Type} & 
\colhead{(days)} & 
\colhead{($M_\odot$)} &
\colhead{(K)} &
\colhead{($R_\odot$)} & 
\colhead{(dex)} & 
\colhead{}} 
\startdata
\multicolumn{8}{c}{EBs with Masses within $3\%$ of 0.9056 $M_\odot$ (GJ 566 A)} \\
HS Aur B          & K0V & 9.82  & $0.879\pm0.017$  & $5200\pm75$     & $0.874\pm0.024$ & ...         & 15 \\
HS Aur A          & G8V & 9.82  & $0.898\pm0.019$  & $5350\pm75$     & $1.004\pm0.024$ & ...         & 15 \\
RT And B          & K3V & 0.63  & $0.907\pm0.017$  & $4880\pm100$    & $0.906\pm0.011$ & ...         & 15 \\
EPIC 219511354 A  & K4V & 13.16 & $0.912\pm0.013$  & $5035\pm153$    & $0.920\pm0.016$ & $0.07\pm0.03$ & 17 \\
V1061 Cyg B       & G8V & 2.37  & $0.932\pm0.007$  & $5297\pm148$    & $0.974\pm0.02$  & ...         &  14 \\
\hline
\multicolumn{8}{c}{EBs with Masses within $3\%$ of 0.6744 $M_\odot$ (GJ 566 B)} \\
NGC 2204-S892 B       & late-KV & 0.45 & $0.662\pm0.005$ & $3940\pm110$ & $0.680\pm0.017$ & ...               & 11 \\ 
IM Vir B              & K7V     & 1.31 & $0.664\pm0.005$ & $4246\pm129$ & $0.681\pm0.013$ & $-0.10\pm0.25$   & 7 \\ 
ASAS J065134-2111.5 B & V       & 8.22 & $0.674\pm0.005$ & $3970\pm110$ & $0.690\pm0.007$ & $0.09\pm0.13$   & 4 \\ 
ASAS J082552-1622.8 B & M0V     & 1.53 & $0.687\pm0.005$ & $4090\pm200$ & $0.699\pm0.014$ & ...               & 3 \\ 
RXJ0239.1-1028 B      & V       & 2.07 & $0.693\pm0.006$ & $4275\pm15$  & $0.703\pm0.002$ & ...               & 6\\  
\hline
\multicolumn{8}{c}{EBs with Masses within $3\%$ of 0.8827 $M_\odot$ (GJ 702 A)} \\
V375 Cep B       & ...  & 1.91  & $0.871\pm0.008$ & $5047\pm177$ & $0.897\pm0.003$ & $0.09\pm0.05$    & 12\\
V362 Pav B       & K    & 2.75  & $0.877\pm0.005$ & $4955\pm209$ & $0.843\pm0.006$ & $0.02\pm0.15$    & 2\\
HS Aur B         & K0V  & 9.82  & $0.879\pm0.017$ & $5200\pm75$  & $0.874\pm0.024$ & ...              & 15 \\
HS Aur A         & G8V  & 9.82  & $0.898\pm0.019$ & $5350\pm75$  & $1.004\pm0.024$ & ...              & 15 \\
RT And B         & K3V  & 0.63  & $0.907\pm0.017$ & $4880\pm100$ & $0.906\pm0.011$ & ...              & 15 \\
\hline
\multicolumn{8}{c}{EBs with Masses within $3\%$ of 0.7319 $M_\odot$ (GJ 702 B)} \\
TIC 17303250 A    & K1V    & 7.12  & $0.714\pm0.004$ & $5140\pm83$  & $0.768\pm0.004$ & $-0.05\pm0.15$ & 10 \\
HD 135671 B       & ...    & 3.66  & $0.721\pm0.009$ & $4660\pm70$  & $0.722\pm0.003$ & $-0.18\pm0.04$ & 5\\
RXJ0239.1-1028 A  & K7     & 2.07  & $0.730\pm0.009$ & $4645\pm20$  & $0.741\pm0.004$ & ...             & 6\\
TYC 6296-2012-1 B & ...    & 11.99 & $0.733\pm0.005$ & $4498\pm105$ & $0.640\pm0.017$ & $0.07\pm0.03$  & 16 \\ 
NGC 2204-S892 A   & late-K & 0.45  & $0.733\pm0.005$ & $4200\pm100$ & $0.719\pm0.014$ & ...             & 11\\
V818 Tau B        & K4V    & 5.61  & $0.743\pm0.002$ & $4295\pm100$ & $0.739\pm0.026$ & $0.05\pm0.05$  & 1\\
WOCS 12009 B      & ...    & 69.73 & $0.748\pm0.005$ & $4710\pm110$ & $0.713\pm0.032$ & $0.03\pm0.04$  & 13\\
TIC 56913729 B    & K3V    & 0.46  & $0.753\pm0.005$ & $4850\pm180$ & $0.765\pm0.002$ & $-0.02\pm0.12$ & 9\\
\enddata
\tablerefs {(1) \cite{brogaard_properties_2021}; (2) \cite{graczyk_surface_2022}; (3) \cite{helminiak_orbital_2011}; (4) \cite{helminiak_orbital_2019}; (5) \cite{helminiak_highresolution_2024}; (6) \cite{lopez-morales_testing_2007}; (7) \cite{morales_absolute_2009}; (8) \cite{marcadon_analysis_2020}; (9) \cite{pan_starspot2024}; (10) \cite{pan_analysis_2025}; (11) \cite{rozyczka_new_2010}; (12) \cite{sandquist_analysis_2013}; (13) \cite{sandquist_k2_2018}; (14) \cite{torres_eclipsing_2006}; (15) \cite{torres_accurate_2010}; (16) \cite{torres_eclipsing_2019}; (17) \cite{torres_eclipsing_2021}}
\end{deluxetable*}

\section{Summary} \label{sec:Summary}

We present new angular diameter measurements of the components of the G8V+K4V visual binary GJ 566 AB and the K0V+K5V visual binary GJ 702 AB, using the MIRC-X \& MYSTIC beam combiners at the six-telescope CHARA Array. In combination with precise Gaia DR3 parallax values, these interferometric measurements translate to stellar radii that are precise to $\le 1.3$\% for GJ 566 A, GJ 702 A and GJ 702 B, and $2.4$\% for GJ 566 B. The measurements are combined with bolometric flux values to determine effective temperatures precise to $\le 1.7$\%. These systems are ideal for testing stellar evolutionary models because they have dynamically determined masses. A new orbital solution for GJ 566 AB and the recently revised orbital solution for GJ 702 AB yield component masses precise to 0.4\%.

The measured radii and temperatures are compared with the predictions of 4 sets of stellar evolutionary models (Lyon, Dartmouth, PARSEC, MIST), all computed at solar metallicity. Over the mass range $0.67 - 0.91$ M$_\odot$, all models predict stellar masses consistent with the measured properties to better than 3 standard deviations, and there are no systematic offsets over all stars.
The models also predict ages that are consistent with previous estimates (e.g., gyrochronology) and that the pairs are coeval, further validating the models.
Despite being historically classified as chromospherically active stars, these stars show no evidence of radius inflation compared to model predictions. 
The adolescent age determined for GJ 566 supports its association with the Ursa Majoris Moving Group; the components are the closest members of any Moving Group with angular diameter measurements.

To investigate radius inflation further, a sample of 21 short-period binary stars (P$<$100 d) with similar metallicities and with masses that are identical to within 3\% are assembled for comparison. These stars are systematically larger, cooler, and often both larger and cooler than the 4 visual binary stars and model predictions. Over this limited mass range, the eclipsing binary stars exhibit classic signatures of radius inflation.

These comparisons suggest that visual binary stars with dynamically determined masses and with physical separations of 10s of au are more useful for testing stellar evolutionary models, as each component can be observed independently and more completely, and the evolution more closely resembles that of single stars. It also suggests caution when using eclipsing binary stars of similar masses. 

\begin{acknowledgments}
We'd like to thank Aatiya Ali and Aylin Garcia Soto for their support of this work, and appreciative of helpful discussions with Colin Kane and Mahir Patel. We thank G. Torres for helpful guidance at the inception of this project. This work is supported by the National Science Foundation Grant AST-2511438. BF acknowledges support from the Georgia State University's Dissertation Grant Program and the 2025 William M. Suttles Fellowship. This work is based upon observations obtained with the Georgia State University Center for High Angular Resolution Astronomy Array at Mount Wilson Observatory.  The CHARA Array is supported by the National Science Foundation under Grant No. AST-2034336 and AST-2407956. Institutional support has been provided from the GSU College of Arts and Sciences, Office of the Provost, and Office of the Vice President for Research and Economic Development. We acknowledge that the CHARA Array is located on the ancestral territory of the Tongva Nation and that the main campus of Georgia State University is on the territory of the Muscogee Nation. We, the authors, acknowledge our obligation to honor the past, present, and future Native residents and cultures of California and Georgia. S.K acknowledges funding for MIRC-X received from the European Research Council (ERC) under the European Union's Horizon 2020 research and innovation programme (Starting Grant No. 639889 and Consolidated Grant No. 101003096) and from STFC Small Award (ST/Y002695/1). J.D.M acknowledges funding for the development of MIRC-X (NASA-XRP NNX16AD43G, NSF-AST 2009489) and MYSTIC (NSF-ATI 1506540, NSF-AST 1909165). C.K.H. acknowledges support from the National Science Foundation (NSF) Graduate Research Fellowship Program (GRFP) under Grant No.~DGE 2146752. This research has made use of the Washington Double Star Catalog maintained at the U.S. Naval Observatory. This research has made use of the Jean-Marie Mariotti Center Aspro and SearchCal services.

\end{acknowledgments}

\facilities{CHARA (MIRC-X and MYSTIC)}

\clearpage

\appendix

\section{Photometry and Energy Distributions}\label{app:phot}

\begin{figure*}[h]
     \centering
     \begin{subfigure}{0.49\textwidth}
         \centering
         \includegraphics[width=\linewidth]{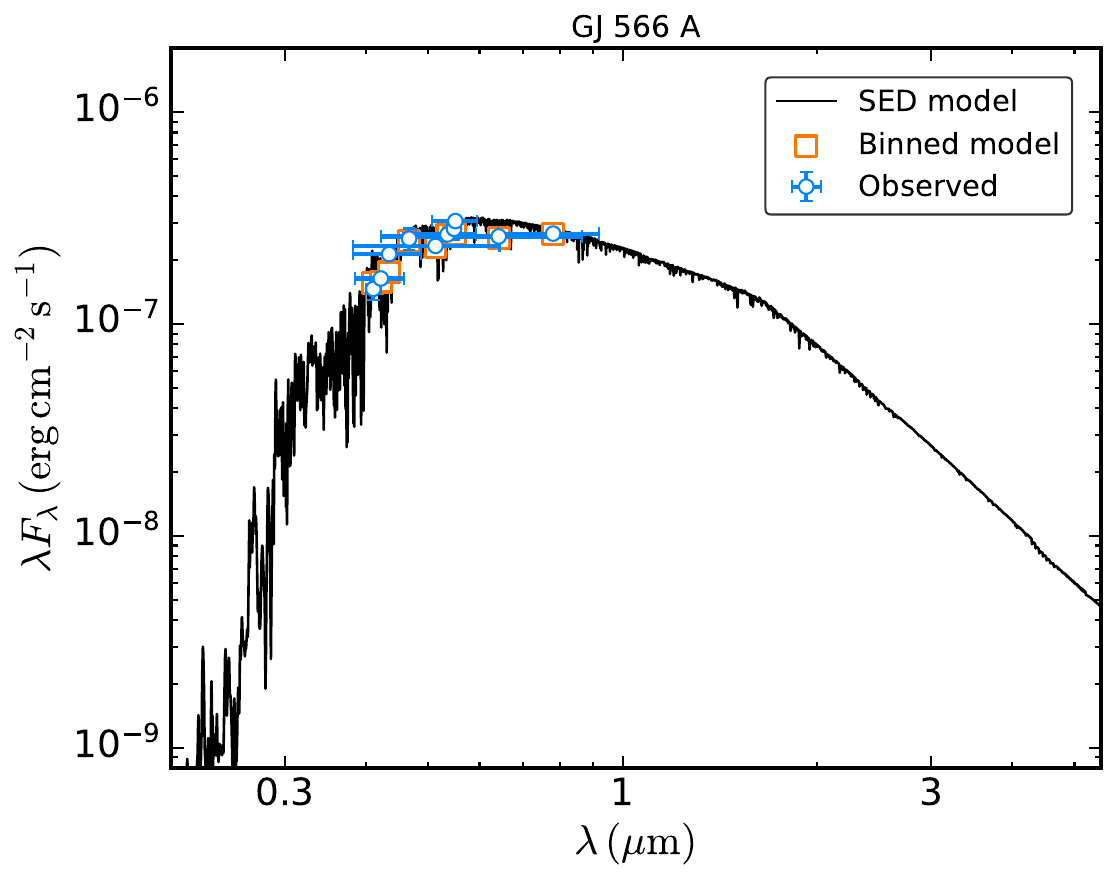}
         \caption{ }
         \label{fig:gj566a-sed}
     \end{subfigure}
     \begin{subfigure}{0.49\textwidth}
         \centering
         \includegraphics[width=\linewidth]{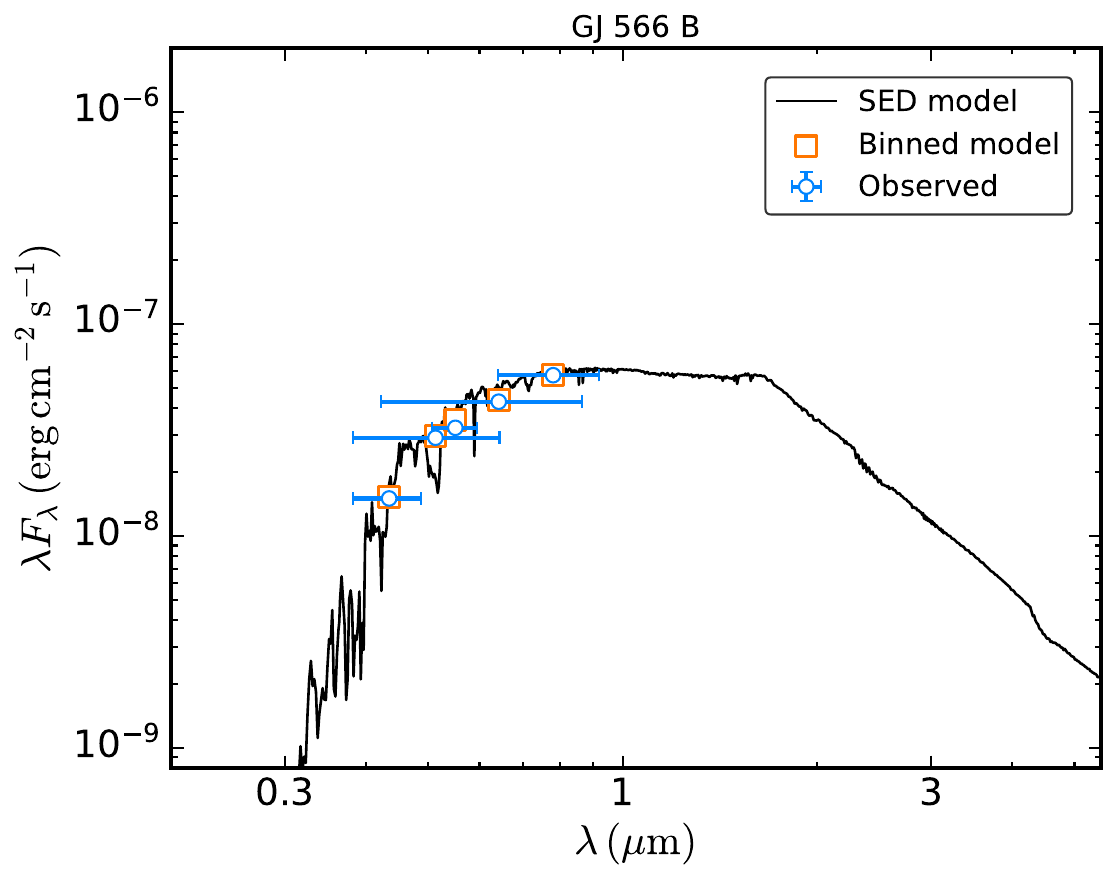}
         \caption{}
         \label{fig:gj566b-sed}
     \end{subfigure}
     \caption{Best fit SED model for (a) GJ 566 A and (b) GJ 566 B. Blue points are observed photometry from Johnson, Gaia DR3, Tycho-2, and Str\"{o}mgren for (a) and Johnson and Gaia DR3 for (b); horizontal bars indicate the filter widths. Orange squares show the synthetic model photometry. }
     \label{fig:gj566sed}
\end{figure*}

\begin{figure*}[h]
     \centering
     \begin{subfigure}{0.49\textwidth}
         \centering
         \includegraphics[width=\linewidth]{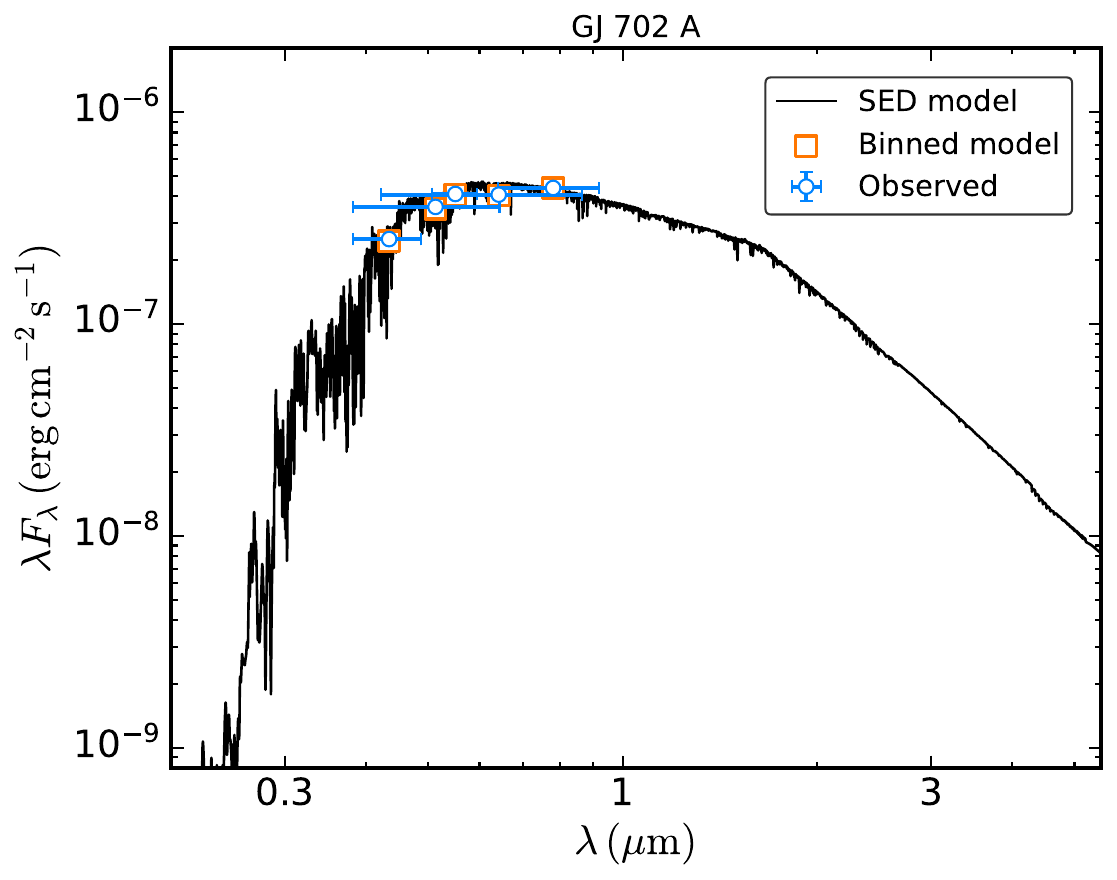}
         \caption{ }
         \label{fig:gj702a-sed}
     \end{subfigure}
     \begin{subfigure}{0.49\textwidth}
         \centering
         \includegraphics[width=\linewidth]{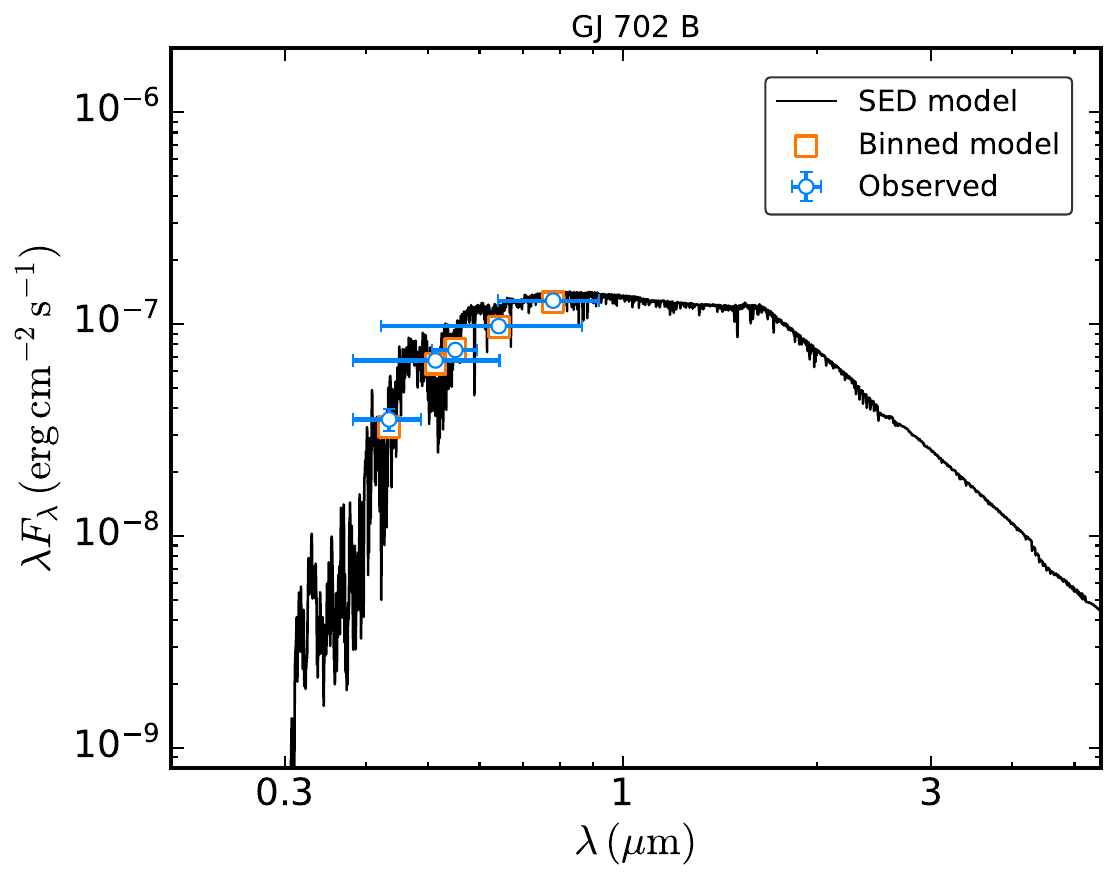}
         \caption{}
         \label{fig:gj702b-sed}
     \end{subfigure}
     \caption{Best fit SED model for (a) GJ 702 A and (b) GJ 702 B. Blue points are observed photometry from Johnson and Gaia DR3 for (a) and (b); horizontal bars indicate the filter widths. Orange squares show the synthetic model photometry.}
     \label{fig:gj702sed}
\end{figure*}

The spatially resolved photometric measurements used to construct the photometric energy distribution of each component are listed in Table \ref{tab:phot-mag}. Since these are bright ($2.0<G<8.0$ mag, $2.0<G_{RP}<3.45$ mag), the reported Gaia magnitudes are corrected for saturation, as recommended \citep{riello_gaia_2021}. The best-fit energy distributions, as calculated by \cite{harada_setting_2024} using the \texttt{ARIADNE} package, are presented in Figure \ref{fig:gj566sed} for GJ 566 A and B and in Figure \ref{fig:gj702sed} for GJ 702 A and B.

\begin{deluxetable*}{lcccc}[h]
\tablecaption{Photometric Magnitudes for GJ 566 A and B and GJ 702 A and B \label{tab:phot-mag}}
\tablehead{
\colhead{Filter} &
\colhead{GJ 566 A} &
\colhead{GJ 566 B} &
\colhead{GJ 702 A} &
\colhead{GJ 702 B}
}
\startdata
Str\"omgren $v$   & $5.781$           & --                & --                & -- \\
Tycho $B_T$       & $5.575 \pm 0.014$ & --                & --                & -- \\
Johnson $B$       & $5.260 \pm 0.018$ & $8.143 \pm 0.043$ & $5.083 \pm 0.014$ & $7.211$ \\
Str\"omgren $b$   & $5.069$           & --                & --                & -- \\
Gaia DR2 $G_{\rm BP}$ & $4.870 \pm 0.004$ & $7.133 \pm 0.003$ & $4.409 \pm 0.005$ & $6.221 \pm 0.004$ \\
Tycho $V_T$       & $4.757 \pm 0.009$ & --                & --                & -- \\
Str\"omgren $y$   & $4.620$           & --                & --                & -- \\
Johnson $V$       & $4.540 \pm 0.010$ & $6.979 \pm 0.037$ & $4.220 \pm 0.010$ & $6.061 \pm 0.020$ \\
Gaia DR2 $G$          & $4.494 \pm 0.003$ & $6.438 \pm 0.003$ & $4.000 \pm 0.003$ & $5.551 \pm 0.003$ \\
Gaia DR2 $G_{\rm RP}$ & $3.944 \pm 0.005$ & $5.615 \pm 0.004$ & $3.396 \pm 0.005$ & $4.736 \pm 0.004$ \\
\enddata
\tablecomments{Str\"omgren magnitudes are reported without uncertainties in the source catalog. The Johnson B magnitude of GJ 702 B is reported without an uncertainty in the source catalog. }
\end{deluxetable*}

\clearpage

\bibliography{zotero-bibtex,new}{}

@UNPUBLISHED{LL:LL-124,
author = {L.~Lindegren}, 
title={{R}e-normalising the astrometric chi-square in {G}aia {D}{R}2},
institution={Lund Observatory},
year={2018},
month={August},
url={https://dms.cosmos.esa.int/COSMOS/doc_fetch.php?id=3757412},
note={GAIA-C3-TN-LU-LL-124},
type={Technical Note} }

@ARTICLE{mahadevan_survey2019,
       author = {{Mahadevan}, Suvrath and {Bender}, Chad F. and {Hambleton}, Kelly and {Fleming}, Scott W. and {Deshpande}, Rohit and {Conroy}, Kyle and {Matijevi{\v{c}}}, Gal and {Hebb}, Leslie and {Roy}, Arpita and {Ak}, Hasan and {Leban}, Bla{\v{z}} and {Pr{\v{s}}a}, Andrej},
        title = "{The SDSS-HET Survey of Kepler Eclipsing Binaries. Description of the Survey and First Results}",
      journal = {\apj},
         year = 2019,
        month = oct,
       volume = {884},
       number = {2},
          eid = {126},
        pages = {126},
          doi = {10.3847/1538-4357/ab3793},
archivePrefix = {arXiv},
       eprint = {2004.02962},
 primaryClass = {astro-ph.SR},
       adsurl = {https://ui.adsabs.harvard.edu/abs/2019ApJ...884..126M}
}

@ARTICLE{gagne_montreal2026,
       author = {{Gagn{\'e}}, Jonathan and {Moranta}, Leslie and {Faherty}, Jacqueline K. and {Curtis}, Jason Lee and {Bickle}, Thomas P. and {Couture}, Dominic and {Chiasson David}, Am{\'e}lie and {Christie}, Katie and {Lambier}, Samantha and {Leclerc}, Elise and {Poliquin}, Livia and {Belzile}, Danika and {Mamajek}, Eric E.},
        title = "{The Montreal Open Clusters and Associations (MOCA) Database: A Census of Nearby Associations, Open Clusters, and Young Substellar Objects within 500 pc of the Sun}",
      journal = {arXiv e-prints},
         year = 2026,
        month = feb,
          eid = {arXiv:2602.15695},
        pages = {arXiv:2602.15695},
          doi = {10.48550/arXiv.2602.15695},
archivePrefix = {arXiv},
       eprint = {2602.15695},
 primaryClass = {astro-ph.SR},
       adsurl = {https://ui.adsabs.harvard.edu/abs/2026arXiv260215695G}
}

@ARTICLE{Han_magnetic_inflation2019,
       author = {{Han}, Eunkyu and {Muirhead}, Philip S. and {Swift}, Jonathan J.},
        title = "{Magnetic Inflation and Stellar Mass. IV. Four Low-mass Kepler Eclipsing Binaries Consistent with Non-magnetic Stellar Evolutionary Models}",
      journal = {\aj},
         year = 2019,
        month = sep,
       volume = {158},
       number = {3},
          eid = {111},
        pages = {111},
          doi = {10.3847/1538-3881/ab2ed7},
archivePrefix = {arXiv},
       eprint = {1907.07180},
 primaryClass = {astro-ph.SR},
       adsurl = {https://ui.adsabs.harvard.edu/abs/2019AJ....158..111H}
}

@ARTICLE{miller_fundamental2020,
       author = {{Miller}, N.~J. and {Maxted}, P.~F.~L. and {Smalley}, B.},
        title = "{Fundamental effective temperature measurements for eclipsing binary stars - I. Development of the method and application to AI Phoenicis}",
      journal = {\mnras},
         year = 2020,
        month = sep,
       volume = {497},
       number = {3},
        pages = {2899-2909},
          doi = {10.1093/mnras/staa2167},
archivePrefix = {arXiv},
       eprint = {2004.04568},
 primaryClass = {astro-ph.SR},
       adsurl = {https://ui.adsabs.harvard.edu/abs/2020MNRAS.497.2899M}
}

@ARTICLE{sebastian_EBLMIX2023,
       author = {{Sebastian}, D. and {Swayne}, M.~I. and {Maxted}, P.~F.~L. and {Triaud}, A.~H.~M.~J. and {Sousa}, S.~G. and {Olofsson}, G. and {Beck}, M. and {Billot}, N. and {Hoyer}, S. and {Gill}, S. and {Heidari}, N. and {Martin}, D.~V. and {Persson}, C.~M. and {Standing}, M.~R. and {Alibert}, Y. and {Alonso}, R. and {Anglada}, G. and {Asquier}, J. and {B{\'a}rczy}, T. and {Barrado}, D. and {Barros}, S.~C.~C. and {Battley}, M.~P. and {Baumjohann}, W. and {Beck}, T. and {Benz}, W. and {Bergomi}, M. and {Boisse}, I. and {Bonfils}, X. and {Brandeker}, A. and {Broeg}, C. and {Cabrera}, J. and {Charnoz}, S. and {Collier Cameron}, A. and {Csizmadia}, Sz and {Davies}, M.~B. and {Deleuil}, M. and {Delrez}, L. and {Demangeon}, O.~D.~S. and {Demory}, B. -O. and {Dransfield}, G. and {Ehrenreich}, D. and {Erikson}, A. and {Fortier}, A. and {Fossati}, L. and {Fridlund}, M. and {Gandolfi}, D. and {Gillon}, M. and {G{\"u}del}, M. and {Hasiba}, J. and {H{\'e}brard}, G. and {Heng}, K. and {Isaak}, K.~G. and {Kiss}, L.~L. and {Kopp}, E. and {Kunovac}, V. and {Laskar}, J. and {Lecavelier des Etangs}, A. and {Lendl}, M. and {Lovis}, C. and {Magrin}, D. and {McCormac}, J. and {Miller}, N.~J. and {Nascimbeni}, V. and {Ottensamer}, R. and {Pagano}, I. and {Pall{\'e}}, E. and {Pepe}, F.~A. and {Peter}, G. and {Piotto}, G. and {Pollacco}, D. and {Queloz}, D. and {Ragazzoni}, R. and {Rando}, N. and {Rauer}, H. and {Ribas}, I. and {Lalitha}, S. and {Santerne}, A. and {Santos}, N.~C. and {Scandariato}, G. and {S{\'e}gransan}, D. and {Simon}, A.~E. and {Smith}, A.~M.~S. and {Steller}, M. and {Szab{\'o}}, Gy M. and {Thomas}, N. and {Udry}, S. and {Van Grootel}, V. and {Walton}, N.~A.},
        title = "{The EBLM project - IX. Five fully convective M-dwarfs, precisely measured with CHEOPS and TESS light curves}",
      journal = {\mnras},
         year = 2023,
        month = mar,
       volume = {519},
       number = {3},
        pages = {3546-3563},
          doi = {10.1093/mnras/stac2565},
archivePrefix = {arXiv},
       eprint = {2209.03128},
 primaryClass = {astro-ph.SR},
       adsurl = {https://ui.adsabs.harvard.edu/abs/2023MNRAS.519.3546S}
}

@ARTICLE{torres_stringent2021,
       author = {{Torres}, Guillermo and {Feiden}, Gregory A. and {Vanderburg}, Andrew and {Curtis}, Jason L.},
        title = "{A Stringent Test of Magnetic Models of Stellar Evolution}",
      journal = {Galaxies},
         year = 2021,
        month = dec,
       volume = {10},
       number = {1},
          eid = {3},
        pages = {3},
          doi = {10.3390/galaxies10010003},
       adsurl = {https://ui.adsabs.harvard.edu/abs/2021Galax..10....3T}
}

@ARTICLE{feiden_chaboyer_magnetic2013,
       author = {{Feiden}, Gregory A. and {Chaboyer}, Brian},
        title = "{Magnetic Inhibition of Convection and the Fundamental Properties of Low-mass Stars. I. Stars with a Radiative Core}",
      journal = {\apj},
         year = 2013,
        month = dec,
       volume = {779},
       number = {2},
          eid = {183},
        pages = {183},
          doi = {10.1088/0004-637X/779/2/183},
archivePrefix = {arXiv},
       eprint = {1309.0033},
 primaryClass = {astro-ph.SR},
       adsurl = {https://ui.adsabs.harvard.edu/abs/2013ApJ...779..183F}
}

@ARTICLE{pan_starspot2024,
       author = {{Pan}, Yang and {Frasca}, Antonio and {Wang}, Jia-Xin and {Fu}, Jian-Ning and {Zhang}, Xiao-Bin},
        title = "{Starspot Coverage on Two New K-type Low-mass Eclipsing Binaries with Radius Inflation}",
      journal = {\aj},
         year = 2024,
        month = dec,
       volume = {168},
       number = {6},
          eid = {253},
        pages = {253},
          doi = {10.3847/1538-3881/ad84f5},
       adsurl = {https://ui.adsabs.harvard.edu/abs/2024AJ....168..253P}
}

@ARTICLE{lopez_morales_correlation2007,
       author = {{L{\'o}pez-Morales}, Mercedes},
        title = "{On the Correlation between the Magnetic Activity Levels, Metallicities, and Radii of Low-Mass Stars}",
      journal = {\apj},
         year = 2007,
        month = may,
       volume = {660},
       number = {1},
        pages = {732-739},
          doi = {10.1086/513142},
archivePrefix = {arXiv},
       eprint = {astro-ph/0701702},
 primaryClass = {astro-ph},
       adsurl = {https://ui.adsabs.harvard.edu/abs/2007ApJ...660..732L}
}

@ARTICLE{olah_magnetic2016,
       author = {{Ol{\'a}h}, K. and {K{\H{o}}v{\'a}ri}, Zs. and {Petrovay}, K. and {Soon}, W. and {Baliunas}, S. and {Koll{\'a}th}, Z. and {Vida}, K.},
        title = "{Magnetic cycles at different ages of stars}",
      journal = {\aap},
         year = 2016,
        month = jun,
       volume = {590},
          eid = {A133},
        pages = {A133},
          doi = {10.1051/0004-6361/201628479},
archivePrefix = {arXiv},
       eprint = {1604.06701},
 primaryClass = {astro-ph.SR},
       adsurl = {https://ui.adsabs.harvard.edu/abs/2016A&A...590A.133O}
}

@ARTICLE{morales_fundamental2022,
       author = {{Morales}, Juan Carlos and {Ribas}, Ignasi and {Gim{\'e}nez}, {\'A}lvaro and {Baroch}, David},
        title = "{Fundamental Properties of Late-Type Stars in Eclipsing Binaries}",
      journal = {Galaxies},
         year = 2022,
        month = sep,
       volume = {10},
       number = {5},
          eid = {98},
        pages = {98},
          doi = {10.3390/galaxies10050098},
archivePrefix = {arXiv},
       eprint = {2212.05797},
 primaryClass = {astro-ph.SR},
       adsurl = {https://ui.adsabs.harvard.edu/abs/2022Galax..10...98M}
}

@ARTICLE{swayne_EBLMXI2024,
       author = {{Swayne}, M.~I. and {Maxted}, P.~F.~L. and {Triaud}, A.~H.~M.~J. and {Sousa}, S.~G. and {Deline}, A. and {Ehrenreich}, D. and {Hoyer}, S. and {Olofsson}, G. and {Boisse}, I. and {Duck}, A. and {Gill}, S. and {Martin}, D. and {McCormac}, J. and {Persson}, C.~M. and {Santerne}, A. and {Sebastian}, D. and {Standing}, M.~R. and {Acu{\~n}a}, L. and {Alibert}, Y. and {Alonso}, R. and {Anglada}, G. and {B{\'a}rczy}, T. and {Barrado Navascues}, D. and {Barros}, S.~C.~C. and {Baumjohann}, W. and {Baycroft}, T.~A. and {Beck}, M. and {Beck}, T. and {Benz}, W. and {Billot}, N. and {Bonfils}, X. and {Borsato}, L. and {Bourrier}, V. and {Brandeker}, A. and {Broeg}, C. and {Carmona}, A. and {Charnoz}, S. and {Collier Cameron}, A. and {Cort{\'e}s-Zuleta}, P. and {Csizmadia}, Sz and {Cubillos}, P.~E. and {Davies}, M.~B. and {Deleuil}, M. and {Delfosse}, X. and {Delrez}, L. and {Demangeon}, O.~D.~S. and {Demory}, B. -O. and {Dransfield}, G. and {Erikson}, A. and {Fortier}, A. and {Forveille}, T. and {Fossati}, L. and {Fridlund}, M. and {Gandolfi}, D. and {Gillon}, M. and {G{\"u}del}, M. and {G{\"u}nther}, M.~N. and {Hara}, N. and {H{\'e}brard}, G. and {Heidari}, N. and {Hellier}, C. and {Helling}, Ch and {Isaak}, K.~G. and {Kerschbaum}, F. and {Kiefer}, F. and {Kiss}, L.~L. and {Kunovac}, V. and {Lalitha}, S. and {Lam}, K.~W.~F. and {Laskar}, J. and {Lecavelier des Etangs}, A. and {Lendl}, M. and {Magrin}, D. and {Marafatto}, L. and {Martioli}, E. and {Miller}, N.~J. and {Mordasini}, C. and {Moutou}, C. and {Nascimbeni}, V. and {Ottensamer}, R. and {Pagano}, I. and {Pall{\'e}}, E. and {Peter}, G. and {Piazza}, D. and {Piotto}, G. and {Pollacco}, D. and {Queloz}, D. and {Ragazzoni}, R. and {Rando}, N. and {Rauer}, H. and {Ribas}, I. and {Santos}, N.~C. and {Scandariato}, G. and {S{\'e}gransan}, D. and {Simon}, A.~E. and {Smith}, A.~M.~S. and {Southworth}, R. and {Stalport}, M. and {Szab{\'o}}, Gy M. and {Thomas}, N. and {Udry}, S. and {Ulmer}, B. and {Van Grootel}, V. and {Venturini}, J. and {Walton}, N.~A. and {Willett}, E. and {Wilson}, T.~G.},
        title = "{The EBLM Project- XI. Mass, radius, and effective temperature measurements for 23 M-dwarf companions to solar-type stars observed with CHEOPS}",
      journal = {\mnras},
         year = 2024,
        month = mar,
       volume = {528},
       number = {4},
        pages = {5703-5722},
          doi = {10.1093/mnras/stad3866},
archivePrefix = {arXiv},
       eprint = {2312.11339},
 primaryClass = {astro-ph.SR},
       adsurl = {https://ui.adsabs.harvard.edu/abs/2024MNRAS.528.5703S}
}

@ARTICLE{cakirli_low_mass_companions2025,
       author = {{{\c{C}}ak{\i}rl{\i}}, {\"O}. and {Hoyman}, B.},
        title = "{The low-mass companions in five totally eclipsing binaries with F/G primaries}",
      journal = {\mnras},
         year = 2025,
        month = jun,
       volume = {540},
       number = {1},
        pages = {612-625},
          doi = {10.1093/mnras/staf780},
       adsurl = {https://ui.adsabs.harvard.edu/abs/2025MNRAS.540..612C}
}

@ARTICLE{andersen_EBs1991,
       author = {{Andersen}, J.},
        title = "{Accurate masses and radii of normal stars}",
      journal = {\aapr},
         year = 1991,
        month = jan,
       volume = {3},
       number = {2},
        pages = {91-126},
          doi = {10.1007/BF00873538},
       adsurl = {https://ui.adsabs.harvard.edu/abs/1991A&ARv...3...91A}
}

@ARTICLE{hillenbrand_assessment2004,
       author = {{Hillenbrand}, Lynne A. and {White}, Russel J.},
        title = "{An Assessment of Dynamical Mass Constraints on Pre-Main-Sequence Evolutionary Tracks}",
      journal = {\apj},
         year = 2004,
        month = apr,
       volume = {604},
       number = {2},
        pages = {741-757},
          doi = {10.1086/382021},
archivePrefix = {arXiv},
       eprint = {astro-ph/0312189},
 primaryClass = {astro-ph},
       adsurl = {https://ui.adsabs.harvard.edu/abs/2004ApJ...604..741H}
}

@dataset{gliese_nearby1995,
       author = {{Gliese}, W. and {Jahreiss}, H.},
        title = "{VizieR Online Data Catalog: Nearby Stars, Preliminary 3rd Version (Gliese+ 1991)}",
 howpublished = {VizieR On-line Data Catalog: V/70A.  Originally published in: Astron. Rechen-Institut, Heidelberg (1991)},
         year = 1995,
        month = nov,
          eid = {V/70A},
       adsurl = {https://ui.adsabs.harvard.edu/abs/1995yCat.5070....0G}
}

@ARTICLE{strassmeier_doppler2023,
       author = {{Strassmeier}, K.~G. and {Carroll}, T.~A. and {Ilyin}, I.~V.},
        title = "{Zeeman Doppler imaging of {\ensuremath{\xi}} Boo A and B}",
      journal = {\aap},
         year = 2023,
        month = jun,
       volume = {674},
          eid = {A118},
        pages = {A118},
          doi = {10.1051/0004-6361/202245664},
archivePrefix = {arXiv},
       eprint = {2305.07470},
 primaryClass = {astro-ph.SR},
       adsurl = {https://ui.adsabs.harvard.edu/abs/2023A&A...674A.118S}
}

@ARTICLE{mason_WDS2001,
       author = {{Mason}, Brian D. and {Wycoff}, Gary L. and {Hartkopf}, William I. and {Douglass}, Geoffrey G. and {Worley}, Charles E.},
        title = "{The 2001 US Naval Observatory Double Star CD-ROM. I. The Washington Double Star Catalog}",
      journal = {\aj},
         year = 2001,
        month = dec,
       volume = {122},
       number = {6},
        pages = {3466-3471},
          doi = {10.1086/323920},
       adsurl = {https://ui.adsabs.harvard.edu/abs/2001AJ....122.3466M}
}

@ARTICLE{strassmeier_lithium2022,
       author = {{Strassmeier}, Klaus G. and {Steffen}, Matthias},
        title = "{On the lithium abundance of the visual binary components {\ensuremath{\xi}} Boo A ( G8V ) and {\ensuremath{\xi}} Boo B ( K5V )}",
      journal = {Astronomische Nachrichten},
         year = 2022,
        month = nov,
       volume = {343},
       number = {9-10},
          eid = {e20220036},
        pages = {e20220036},
          doi = {10.1002/asna.20220036},
archivePrefix = {arXiv},
       eprint = {2211.06336},
 primaryClass = {astro-ph.SR},
       adsurl = {https://ui.adsabs.harvard.edu/abs/2022AN....34320036S}
}

@ARTICLE{gruner_wide_binaries2023,
       author = {{Gruner}, D. and {Barnes}, S.~A. and {Janes}, K.~A.},
        title = "{Wide binaries demonstrate the consistency of rotational evolution between open cluster and field stars}",
      journal = {\aap},
         year = 2023,
        month = jul,
       volume = {675},
          eid = {A180},
        pages = {A180},
          doi = {10.1051/0004-6361/202346590},
archivePrefix = {arXiv},
       eprint = {2307.10836},
 primaryClass = {astro-ph.SR},
       adsurl = {https://ui.adsabs.harvard.edu/abs/2023A&A...675A.180G}
}

@ARTICLE{dopcke_UMa2019,
       author = {{Dopcke}, Gustavo and {Porto de Mello}, G.~F. and {Sneden}, C.},
        title = "{The Ursa Major Moving Group: a chronochemokinematic analysis}",
      journal = {\mnras},
         year = 2019,
        month = may,
       volume = {485},
       number = {3},
        pages = {4375-4388},
          doi = {10.1093/mnras/stz631},
       adsurl = {https://ui.adsabs.harvard.edu/abs/2019MNRAS.485.4375D}
}

@ARTICLE{king_UMa2003,
       author = {{King}, Jeremy R. and {Villarreal}, Adam R. and {Soderblom}, David R. and {Gulliver}, Austin F. and {Adelman}, Saul J.},
        title = "{Stellar Kinematic Groups. II. A Reexamination of the Membership, Activity, and Age of the Ursa Major Group}",
      journal = {\aj},
         year = 2003,
        month = apr,
       volume = {125},
       number = {4},
        pages = {1980-2017},
          doi = {10.1086/368241},
       adsurl = {https://ui.adsabs.harvard.edu/abs/2003AJ....125.1980K}
}

@ARTICLE{jones_UMa2015,
       author = {{Jones}, Jeremy and {White}, R.~J. and {Boyajian}, T. and {Schaefer}, G. and {Baines}, E. and {Ireland}, M. and {Patience}, J. and {ten Brummelaar}, T. and {McAlister}, H. and {Ridgway}, S.~T. and {Sturmann}, J. and {Sturmann}, L. and {Turner}, N. and {Farrington}, C. and {Goldfinger}, P.~J.},
        title = "{The Ages of A-Stars. I. Interferometric Observations and Age Estimates for Stars in the Ursa Major Moving Group}",
      journal = {\apj},
         year = 2015,
        month = nov,
       volume = {813},
       number = {1},
          eid = {58},
        pages = {58},
          doi = {10.1088/0004-637X/813/1/58},
archivePrefix = {arXiv},
       eprint = {1508.05643},
 primaryClass = {astro-ph.SR},
       adsurl = {https://ui.adsabs.harvard.edu/abs/2015ApJ...813...58J}
}

@ARTICLE{popper_orbits1997,
       author = {{Popper}, Daniel M.},
        title = "{Orbits of detached main-sequence eclipsing binaries of types late F to K. II. UV leonis, UV piscium, and BH virginis.}",
      journal = {\aj},
         year = 1997,
        month = sep,
       volume = {114},
        pages = {1195},
          doi = {10.1086/118552},
       adsurl = {https://ui.adsabs.harvard.edu/abs/1997AJ....114.1195P}
}

@ARTICLE{kraus_massradius_2011ApJ...728...48K,
       author = {{Kraus}, Adam L. and {Tucker}, Roy A. and {Thompson}, Michael I. and {Craine}, Eric R. and {Hillenbrand}, Lynne A.},
        title = "{The Mass-Radius(-Rotation?) Relation for Low-mass Stars}",
      journal = {\apj},
         year = 2011,
        month = feb,
       volume = {728},
       number = {1},
          eid = {48},
        pages = {48},
          doi = {10.1088/0004-637X/728/1/48},
archivePrefix = {arXiv},
       eprint = {1011.2757},
 primaryClass = {astro-ph.SR},
       adsurl = {https://ui.adsabs.harvard.edu/abs/2011ApJ...728...48K}
}

@ARTICLE{dittman_discovery2017,
       author = {{Dittmann}, Jason A. and {Irwin}, Jonathan M. and {Charbonneau}, David and {Berta-Thompson}, Zachory K. and {Newton}, Elisabeth R. and {Latham}, David W. and {Latham}, Christian A. and {Esquerdo}, Gilbert and {Berlind}, Perry and {Calkins}, Michael L.},
        title = "{Discovery and Precise Characterization by the MEarth Project of LP 661-13, an Eclipsing Binary Consisting of Two Fully Convective Low-mass Stars}",
      journal = {\apj},
         year = 2017,
        month = feb,
       volume = {836},
       number = {1},
          eid = {124},
        pages = {124},
          doi = {10.3847/1538-4357/836/1/124},
archivePrefix = {arXiv},
       eprint = {1609.03591},
 primaryClass = {astro-ph.SR},
       adsurl = {https://ui.adsabs.harvard.edu/abs/2017ApJ...836..124D}
}

@ARTICLE{cruz_detached2022,
       author = {{Cruz}, Patricia and {Aguilar}, John F. and {Garrido}, Hern{\'a}n E. and {Diaz}, Marcos P. and {Solano}, Enrique},
        title = "{Detached eclipsing binaries from the Kepler field: radii and photometric masses of components in short-period systems}",
      journal = {\mnras},
         year = 2022,
        month = sep,
       volume = {515},
       number = {1},
        pages = {1416-1429},
          doi = {10.1093/mnras/stac1707},
archivePrefix = {arXiv},
       eprint = {2206.08708},
 primaryClass = {astro-ph.SR},
       adsurl = {https://ui.adsabs.harvard.edu/abs/2022MNRAS.515.1416C}
}

@ARTICLE{baliunas_chromospheric1995,
       author = {{Baliunas}, S.~L. and {Donahue}, R.~A. and {Soon}, W.~H. and {Horne}, J.~H. and {Frazer}, J. and {Woodard-Eklund}, L. and {Bradford}, M. and {Rao}, L.~M. and {Wilson}, O.~C. and {Zhang}, Q. and {Bennett}, W. and {Briggs}, J. and {Carroll}, S.~M. and {Duncan}, D.~K. and {Figueroa}, D. and {Lanning}, H.~H. and {Misch}, T. and {Mueller}, J. and {Noyes}, R.~W. and {Poppe}, D. and {Porter}, A.~C. and {Robinson}, C.~R. and {Russell}, J. and {Shelton}, J.~C. and {Soyumer}, T. and {Vaughan}, A.~H. and {Whitney}, J.~H.},
        title = "{Chromospheric Variations in Main-Sequence Stars. II.}",
      journal = {\apj},
         year = 1995,
        month = jan,
       volume = {438},
        pages = {269},
          doi = {10.1086/175072},
       adsurl = {https://ui.adsabs.harvard.edu/abs/1995ApJ...438..269B}
}

@ARTICLE{colliercameron_2009,
       author = {{Collier Cameron}, A. and {Davidson}, V.~A. and {Hebb}, L. and {Skinner}, G. and {Anderson}, D.~R. and {Christian}, D.~J. and {Clarkson}, W.~I. and {Enoch}, B. and {Irwin}, J. and {Joshi}, Y. and {Haswell}, C.~A. and {Hellier}, C. and {Horne}, K.~D. and {Kane}, S.~R. and {Lister}, T.~A. and {Maxted}, P.~F.~L. and {Norton}, A.~J. and {Parley}, N. and {Pollacco}, D. and {Ryans}, R. and {Scholz}, A. and {Skillen}, I. and {Smalley}, B. and {Street}, R.~A. and {West}, R.~G. and {Wilson}, D.~M. and {Wheatley}, P.~J.},
        title = "{The main-sequence rotation-colour relation in the Coma Berenices open cluster}",
      journal = {\mnras},
         year = 2009,
        month = nov,
       volume = {400},
       number = {1},
        pages = {451-462},
          doi = {10.1111/j.1365-2966.2009.15476.x},
archivePrefix = {arXiv},
       eprint = {0908.0189},
 primaryClass = {astro-ph.SR},
       adsurl = {https://ui.adsabs.harvard.edu/abs/2009MNRAS.400..451C}
}

@ARTICLE{noyes_1984,
       author = {{Noyes}, R.~W. and {Hartmann}, L.~W. and {Baliunas}, S.~L. and {Duncan}, D.~K. and {Vaughan}, A.~H.},
        title = "{Rotation, convection, and magnetic activity in lower main-sequence stars.}",
      journal = {\apj},
         year = 1984,
        month = apr,
       volume = {279},
        pages = {763-777},
          doi = {10.1086/161945},
       adsurl = {https://ui.adsabs.harvard.edu/abs/1984ApJ...279..763N}
}

@ARTICLE{jorgensen_lindegren2005,
       author = {{J{\o}rgensen}, B.~R. and {Lindegren}, L.},
        title = "{Determination of stellar ages from isochrones: Bayesian estimation versus isochrone fitting}",
      journal = {\aap},
         year = 2005,
        month = jun,
       volume = {436},
       number = {1},
        pages = {127-143},
          doi = {10.1051/0004-6361:20042185},
       adsurl = {https://ui.adsabs.harvard.edu/abs/2005A&A...436..127J}
}

@ARTICLE{ligi_radii2016,
       author = {{Ligi}, R. and {Creevey}, O. and {Mourard}, D. and {Crida}, A. and {Lagrange}, A.-M. and {Nardetto}, N. and {Perraut}, K. and {Schultheis}, M. and {Tallon-Bosc}, I. and {ten Brummelaar}, T.},
        title = "{Radii, masses, and ages of 18 bright stars using interferometry and new estimations of exoplanetary parameters}",
      journal = {\aap},
         year = 2016,
        month = feb,
       volume = {586},
          eid = {A94},
        pages = {A94},
          doi = {10.1051/0004-6361/201527054},
archivePrefix = {arXiv},
       eprint = {1511.03197},
 primaryClass = {astro-ph.SR},
       adsurl = {https://ui.adsabs.harvard.edu/abs/2016A&A...586A..94L}
}

@ARTICLE{gagne_oceanus2023,
       author = {{Gagn{\'e}}, Jonathan and {Moranta}, Leslie and {Faherty}, Jacqueline K. and {Kiman}, Rocio and {Couture}, Dominic and {Larochelle}, Arnaud Ren{\'e} and {Popinchalk}, Mark and {Morrone}, Daniella},
        title = "{The Oceanus Moving Group: A New 500 Myr Old Host for the Nearest Brown Dwarf}",
      journal = {\apj},
         year = 2023,
        month = mar,
       volume = {945},
       number = {2},
          eid = {119},
        pages = {119},
          doi = {10.3847/1538-4357/acb8b7},
archivePrefix = {arXiv},
       eprint = {2208.00070},
 primaryClass = {astro-ph.SR},
       adsurl = {https://ui.adsabs.harvard.edu/abs/2023ApJ...945..119G}
}

@ARTICLE{luhman_discovery2014,
       author = {{Luhman}, K.~L.},
        title = "{Discovery of a \raisebox{-0.5ex}\textasciitilde250 K Brown Dwarf at 2 pc from the Sun}",
      journal = {\apjl},
         year = 2014,
        month = may,
       volume = {786},
       number = {2},
          eid = {L18},
        pages = {L18},
          doi = {10.1088/2041-8205/786/2/L18},
archivePrefix = {arXiv},
       eprint = {1404.6501},
 primaryClass = {astro-ph.GA},
       adsurl = {https://ui.adsabs.harvard.edu/abs/2014ApJ...786L..18L}
}

@ARTICLE{schofer_CARMENES2022,
       author = {{Sch{\"o}fer}, P. and {Jeffers}, S.~V. and {Reiners}, A. and {Zechmeister}, M. and {Fuhrmeister}, B. and {Lafarga}, M. and {Ribas}, I. and {Quirrenbach}, A. and {Amado}, P.~J. and {Caballero}, J.~A. and {Anglada-Escud{\'e}}, G. and {Bauer}, F.~F. and {B{\'e}jar}, V.~J.~S. and {Cort{\'e}s-Contreras}, M. and {Alonso}, E. D{\'\i}ez and {Dreizler}, S. and {Guenther}, E.~W. and {Herbort}, O. and {Johnson}, E.~N. and {Kaminski}, A. and {K{\"u}rster}, M. and {Montes}, D. and {Morales}, J.~C. and {Pedraz}, S. and {Tal-Or}, L.},
        title = "{The CARMENES search for exoplanets around M dwarfs. Rotational variation in activity indicators of Ross 318, YZ CMi, TYC 3529-1437-1, and EV Lac}",
      journal = {\aap},
         year = 2022,
        month = jul,
       volume = {663},
          eid = {A68},
        pages = {A68},
          doi = {10.1051/0004-6361/201936102},
archivePrefix = {arXiv},
       eprint = {2204.11685},
 primaryClass = {astro-ph.SR},
       adsurl = {https://ui.adsabs.harvard.edu/abs/2022A&A...663A..68S}
}

@ARTICLE{spejcher_moderate_spin-orbit2026,
       author = {{Spejcher}, Becca and {Martin}, David V. and {Pandina}, Jake and {Zhang}, Andy and {Ammons}, Max and {Tubthong}, Wata and {Triaud}, Amaury and {Sethi}, Ritika and {Vowell}, Noah and {Barker}, Adrian and {Maxted}, Pierre and {Duck}, Alison and {Summers}, Shelby and {Bouchy}, Fran{\c{c}}ois and {Lendl}, Monika and {Marmier}, Maxime and {Tewes}, Malte and {Udry}, St{\'e}phane},
        title = "{The EBLM project ─ XVI. Moderate spin─orbit misalignment of the low-mass eclipsing binary EBLM J0021-16}",
      journal = {\mnras},
         year = 2026,
        month = feb,
       volume = {545},
       number = {4},
          eid = {staf2236},
        pages = {staf2236},
          doi = {10.1093/mnras/staf2236},
archivePrefix = {arXiv},
       eprint = {2509.21517},
 primaryClass = {astro-ph.SR},
       adsurl = {https://ui.adsabs.harvard.edu/abs/2026MNRAS.545f2236S}
}

@ARTICLE{kochukhov_shylyak_magnetic2019,
       author = {{Kochukhov}, Oleg and {Shulyak}, Denis},
        title = "{Magnetic Field of the Eclipsing M-dwarf Binary YY Gem}",
      journal = {\apj},
         year = 2019,
        month = mar,
       volume = {873},
       number = {1},
          eid = {69},
        pages = {69},
          doi = {10.3847/1538-4357/ab06c5},
archivePrefix = {arXiv},
       eprint = {1902.04157},
 primaryClass = {astro-ph.SR},
       adsurl = {https://ui.adsabs.harvard.edu/abs/2019ApJ...873...69K}
}

@article{boetticher_eblm_2019,
	title = {The {EBLM} {Project} - {V}. {Physical} properties of ten fully convective, very-low-mass stars},
	volume = {625},
	copyright = {© ESO 2019},
	issn = {0004-6361, 1432-0746},
	url = {https://www.aanda.org/articles/aa/abs/2019/05/aa34539-18/aa34539-18.html},
	doi = {10.1051/0004-6361/201834539},
	language = {en},
	urldate = {2025-09-03},
	journal = {Astronomy \& Astrophysics},
	publisher = {EDP Sciences},
	author = {von Boetticher, Alexander and Triaud, Amaury H. M. J. and Queloz, Didier and Gill, Sam and Maxted, Pierre F. L. and Almleaky, Yaseen and Anderson, David R. and Bouchy, François and Burdanov, Artem and Cameron, Andrew Collier and Delrez, Laetitia and Ducrot, Elsa and Faedi, Francesca and Gillon, Michaël and Chew, Yilen Gómez Maqueo and Hebb, Leslie and Hellier, Coel and Jehin, Emmanuël and Lendl, Monika and Marmier, Maxime and Martin, David V. and McCormac, James and Pepe, Francesco and Pollacco, Don and Ségransan, Damien and Smalley, Barry and Thompson, Samantha and Turner, Oliver and Udry, Stéphane and Grootel, Valérie Van and West, Richard},
	month = may,
	year = {2019},
	pages = {A150},
}

@article{gray_contributions_2003,
	title = {Contributions to the {Nearby} {Stars} ({NStars}) {Project}: {Spectroscopy} of {Stars} {Earlier} than {M0} within 40 {Parsecs}: {The} {Northern} {Sample}. {I}.},
	volume = {126},
	issn = {0004-6256},
	shorttitle = {Contributions to the {Nearby} {Stars} ({NStars}) {Project}},
	url = {https://ui.adsabs.harvard.edu/abs/2003AJ....126.2048G},
	doi = {10.1086/378365},
	urldate = {2026-10-04},
	journal = {The Astronomical Journal},
	publisher = {IOP},
	author = {Gray, R. O. and Corbally, C. J. and Garrison, R. F. and McFadden, M. T. and Robinson, P. E.},
	month = oct,
	year = {2003},
	note = {ADS Bibcode: 2003AJ....126.2048G},
	pages = {2048--2059},
}

@article{riello_gaia_2021,
	title = {Gaia {Early} {Data} {Release} 3 - {Photometric} content and validation},
	volume = {649},
	copyright = {© ESO 2021},
	issn = {0004-6361, 1432-0746},
	url = {https://www.aanda.org/articles/aa/abs/2021/05/aa39587-20/aa39587-20.html},
	doi = {10.1051/0004-6361/202039587},
	language = {en},
	urldate = {2026-08-16},
	journal = {Astronomy \& Astrophysics},
	publisher = {EDP Sciences},
	author = {Riello, M. and Angeli, F. De and Evans, D. W. and Montegriffo, P. and Carrasco, J. M. and Busso, G. and Palaversa, L. and Burgess, P. W. and Diener, C. and Davidson, M. and Rowell, N. and Fabricius, C. and Jordi, C. and Bellazzini, M. and Pancino, E. and Harrison, D. L. and Cacciari, C. and Leeuwen, F. van and Hambly, N. C. and Hodgkin, S. T. and Osborne, P. J. and Altavilla, G. and Barstow, M. A. and Brown, A. G. A. and Castellani, M. and Cowell, S. and Luise, F. De and Gilmore, G. and Giuffrida, G. and Hidalgo, S. and Holland, G. and Marinoni, S. and Pagani, C. and Piersimoni, A. M. and Pulone, L. and Ragaini, S. and Rainer, M. and Richards, P. J. and Sanna, N. and Walton, N. A. and Weiler, M. and Yoldas, A.},
	month = may,
	year = {2021},
	pages = {A3},
}

@inproceedings{ireland_sensitive_2008,
	title = {Sensitive visible interferometry with {PAVO}},
	volume = {7013},
	url = {https://ui.adsabs.harvard.edu/abs/2008SPIE.7013E..24I},
	doi = {10.1117/12.788386},
	urldate = {2025-10-02},
	booktitle = {Optical and {Infrared} {Interferometry}},
	author = {Ireland, Michael J. and Mérand, Antoine and ten Brummelaar, Theo A. and Tuthill, Peter G. and Schaefer, Gail H. and Turner, Nils H. and Sturmann, Judit and Sturmann, Laszlo and McAlister, Harold A.},
	month = jul,
	year = {2008},
	note = {ADS Bibcode: 2008SPIE.7013E..24I},
	pages = {701324},
}

@inproceedings{clausen_0711_1999,
	title = {Do 0.7-1.1 {Msun} {Eclipsing} {Binaries} {Pose} a {Problem} for {Current} {Stellar} {Evolutionary} {Models}?},
	volume = {173},
	url = {https://ui.adsabs.harvard.edu/abs/1999ASPC..173..265C},
	urldate = {2025-09-03},
	booktitle = {Theory and {Test} of {Connective} {Energy} {Transport}},
	author = {Clausen, J. V. and Baraffe, I. and Claret, A. and Vandenberg, D. A.},
	month = jan,
	year = {1999},
	note = {ADS Bibcode: 1999ASPC..173..265C},
	pages = {265},
}

@article{boyajian_stellar_2012b,
	title = {Stellar {Diameters} and {Temperatures}. {II}. {Main}-sequence {K}- and {M}-stars},
	volume = {757},
	issn = {0004-637X},
	url = {https://ui.adsabs.harvard.edu/abs/2012ApJ...757..112B},
	doi = {10.1088/0004-637X/757/2/112},
	urldate = {2025-06-03},
	journal = {The Astrophysical Journal},
	publisher = {IOP},
	author = {Boyajian, Tabetha S. and von Braun, Kaspar and van Belle, Gerard and McAlister, Harold A. and ten Brummelaar, Theo A. and Kane, Stephen R. and Muirhead, Philip S. and Jones, Jeremy and White, Russel and Schaefer, Gail and Ciardi, David and Henry, Todd and López-Morales, Mercedes and Ridgway, Stephen and Gies, Douglas and Jao, Wei-Chun and Rojas-Ayala, Bárbara and Parks, J. Robert and Sturmann, Laszlo and Sturmann, Judit and Turner, Nils H. and Farrington, Chris and Goldfinger, P. J. and Berger, David H.},
	month = oct,
	year = {2012},
	note = {ADS Bibcode: 2012ApJ...757..112B
tex.ids= boyajianStellarDiametersTemperatures2012a},
	pages = {112},
}

@article{boyajian_stellar_2012a,
	title = {Stellar {Diameters} and {Temperatures}. {I}. {Main}-sequence {A}, {F}, and {G} {Stars}},
	volume = {746},
	issn = {0004-637X},
	url = {https://ui.adsabs.harvard.edu/abs/2012ApJ...746..101B},
	doi = {10.1088/0004-637X/746/1/101},
	urldate = {2025-06-03},
	journal = {The Astrophysical Journal},
	publisher = {IOP},
	author = {Boyajian, Tabetha S. and McAlister, Harold A. and van Belle, Gerard and Gies, Douglas R. and ten Brummelaar, Theo A. and von Braun, Kaspar and Farrington, Chris and Goldfinger, P. J. and O'Brien, David and Parks, J. Robert and Richardson, Noel D. and Ridgway, Stephen and Schaefer, Gail and Sturmann, Laszlo and Sturmann, Judit and Touhami, Yamina and Turner, Nils H. and White, Russel},
	month = feb,
	year = {2012},
	note = {ADS Bibcode: 2012ApJ...746..101B},
	pages = {101},
}

@article{baines_vintage_2025,
	title = {Vintage {NPOI}: {New} and {Updated} {Angular} {Diameters} for 145 {Stars}},
	volume = {169},
	issn = {0004-6256},
	shorttitle = {Vintage {NPOI}},
	url = {https://ui.adsabs.harvard.edu/abs/2025AJ....169..293B},
	doi = {10.3847/1538-3881/adc930},
	urldate = {2026-06-09},
	journal = {The Astronomical Journal},
	publisher = {IOP},
	author = {Baines, Ellyn K. and Clark, James H. and Kingsley, Bradley I. and Schmitt, Henrique R. and Stone, Jordan M.},
	month = jun,
	year = {2025},
	note = {ADS Bibcode: 2025AJ....169..293B},
	pages = {293},
}

@article{marcadon_analysis_2020,
	title = {Analysis of eclipsing binaries in multiple stellar systems: the case of {V1200} {Centauri}},
	volume = {499},
	issn = {0035-8711},
	shorttitle = {Analysis of eclipsing binaries in multiple stellar systems},
	url = {https://doi.org/10.1093/mnras/staa3040},
	doi = {10.1093/mnras/staa3040},
	number = {2},
	urldate = {2026-04-18},
	journal = {Monthly Notices of the Royal Astronomical Society},
	author = {Marcadon, F and Hełminiak, K G and Marques, J P and Pawłaszek, R and Sybilski, P and Kozłowski, S K and Ratajczak, M and Konacki, M},
	month = oct,
	year = {2020},
	pages = {3019--3033},
}

@article{pan_starspot_2024,
	title = {Starspot {Coverage} on {Two} {New} {K}-type {Low}-mass {Eclipsing} {Binaries} with {Radius} {Inflation}},
	volume = {168},
	issn = {1538-3881},
	url = {https://doi.org/10.3847/1538-3881/ad84f5},
	doi = {10.3847/1538-3881/ad84f5},
	language = {en},
	number = {6},
	urldate = {2026-04-18},
	journal = {The Astronomical Journal},
	publisher = {The American Astronomical Society},
	author = {Pan, Yang and Frasca, Antonio and Wang, Jia-Xin and Fu, Jian-Ning and Zhang, Xiao-Bin},
	month = nov,
	year = {2024},
	pages = {253},
}

@article{luck_abundances_2017,
	title = {Abundances in the {Local} {Region} {II}: {F}, {G}, and {K} {Dwarfs} and {Subgiants}},
	volume = {153},
	issn = {0004-6256},
	shorttitle = {Abundances in the {Local} {Region} {II}},
	url = {https://ui.adsabs.harvard.edu/abs/2017AJ....153...21L},
	doi = {10.3847/1538-3881/153/1/21},
	urldate = {2026-04-17},
	journal = {The Astronomical Journal},
	publisher = {IOP},
	author = {Luck, R. Earle},
	month = jan,
	year = {2017},
	note = {ADS Bibcode: 2017AJ....153...21L},
	pages = {21},
}

@article{morris_stellar_2019,
	title = {Stellar {Properties} of {Active} {G} and {K} {Stars}: {Exploring} the {Connection} between {Starspots} and {Chromospheric} {Activity}},
	volume = {158},
	issn = {0004-6256},
	shorttitle = {Stellar {Properties} of {Active} {G} and {K} {Stars}},
	url = {https://ui.adsabs.harvard.edu/abs/2019AJ....158..101M},
	doi = {10.3847/1538-3881/ab2e04},
	urldate = {2026-04-17},
	journal = {The Astronomical Journal},
	publisher = {IOP},
	author = {Morris, Brett M. and Curtis, Jason L. and Sakari, Charli and Hawley, Suzanne L. and Agol, Eric},
	month = sep,
	year = {2019},
	note = {ADS Bibcode: 2019AJ....158..101M},
	pages = {101},
}

@article{stefan_uber_1879,
	title = {Uber die {Beziehung} zwischen der {Warmestrahlung} und der {Temperatur} ({About} the relationship between the thermal radiation and temperature])},
	volume = {LXXI},
	journal = {Sitzungsberichte der Kaiserlichen Akademie der Wissenschaften (Wien)},
	author = {Stefan, J},
	year = {1879},
	pages = {391--429},
}

@article{an_orbits_2025,
	title = {Orbits and {Masses} for 156 {Companions} from {Combined} {Astrometry} and {Radial} {Velocities}, and a {Validation} of {Gaia} {Non}-single-star {Solutions}},
	volume = {280},
	issn = {0067-0049},
	url = {https://ui.adsabs.harvard.edu/abs/2025ApJS..280...61A},
	doi = {10.3847/1538-4365/adfa99},
	urldate = {2026-04-16},
	journal = {The Astrophysical Journal Supplement Series},
	publisher = {IOP},
	author = {An, Qier and Brandt, Timothy D. and Brandt, G. Mirek and Venner, Alexander},
	month = oct,
	year = {2025},
	note = {ADS Bibcode: 2025ApJS..280...61A},
	pages = {61},
}

@article{boltzmann_ableitung_1884,
	title = {Ableitung des {Stefan}'schen {Gesetzes}, betreffend die {Abhängigkeit} der {Wärmestrahlung} von der {Temperatur} aus der electromagnetischen {Lichttheorie}},
	volume = {258},
	issn = {0003-3804},
	url = {https://ui.adsabs.harvard.edu/abs/1884AnP...258..291B},
	doi = {10.1002/andp.18842580616},
	urldate = {2026-04-16},
	journal = {Annalen der Physik},
	author = {Boltzmann, Ludwig},
	month = jan,
	year = {1884},
	note = {ADS Bibcode: 1884AnP...258..291B},
	pages = {291--294},
}

@article{brandt_orvara_2021,
	title = {orvara: {An} {Efficient} {Code} to {Fit} {Orbits} {Using} {Radial} {Velocity}, {Absolute}, and/or {Relative} {Astrometry}},
	volume = {162},
	issn = {1538-3881},
	shorttitle = {orvara},
	url = {https://doi.org/10.3847/1538-3881/ac042e},
	doi = {10.3847/1538-3881/ac042e},
	language = {en},
	number = {5},
	urldate = {2026-04-15},
	journal = {The Astronomical Journal},
	publisher = {The American Astronomical Society},
	author = {Brandt, Timothy D. and Dupuy, Trent J. and Li, Yiting and Brandt, G. Mirek and Zeng, Yunlin and Michalik, Daniel and Bardalez Gagliuffi, Daniella C. and Raposo-Pulido, Virginia},
	month = oct,
	year = {2021},
	pages = {186},
}

@article{rosenthal_california_2021,
	title = {The {California} {Legacy} {Survey}. {I}. {A} {Catalog} of 178 {Planets} from {Precision} {Radial} {Velocity} {Monitoring} of 719 {Nearby} {Stars} over {Three} {Decades}},
	volume = {255},
	issn = {0067-0049},
	url = {https://doi.org/10.3847/1538-4365/abe23c},
	doi = {10.3847/1538-4365/abe23c},
	language = {en},
	number = {1},
	urldate = {2026-04-15},
	journal = {The Astrophysical Journal Supplement Series},
	publisher = {The American Astronomical Society},
	author = {Rosenthal, Lee J. and Fulton, Benjamin J. and Hirsch, Lea A. and Isaacson, Howard T. and Howard, Andrew W. and Dedrick, Cayla M. and Sherstyuk, Ilya A. and Blunt, Sarah C. and Petigura, Erik A. and Knutson, Heather A. and Behmard, Aida and Chontos, Ashley and Crepp, Justin R. and Crossfield, Ian J. M. and Dalba, Paul A. and Fischer, Debra A. and Henry, Gregory W. and Kane, Stephen R. and Kosiarek, Molly and Marcy, Geoffrey W. and Rubenzahl, Ryan A. and Weiss, Lauren M. and Wright, Jason T.},
	month = jul,
	year = {2021},
	pages = {8},
}

@inproceedings{radovan_automated_2014,
	title = {The automated planet finder at {Lick} {Observatory}},
	volume = {9145},
	url = {https://www.spiedigitallibrary.org/conference-proceedings-of-spie/9145/91452B/The-automated-planet-finder-at-Lick-Observatory/10.1117/12.2057310.full},
	doi = {10.1117/12.2057310},
	urldate = {2026-04-15},
	booktitle = {Ground-based and {Airborne} {Telescopes} {V}},
	publisher = {SPIE},
	author = {Radovan, Matt V. and Lanclos, Kyle and Holden, Bradford P. and Kibrick, Robert I. and Allen, S. L. and Deich, William T. S. and Rivera, Eugenio and Burt, Jennifer and Fulton, Benjamin and Butler, Paul and Vogt, Steven S.},
	month = jul,
	year = {2014},
	pages = {842--853},
}

@inproceedings{vogt_hires_1994,
	title = {{HIRES}: the high-resolution echelle spectrometer on the {Keck} 10-m {Telescope}},
	volume = {2198},
	shorttitle = {{HIRES}},
	url = {https://www.spiedigitallibrary.org/conference-proceedings-of-spie/2198/0000/HIRES--the-high-resolution-echelle-spectrometer-on-the-Keck/10.1117/12.176725.full},
	doi = {10.1117/12.176725},
	urldate = {2026-04-15},
	booktitle = {Instrumentation in {Astronomy} {VIII}},
	publisher = {SPIE},
	author = {Vogt, Steven S. and Allen, Steven L. and Bigelow, Bruce C. and Bresee, L. and Brown, William E. and Cantrall, T. and Conrad, Albert and Couture, M. and Delaney, C. and Epps, Harland W. and Hilyard, Darrie and Hilyard, David F. and Horn, E. and Jern, Neal and Kanto, D. and Keane, Michael J. and Kibrick, Robert I. and Lewis, James W. and Osborne, Jack and Pardeilhan, Gerard H. and Pfister, Terry and Ricketts, T. and Robinson, Lloyd B. and Stover, Richard J. and Tucker, Dean and Ward, Janice M. and Wei, Mingzhi},
	month = jun,
	year = {1994},
	pages = {362--375},
}

@article{vogt_lick_1987,
	title = {{THE} {LICK} {OBSERVATORY} {HAMILTON} {ECHELLE} {SPECTROMETER}.},
	volume = {99},
	issn = {1538-3873},
	url = {https://iopscience.iop.org/article/10.1086/132107},
	doi = {10.1086/132107},
	language = {en},
	number = {621},
	urldate = {2026-04-15},
	journal = {Publications of the Astronomical Society of the Pacific},
	publisher = {IOP Publishing},
	author = {Vogt, Steven S.},
	month = nov,
	year = {1987},
	pages = {1214},
}

@article{chabrier_evolution_2007,
	title = {Evolution of low-mass star and brown dwarf eclipsing binaries},
	volume = {472},
	issn = {0004-6361},
	url = {https://ui.adsabs.harvard.edu/abs/2007A&A...472L..17C},
	doi = {10.1051/0004-6361:20077702},
	urldate = {2026-04-15},
	journal = {Astronomy and Astrophysics},
	publisher = {EDP},
	author = {Chabrier, G. and Gallardo, J. and Baraffe, I.},
	month = sep,
	year = {2007},
	note = {ADS Bibcode: 2007A\&A...472L..17C},
	pages = {L17--L20},
}

@article{nguyen_parsec_2025,
	title = {{PARSEC} {V2}.0: {Rotating} tracks and isochrones for seven additional metallicities in the range {Z} = 0.0001─0.03},
	volume = {701},
	issn = {0004-6361},
	shorttitle = {{PARSEC} {V2}.0},
	url = {https://ui.adsabs.harvard.edu/abs/2025A&A...701A.258N},
	doi = {10.1051/0004-6361/202556005},
	urldate = {2026-04-15},
	journal = {Astronomy and Astrophysics},
	publisher = {EDP},
	author = {Nguyen, C. T. and Costa, G. and Bressan, A. and Girardi, L. and Cescutti, G. and Korn, A. J. and Volpato, G. and Chen, Y. and Pastorelli, G. and Trabucchi, M. and Shepherd, K. G. and Ettorre, G. and Zaggia, S.},
	month = sep,
	year = {2025},
	note = {ADS Bibcode: 2025A\&A...701A.258N},
	pages = {A258},
}

@misc{li_century_2026,
	title = {A {Century} of {Radial} {Velocity} and {Astrometric} {Monitoring} of 70 {Oph} {AB}: {New} {PFS} {Data} and {Constraints} on {Planetary} {Companions}},
	shorttitle = {A {Century} of {Radial} {Velocity} and {Astrometric} {Monitoring} of 70 {Oph} {AB}},
	url = {http://arxiv.org/abs/2603.20044},
	doi = {10.48550/arXiv.2603.20044},
	urldate = {2026-03-24},
	publisher = {arXiv},
	author = {Li, Yiting and Meyer, Michael R. and D'Angiolillo, Skylar and Kane, Stephen R. and Butler, R. Paul and Shectman, Stephen A. and Mamajek, Eric E. and Teske, Johanna and Lubin, Jack and Robertson, Paul and Christiansen, Jessie L. and Isaacson, Howard and Harada, Caleb K. and Holden, Bradford and Cochran, William D. and Endl, Michael and Burt, Jennifer and Becker, Juliette and Jankowski, Alyssa and Tuthill, Peter and Clark, Catherine A. and Roettenbacher, Rachael M. and Nielsen, Eric and Bendek, Eduardo and Tokadjian, Armen and Roberson, William and Kratter, Kaitlin M. and Bergin, Edwin and Osip, Dave and Crane, Jeffrey D. and Davis, Alex and Vasisht, Gautam},
	month = mar,
	year = {2026},
	note = {arXiv:2603.20044 [astro-ph]},
}

@article{harada_sporeshwo_2025,
	title = {{SPORES}-{HWO}. {II}. {Companion} {Mass} {Limits} and {Updated} {Planet} {Properties} for 120 {Future} {Exoplanet} {Imaging} {Targets} from 35 yr of {Precise} {Doppler} {Monitoring}},
	volume = {170},
	issn = {0004-6256},
	url = {https://ui.adsabs.harvard.edu/abs/2025AJ....170..343H},
	doi = {10.3847/1538-3881/ae0b62},
	urldate = {2026-03-05},
	journal = {The Astronomical Journal},
	publisher = {IOP},
	author = {Harada, Caleb K. and Dressing, Courtney D. and Turtelboom, Emma V. and Kane, Stephen R. and Blunt, Sarah and Dietrich, Jamie and Hinkel, Natalie R. and Li, Zhexing and Mamajek, Eric and Rice, Malena and Tuchow, Noah W. and Wittenmyer, Robert A. and Chin, Christopher and Fernandez, Aidan and Kulkarni, Shivani and Lin, Emerald and Liu, Nykole and Liu, Remy and Nathan, Abhi and Zbriger, Adam},
	month = dec,
	year = {2025},
	note = {ADS Bibcode: 2025AJ....170..343H},
	pages = {343},
}

@article{mamajek_provisional_2024,
	title = {Provisional {NASA} {ExEP} {Mission} {Target} {Star} {List} for the {Habitable} {Worlds} {Observatory}},
	volume = {56},
	issn = {2330-9458, 0002-7537},
	url = {https://baas.aas.org/pub/2024n4i628p17/release/1},
	doi = {10.3847/25c2cfeb.e7ccaa3b},
	language = {en},
	number = {4},
	urldate = {2026-03-05},
	journal = {Bulletin of the AAS},
	author = {Mamajek, Eric and Stapelfeldt, Karl},
	month = apr,
	year = {2024},
}

@article{pan_analysis_2025,
	title = {Analysis of {Six} {Long}-period {Low}-mass {Eclipsing} {Binaries}},
	volume = {170},
	issn = {0004-6256},
	url = {https://ui.adsabs.harvard.edu/abs/2025AJ....170..252P},
	doi = {10.3847/1538-3881/ae04ed},
	urldate = {2026-02-05},
	journal = {The Astronomical Journal},
	publisher = {IOP},
	author = {Pan, Yang and Frasca, Antonio and Wang, Jiaxin and Fu, Jianning and Zhang, Xiaobin},
	month = nov,
	year = {2025},
	note = {ADS Bibcode: 2025AJ....170..252P},
	pages = {252},
}

@article{brogaard_properties_2021,
	title = {Properties of the {Hyades}, the eclipsing binary {HD} 27130, and the oscillating red giant ɛ {Tauri}},
	volume = {645},
	issn = {0004-6361},
	url = {https://ui.adsabs.harvard.edu/abs/2021A&A...645A..25B},
	doi = {10.1051/0004-6361/202039250},
	urldate = {2026-02-05},
	journal = {Astronomy and Astrophysics},
	publisher = {EDP},
	author = {Brogaard, K. and Pakštienė, E. and Grundahl, F. and Mikolaitis, Š. and Tautvaišienė, G. and Slumstrup, D. and Talens, G. J. J. and VandenBerg, D. A. and Miglio, A. and Arentoft, T. and Kjeldsen, H. and Janulis, R. and Drazdauskas, A. and Marchini, A. and Minkevičiūtė, R. and Stonkutė, E. and Bagdonas, V. and Fredslund Andersen, M. and Jessen-Hansen, J. and Pallé, P. L. and Dorval, P. and Snellen, I. A. G. and Otten, G. P. P. L. and White, T. R.},
	month = jan,
	year = {2021},
	note = {ADS Bibcode: 2021A\&A...645A..25B},
	pages = {A25},
}

@article{sandquist_k2_2018,
	title = {The {K2} {M67} {Study}: {A} {Curiously} {Young} {Star} in an {Eclipsing} {Binary} in an {Old} {Open} {Cluster}},
	volume = {155},
	issn = {0004-6256},
	shorttitle = {The {K2} {M67} {Study}},
	url = {https://ui.adsabs.harvard.edu/abs/2018AJ....155..152S},
	doi = {10.3847/1538-3881/aab0ff},
	urldate = {2026-02-05},
	journal = {The Astronomical Journal},
	publisher = {IOP},
	author = {Sandquist, Eric L. and Mathieu, Robert D. and Quinn, Samuel N. and Pollack, Maxwell L. and Latham, David W. and Brown, Timothy M. and Esselstein, Rebecca and Aigrain, Suzanne and Parviainen, Hannu and Vanderburg, Andrew and Stello, Dennis and Somers, Garrett and Pinsonneault, Marc H. and Tayar, Jamie and Orosz, Jerome A. and Bedin, Luigi R. and Libralato, Mattia and Malavolta, Luca and Nardiello, Domenico},
	month = apr,
	year = {2018},
	note = {ADS Bibcode: 2018AJ....155..152S},
	pages = {152},
}

@article{torres_eclipsing_2019,
	title = {Eclipsing {Binaries} in the {Open} {Cluster} {Ruprecht} 147. {II}. {Epic} 219568666},
	volume = {887},
	issn = {0004-637X},
	url = {https://ui.adsabs.harvard.edu/abs/2019ApJ...887..109T},
	doi = {10.3847/1538-4357/ab54c5},
	urldate = {2026-02-05},
	journal = {The Astrophysical Journal},
	publisher = {IOP},
	author = {Torres, Guillermo and Vanderburg, Andrew and Curtis, Jason L. and Ciardi, David and Kraus, Adam L. and Rizzuto, Aaron C. and Ireland, Michael J. and Lund, Michael B. and Christiansen, Jessie L. and Beichman, Charles A.},
	month = dec,
	year = {2019},
	note = {ADS Bibcode: 2019ApJ...887..109T},
	pages = {109},
}

@article{helminiak_highresolution_2024,
	title = {High-resolution spectroscopy of detached eclipsing binaries during total eclipses},
	volume = {691},
	issn = {0004-6361},
	url = {https://ui.adsabs.harvard.edu/abs/2024A&A...691A.170H},
	doi = {10.1051/0004-6361/202450607},
	urldate = {2026-02-05},
	journal = {Astronomy and Astrophysics},
	publisher = {EDP},
	author = {Hełminiak, K. G. and Olszewska, J. M. and Puciata-Mroczynska, M. and Pawar, T.},
	month = nov,
	year = {2024},
	note = {ADS Bibcode: 2024A\&A...691A.170H},
	pages = {A170},
}

@article{sandquist_analysis_2013,
	title = {Analysis of {Detached} {Eclipsing} {Binaries} near the {Turnoff} of the {Open} {Cluster} {NGC} 7142},
	volume = {146},
	issn = {0004-6256},
	url = {https://ui.adsabs.harvard.edu/abs/2013AJ....146...40S},
	doi = {10.1088/0004-6256/146/2/40},
	urldate = {2026-02-05},
	journal = {The Astronomical Journal},
	publisher = {IOP},
	author = {Sandquist, Eric L. and Shetrone, Matthew and Serio, Andrew W. and Orosz, Jerome},
	month = aug,
	year = {2013},
	note = {ADS Bibcode: 2013AJ....146...40S},
	pages = {40},
}

@misc{southworth_debcat_2014,
	title = {The {DEBCat} detached eclipsing binary catalogue},
	url = {http://arxiv.org/abs/1411.1219},
	doi = {10.48550/arXiv.1411.1219},
	urldate = {2026-02-03},
	publisher = {arXiv},
	author = {Southworth, John},
	month = nov,
	year = {2014},
	note = {arXiv:1411.1219 [astro-ph]},
}

@article{graczyk_surface_2022,
	title = {Surface brightness-colour relations of dwarf stars from detached eclipsing binaries - {I}. {Calibrating} sample},
	volume = {666},
	copyright = {© D. Graczyk et al. 2022},
	issn = {0004-6361, 1432-0746},
	url = {https://www.aanda.org/articles/aa/abs/2022/10/aa44122-22/aa44122-22.html},
	doi = {10.1051/0004-6361/202244122},
	language = {en},
	urldate = {2026-02-03},
	journal = {Astronomy \& Astrophysics},
	publisher = {EDP Sciences},
	author = {Graczyk, D. and Pietrzyński, G. and Galan, C. and Southworth, J. and Gieren, W. and Kałuszyński, M. and Zgirski, B. and Gallenne, A. and Górski, M. and Hajdu, G. and Karczmarek, P. and Kervella, P. and Maxted, P. F. L. and Nardetto, N. and Narloch, W. and Pilecki, B. and Pych, W. and Garcia, G. Rojas and Storm, J. and Suchomska, K. and Taormina, M. and Wielgórski, P.},
	month = oct,
	year = {2022},
	pages = {A128},
}

@article{helminiak_orbital_2011,
	title = {Orbital and physical parameters of eclipsing binaries from the {All}-{Sky} {Automated} {Survey} catalogue. {II}. {Two} spotted {M} {\textless} 1 {M}\_⊙ systems at different evolutionary stages},
	volume = {526},
	issn = {0004-6361},
	url = {https://ui.adsabs.harvard.edu/abs/2011A&A...526A..29H},
	doi = {10.1051/0004-6361/200913336},
	urldate = {2026-02-03},
	journal = {Astronomy and Astrophysics},
	publisher = {EDP},
	author = {Hełminiak, K. G. and Konacki, M.},
	month = feb,
	year = {2011},
	note = {ADS Bibcode: 2011A\&A...526A..29H},
	pages = {A29},
}

@article{valenti_spectroscopic_2005,
	title = {Spectroscopic {Properties} of {Cool} {Stars} ({SPOCS}). {I}. 1040 {F}, {G}, and {K} {Dwarfs} from {Keck}, {Lick}, and {AAT} {Planet} {Search} {Programs}},
	volume = {159},
	issn = {0067-0049},
	url = {https://ui.adsabs.harvard.edu/abs/2005ApJS..159..141V},
	doi = {10.1086/430500},
	urldate = {2026-01-29},
	journal = {The Astrophysical Journal Supplement Series},
	publisher = {IOP},
	author = {Valenti, Jeff A. and Fischer, Debra A.},
	month = jul,
	year = {2005},
	note = {ADS Bibcode: 2005ApJS..159..141V},
	pages = {141--166},
}

@article{vines_ariadne_2022,
	title = {ariadne: measuring accurate and precise stellar parameters through {SED} fitting},
	volume = {513},
	issn = {0035-8711},
	shorttitle = {ariadne},
	url = {https://doi.org/10.1093/mnras/stac956},
	doi = {10.1093/mnras/stac956},
	number = {2},
	urldate = {2026-01-29},
	journal = {Monthly Notices of the Royal Astronomical Society},
	author = {Vines, Jose I and Jenkins, James S},
	month = jun,
	year = {2022},
	pages = {2719--2731},
}

@article{colliercameron_mainsequence_2009,
	title = {The main-sequence rotation-colour relation in the {Coma} {Berenices} open cluster},
	volume = {400},
	issn = {0035-8711},
	url = {https://ui.adsabs.harvard.edu/abs/2009MNRAS.400..451C},
	doi = {10.1111/j.1365-2966.2009.15476.x},
	urldate = {2025-10-22},
	journal = {Monthly Notices of the Royal Astronomical Society},
	publisher = {OUP},
	author = {Collier Cameron, A. and Davidson, V. A. and Hebb, L. and Skinner, G. and Anderson, D. R. and Christian, D. J. and Clarkson, W. I. and Enoch, B. and Irwin, J. and Joshi, Y. and Haswell, C. A. and Hellier, C. and Horne, K. D. and Kane, S. R. and Lister, T. A. and Maxted, P. F. L. and Norton, A. J. and Parley, N. and Pollacco, D. and Ryans, R. and Scholz, A. and Skillen, I. and Smalley, B. and Street, R. A. and West, R. G. and Wilson, D. M. and Wheatley, P. J.},
	month = nov,
	year = {2009},
	note = {ADS Bibcode: 2009MNRAS.400..451C},
	pages = {451--462},
}

@article{helminiak_orbital_2019,
	title = {Orbital and physical parameters of eclipsing binaries from the {All}-{Sky} {Automated} {Survey} catalogue - {X}. {Three} high-contrast systems with secondaries detected with {IR} spectroscopy},
	volume = {622},
	copyright = {© ESO 2019},
	issn = {0004-6361, 1432-0746},
	url = {https://www.aanda.org/articles/aa/abs/2019/02/aa32482-17/aa32482-17.html},
	doi = {10.1051/0004-6361/201732482},
	language = {en},
	urldate = {2025-10-08},
	journal = {Astronomy \& Astrophysics},
	publisher = {EDP Sciences},
	author = {Hełminiak, K. G. and Tokovinin, A. and Niemczura, E. and Pawłaszek, R. and Yanagisawa, K. and Brahm, R. and Espinoza, N. and Ukita, N. and Kambe, E. and Ratajczak, M. and Hempel, M. and Jordán, A. and Konacki, M. and Sybilski, P. and Kozłowski, S. K. and Litwicki, M. and Tamura, M.},
	month = feb,
	year = {2019},
	pages = {A114},
}

@misc{lopez-morales_testing_2007,
	address = {eprint: arXiv:astro-ph/0603748},
	title = {Testing {Low}-{Mass} {Stellar} {Models}: {Three} {New} {Detached} {Eclipsing} {Binaries} below 0.{75Msun}},
	shorttitle = {Testing {Low}-{Mass} {Stellar} {Models}},
	url = {https://ui.adsabs.harvard.edu/abs/2007ASPC..362...26L},
	doi = {10.48550/arXiv.astro-ph/0603748},
	urldate = {2025-09-16},
	publisher = {arXiv},
	author = {López-Morales, M. and Shaw, J. S.},
	month = jun,
	year = {2007},
	note = {Conference Name: The Seventh Pacific Rim Conference on Stellar Astrophysics
Volume: 362
ADS Bibcode: 2007ASPC..362...26L},
}

@article{stassun_dynamical_2004,
	title = {Dynamical {Mass} {Constraints} on {Low}-{Mass} {Pre}-{Main}-{Sequence} {Stellar} {Evolutionary} {Tracks}: {An} {Eclipsing} {Binary} in {Orion} with a 1.0 {Msolar} {Primary} and a 0.7 {Msolar} {Secondary}},
	volume = {151},
	issn = {0067-0049},
	shorttitle = {Dynamical {Mass} {Constraints} on {Low}-{Mass} {Pre}-{Main}-{Sequence} {Stellar} {Evolutionary} {Tracks}},
	url = {https://ui.adsabs.harvard.edu/abs/2004ApJS..151..357S},
	doi = {10.1086/382353},
	urldate = {2025-09-16},
	journal = {The Astrophysical Journal Supplement Series},
	author = {Stassun, Keivan G. and Mathieu, Robert D. and Vaz, Luiz Paulo R. and Stroud, Nicholas and Vrba, Frederick J.},
	month = apr,
	year = {2004},
	note = {ADS Bibcode: 2004ApJS..151..357S},
	pages = {357--385},
}

@article{morales_absolute_2009,
	title = {Absolute {Dimensions} of the {G7}+{K7} {Eclipsing} {Binary} {Star} {IM} {Virginis}: {Discrepancies} with {Stellar} {Evolution} {Models}},
	volume = {707},
	issn = {0004-637X},
	shorttitle = {Absolute {Dimensions} of the {G7}+{K7} {Eclipsing} {Binary} {Star} {IM} {Virginis}},
	url = {https://ui.adsabs.harvard.edu/abs/2009ApJ...707..671M},
	doi = {10.1088/0004-637X/707/1/671},
	urldate = {2025-09-11},
	journal = {The Astrophysical Journal},
	author = {Morales, Juan Carlos and Torres, Guillermo and Marschall, Laurence A. and Brehm, William},
	month = dec,
	year = {2009},
	note = {ADS Bibcode: 2009ApJ...707..671M},
	pages = {671--685},
}

@misc{rozyczka_new_2010,
	title = {A {New} {Lower} {Main} {Sequence} {Eclipsing} {Binary} with {Detached} {Components}},
	url = {http://arxiv.org/abs/0910.2543},
	doi = {10.48550/arXiv.0910.2543},
	urldate = {2025-09-10},
	publisher = {arXiv},
	author = {Rozyczka, M. and Kaluzny, J. and Pietrukowicz, P. and Pych, W. and Mazur, B. and Catelan, M. and Thompson, I. B.},
	month = jan,
	year = {2010},
	note = {arXiv:0910.2543 [astro-ph]},
}

@article{dotter_dartmouth_2008,
	title = {The {Dartmouth} {Stellar} {Evolution} {Database}},
	volume = {178},
	issn = {0067-0049},
	url = {https://ui.adsabs.harvard.edu/abs/2008ApJS..178...89D},
	doi = {10.1086/589654},
	urldate = {2025-09-08},
	journal = {The Astrophysical Journal Supplement Series},
	author = {Dotter, Aaron and Chaboyer, Brian and Jevremović, Darko and Kostov, Veselin and Baron, E. and Ferguson, Jason W.},
	month = sep,
	year = {2008},
	note = {ADS Bibcode: 2008ApJS..178...89D},
	pages = {89--101},
}

@article{white_interferometric_2018,
	title = {Interferometric diameters of five evolved intermediate-mass planet-hosting stars measured with {PAVO} at the {CHARA} {Array}},
	volume = {477},
	issn = {0035-8711},
	url = {https://ui.adsabs.harvard.edu/abs/2018MNRAS.477.4403W},
	doi = {10.1093/mnras/sty898},
	urldate = {2025-09-04},
	journal = {Monthly Notices of the Royal Astronomical Society},
	author = {White, T. R. and Huber, D. and Mann, A. W. and Casagrande, L. and Grunblatt, S. K. and Justesen, A. B. and Silva Aguirre, V. and Bedding, T. R. and Ireland, M. J. and Schaefer, G. H. and Tuthill, P. G.},
	month = jul,
	year = {2018},
	note = {ADS Bibcode: 2018MNRAS.477.4403W},
	pages = {4403--4413},
}

@article{rabus_discontinuity_2019,
	title = {A discontinuity in the {Teff}–radius relation of {M}-dwarfs},
	volume = {484},
	issn = {0035-8711},
	url = {https://doi.org/10.1093/mnras/sty3430},
	doi = {10.1093/mnras/sty3430},
	number = {2},
	urldate = {2025-09-04},
	journal = {Monthly Notices of the Royal Astronomical Society},
	author = {Rabus, Markus and Lachaume, Régis and Jordán, Andrés and Brahm, Rafael and Boyajian, Tabetha and von Braun, Kaspar and Espinoza, Néstor and Berger, Jean-Philippe and Le Bouquin, Jean-Baptiste and Absil, Olivier},
	month = apr,
	year = {2019},
	pages = {2674--2683},
}

@article{baines_33_2023,
	title = {33 {New} {Stellar} {Angular} {Diameters} from the {NPOI}, and {Nearly} 180 {NPOI} {Diameters} as an {Ensemble}},
	volume = {166},
	issn = {0004-6256},
	url = {https://ui.adsabs.harvard.edu/abs/2023AJ....166..268B},
	doi = {10.3847/1538-3881/ad08be},
	urldate = {2025-09-04},
	journal = {The Astronomical Journal},
	author = {Baines, Ellyn K. and Clark, III, James H. and Schmitt, Henrique R. and Stone, Jordan M. and von Braun, Kaspar},
	month = dec,
	year = {2023},
	note = {ADS Bibcode: 2023AJ....166..268B},
	pages = {268},
}

@article{karovicova_fundamental_2022,
	title = {Fundamental stellar parameters of benchmark stars from {CHARA} interferometry. {II}. {Dwarf} stars},
	volume = {658},
	issn = {0004-6361},
	url = {https://ui.adsabs.harvard.edu/abs/2022A&A...658A..47K},
	doi = {10.1051/0004-6361/202141833},
	urldate = {2025-09-04},
	journal = {Astronomy and Astrophysics},
	author = {Karovicova, I. and White, T. R. and Nordlander, T. and Casagrande, L. and Ireland, M. and Huber, D.},
	month = feb,
	year = {2022},
	note = {ADS Bibcode: 2022A\&A...658A..47K},
	pages = {A47},
}

@book{vonbraun_extrasolar_2017,
	title = {Extrasolar {Planets} and {Their} {Host} {Stars}},
	url = {https://ui.adsabs.harvard.edu/abs/2017ephs.book.....V},
	doi = {10.1007/978-3-319-61198-3},
	urldate = {2025-09-04},
	author = {von Braun, Kaspar and Boyajian, Tabetha},
	month = jan,
	year = {2017},
	note = {Publication Title: Extrasolar Planets and Their Host Stars
ADS Bibcode: 2017ephs.book.....V},
}

@article{segransan_first_2003,
	title = {First radius measurements of very low mass stars with the {VLTI}},
	volume = {397},
	copyright = {© ESO, 2003},
	issn = {0004-6361, 1432-0746},
	url = {https://www.aanda.org/articles/aa/abs/2003/03/aaej152/aaej152.html},
	doi = {10.1051/0004-6361:20021714},
	language = {en},
	number = {3},
	urldate = {2025-09-04},
	journal = {Astronomy \& Astrophysics},
	publisher = {EDP Sciences},
	author = {Ségransan, D. and Kervella, P. and Forveille, T. and Queloz, D.},
	month = jan,
	year = {2003},
	pages = {L5--L8},
}

@article{tofflemire_lowmass_2023,
	title = {A {Low}-mass, {Pre}-main-sequence {Eclipsing} {Binary} in the 40 {Myr} {Columba} {Association}—{Fundamental} {Stellar} {Parameters} and {Modeling} the {Effect} of {Star} {Spots}},
	volume = {165},
	issn = {1538-3881},
	url = {https://dx.doi.org/10.3847/1538-3881/aca60f},
	doi = {10.3847/1538-3881/aca60f},
	language = {en},
	number = {2},
	urldate = {2025-09-03},
	journal = {The Astronomical Journal},
	publisher = {The American Astronomical Society},
	author = {Tofflemire, Benjamin M. and Kraus, Adam L. and Mann, Andrew W. and Newton, Elisabeth R. and Gully-Santiago, Michael A. and Vanderburg, Andrew and Waalkes, William C. and Berta-Thompson, Zachory K. and Collins, Kevin I. and Collins, Karen A. and Nielsen, Louise D. and Bouchy, François and Ziegler, Carl and Briceño, César and Law, Nicholas M.},
	month = jan,
	year = {2023},
	pages = {46},
}

@article{jones_physics_2020,
	title = {Physics of {Eclipsing} {Binaries}. {IV}. {The} {Impact} of {Interstellar} {Extinction} on the {Light} {Curves} of {Eclipsing} {Binaries}},
	volume = {247},
	issn = {0067-0049},
	url = {https://dx.doi.org/10.3847/1538-4365/ab7927},
	doi = {10.3847/1538-4365/ab7927},
	language = {en},
	number = {2},
	urldate = {2025-09-03},
	journal = {The Astrophysical Journal Supplement Series},
	publisher = {The American Astronomical Society},
	author = {Jones, David and Conroy, Kyle E. and Horvat, Martin and Giammarco, Joseph and Kochoska, Angela and Pablo, Herbert and Brown, Alex J. and Sowicka, Paulina and Prša, Andrej},
	month = apr,
	year = {2020},
	pages = {63},
}

@article{prsa_influence_2005,
	title = {Influence of {Interstellar} and {Atmospheric} {Extinction} on {Light} {Curves} of {Eclipsing} {Binaries}},
	volume = {296},
	issn = {1572-946X},
	url = {https://doi.org/10.1007/s10509-005-4841-0},
	doi = {10.1007/s10509-005-4841-0},
	language = {en},
	number = {1},
	urldate = {2025-09-03},
	journal = {Astrophysics and Space Science},
	author = {Prša, A. and Zwitter, T.},
	month = apr,
	year = {2005},
	pages = {315--320},
}

@article{barnes_ages_2007,
	title = {Ages for {Illustrative} {Field} {Stars} {Using} {Gyrochronology}: {Viability}, {Limitations}, and {Errors}},
	volume = {669},
	issn = {0004-637X},
	shorttitle = {Ages for {Illustrative} {Field} {Stars} {Using} {Gyrochronology}},
	url = {https://iopscience.iop.org/article/10.1086/519295/meta},
	doi = {10.1086/519295},
	language = {en},
	number = {2},
	urldate = {2025-09-03},
	journal = {The Astrophysical Journal},
	publisher = {IOP Publishing},
	author = {Barnes, Sydney A.},
	month = nov,
	year = {2007},
	pages = {1167},
}

@article{vos_absolute_2012,
	title = {Absolute dimensions of solar-type eclipsing binaries - {EF} {Aquarii}: a {G0} test for stellar evolution models},
	volume = {540},
	copyright = {© ESO, 2012},
	issn = {0004-6361, 1432-0746},
	shorttitle = {Absolute dimensions of solar-type eclipsing binaries - {EF} {Aquarii}},
	url = {https://www.aanda.org/articles/aa/abs/2012/04/aa18606-11/aa18606-11.html},
	doi = {10.1051/0004-6361/201118606},
	language = {en},
	urldate = {2025-09-03},
	journal = {Astronomy \& Astrophysics},
	publisher = {EDP Sciences},
	author = {Vos, J. and Clausen, J. V. and Jørgensen, U. G. and Østensen, R. H. and Claret, A. and Hillen, M. and Exter, K.},
	month = apr,
	year = {2012},
	pages = {A64},
}

@article{clausen_absolute_2009,
	title = {Absolute dimensions of solar-type eclipsing binaries - {II}. {V636} {Centauri}: {A} 1.05 primary with an active, cool, oversize 0.85 secondary},
	volume = {502},
	copyright = {© ESO, 2009},
	issn = {0004-6361, 1432-0746},
	shorttitle = {Absolute dimensions of solar-type eclipsing binaries - {II}. {V636} {Centauri}},
	url = {https://www.aanda.org/articles/aa/abs/2009/28/aa12362-09/aa12362-09.html},
	doi = {10.1051/0004-6361/200912362},
	language = {en},
	number = {1},
	urldate = {2025-09-03},
	journal = {Astronomy \& Astrophysics},
	publisher = {EDP Sciences},
	author = {Clausen, J. V. and Bruntt, H. and Claret, A. and Larsen, A. and Andersen, J. and Nordström, B. and Giménez, A.},
	month = jul,
	year = {2009},
	pages = {253--265},
}

@article{somers_spots_2020,
	title = {The {SPOTS} {Models}: {A} {Grid} of {Theoretical} {Stellar} {Evolution} {Tracks} and {Isochrones} for {Testing} the {Effects} of {Starspots} on {Structure} and {Colors}},
	volume = {891},
	issn = {0004-637X},
	shorttitle = {The {SPOTS} {Models}},
	url = {https://dx.doi.org/10.3847/1538-4357/ab722e},
	doi = {10.3847/1538-4357/ab722e},
	language = {en},
	number = {1},
	urldate = {2025-09-03},
	journal = {The Astrophysical Journal},
	publisher = {The American Astronomical Society},
	author = {Somers, Garrett and Cao, Lyra and Pinsonneault, Marc H.},
	month = mar,
	year = {2020},
	pages = {29},
}

@article{morrell_exploring_2019,
	title = {Exploring the {M}-dwarf {Luminosity}–{Temperature}–{Radius} relationships using {Gaia} {DR2}},
	volume = {489},
	issn = {0035-8711},
	url = {https://doi.org/10.1093/mnras/stz2242},
	doi = {10.1093/mnras/stz2242},
	number = {2},
	urldate = {2025-09-03},
	journal = {Monthly Notices of the Royal Astronomical Society},
	author = {Morrell, Sam and Naylor, Tim},
	month = oct,
	year = {2019},
	pages = {2615--2633},
}

@article{torres_eclipsing_2006,
	title = {The {Eclipsing} {Binary} {V1061} {Cygni}: {Confronting} {Stellar} {Evolution} {Models} for {Active} and {Inactive} {Solar}-{Type} {Stars}},
	volume = {640},
	issn = {0004-637X},
	shorttitle = {The {Eclipsing} {Binary} {V1061} {Cygni}},
	url = {https://iopscience.iop.org/article/10.1086/500188/meta},
	doi = {10.1086/500188},
	language = {en},
	number = {2},
	urldate = {2025-09-03},
	journal = {The Astrophysical Journal},
	publisher = {IOP Publishing},
	author = {Torres, Guillermo and Lacy, Claud H. and Marschall, Laurence A. and Sheets, Holly A. and Mader, Jeff A.},
	month = apr,
	year = {2006},
	pages = {1018},
}

@article{costa_multiple_2019,
	title = {Multiple stellar populations in {NGC} 1866. {New} clues from {Cepheids} and colour-magnitude diagram},
	volume = {631},
	issn = {0004-6361},
	url = {https://ui.adsabs.harvard.edu/abs/2019A&A...631A.128C},
	doi = {10.1051/0004-6361/201936409},
	urldate = {2025-08-25},
	journal = {Astronomy and Astrophysics},
	author = {Costa, Guglielmo and Girardi, Léo and Bressan, Alessandro and Chen, Yang and Goudfrooij, Paul and Marigo, Paola and Rodrigues, Thaíse S. and Lanza, Antonio},
	month = nov,
	year = {2019},
	note = {ADS Bibcode: 2019A\&A...631A.128C},
	pages = {A128},
}

@article{costa_mixing_2019,
	title = {Mixing by overshooting and rotation in intermediate-mass stars},
	volume = {485},
	issn = {0035-8711},
	url = {https://ui.adsabs.harvard.edu/abs/2019MNRAS.485.4641C},
	doi = {10.1093/mnras/stz728},
	urldate = {2025-08-25},
	journal = {Monthly Notices of the Royal Astronomical Society},
	author = {Costa, Guglielmo and Girardi, Léo and Bressan, Alessandro and Marigo, Paola and Rodrigues, Thaíse S. and Chen, Yang and Lanza, Antonio and Goudfrooij, Paul},
	month = jun,
	year = {2019},
	note = {ADS Bibcode: 2019MNRAS.485.4641C},
	pages = {4641--4657},
}

@article{choi_mesa_2016,
	title = {Mesa {Isochrones} and {Stellar} {Tracks} ({MIST}). {I}. {Solar}-scaled {Models}},
	volume = {823},
	issn = {0004-637X},
	url = {https://ui.adsabs.harvard.edu/abs/2016ApJ...823..102C},
	doi = {10.3847/0004-637X/823/2/102},
	urldate = {2025-08-25},
	journal = {The Astrophysical Journal},
	author = {Choi, Jieun and Dotter, Aaron and Conroy, Charlie and Cantiello, Matteo and Paxton, Bill and Johnson, Benjamin D.},
	month = jun,
	year = {2016},
	note = {ADS Bibcode: 2016ApJ...823..102C},
	pages = {102},
}

@article{evans_orbit_2024,
	title = {The {Orbit} and {Dynamical} {Mass} of {Polaris}: {Observations} with the {CHARA} {Array}},
	volume = {971},
	issn = {0004-637X},
	shorttitle = {The {Orbit} and {Dynamical} {Mass} of {Polaris}},
	url = {https://dx.doi.org/10.3847/1538-4357/ad5e7a},
	doi = {10.3847/1538-4357/ad5e7a},
	language = {en},
	number = {2},
	urldate = {2025-08-19},
	journal = {The Astrophysical Journal},
	publisher = {The American Astronomical Society},
	author = {Evans, Nancy Remage and Schaefer, Gail H. and Gallenne, Alexandre and Torres, Guillermo and Horch, Elliott P. and Anderson, Richard I. and Monnier, John D. and Roettenbacher, Rachael M. and Baron, Fabien and Anugu, Narsireddy and Davidson, James W. and Kervella, Pierre and Bras, Garance and Proffitt, Charles and Mérand, Antoine and Karovska, Margarita and Jones, Jeremy and Lanthermann, Cyprien and Kraus, Stefan and Codron, Isabelle and Bond, Howard E. and Viviani, Giordano},
	month = aug,
	year = {2024},
	pages = {190},
}

@article{monnier_imaging_2007,
	title = {Imaging the {Surface} of {Altair}},
	volume = {317},
	url = {https://www.science.org/doi/10.1126/science.1143205},
	doi = {10.1126/science.1143205},
	number = {5836},
	urldate = {2025-08-19},
	journal = {Science},
	publisher = {American Association for the Advancement of Science},
	author = {Monnier, John D. and Zhao, M. and Pedretti, E. and Thureau, N. and Ireland, M. and Muirhead, P. and Berger, J.-P. and Millan-Gabet, R. and Van Belle, G. and ten Brummelaar, T. and McAlister, H. and Ridgway, S. and Turner, N. and Sturmann, L. and Sturmann, J. and Berger, D.},
	month = jul,
	year = {2007},
	pages = {342--345},
}

@article{nguyen_parsec_2022,
	title = {{PARSEC} {V2}.0: {Stellar} tracks and isochrones of low- and intermediate-mass stars with rotation},
	volume = {665},
	issn = {0004-6361},
	shorttitle = {{PARSEC} {V2}.0},
	url = {https://ui.adsabs.harvard.edu/abs/2022A&A...665A.126N},
	doi = {10.1051/0004-6361/202244166},
	urldate = {2025-08-15},
	journal = {Astronomy and Astrophysics},
	publisher = {EDP},
	author = {Nguyen, C. T. and Costa, G. and Girardi, L. and Volpato, G. and Bressan, A. and Chen, Y. and Marigo, P. and Fu, X. and Goudfrooij, P.},
	month = sep,
	year = {2022},
	note = {ADS Bibcode: 2022A\&A...665A.126N},
	pages = {A126},
}

@article{hog_tycho2_2000,
	title = {The {Tycho}-2 catalogue of the 2.5 million brightest stars},
	volume = {355},
	issn = {0004-6361},
	url = {https://ui.adsabs.harvard.edu/abs/2000A&A...355L..27H},
	urldate = {2025-07-11},
	journal = {Astronomy and Astrophysics},
	author = {Høg, E. and Fabricius, C. and Makarov, V. V. and Urban, S. and Corbin, T. and Wycoff, G. and Bastian, U. and Schwekendiek, P. and Wicenec, A.},
	month = mar,
	year = {2000},
	note = {ADS Bibcode: 2000A\&A...355L..27H},
	pages = {L27--L30},
}

@article{eggenberger_analysis_2008,
	title = {Analysis of 70 {Ophiuchi} {AB} including seismic constraints},
	volume = {482},
	issn = {0004-6361},
	url = {https://ui.adsabs.harvard.edu/abs/2008A&A...482..631E},
	doi = {10.1051/0004-6361:20078624},
	urldate = {2025-07-01},
	journal = {Astronomy and Astrophysics},
	author = {Eggenberger, P. and Miglio, A. and Carrier, F. and Fernandes, J. and Santos, N. C.},
	month = may,
	year = {2008},
	note = {ADS Bibcode: 2008A\&A...482..631E},
	pages = {631--638},
}

@misc{markwardt_nonlinear_2009,
	address = {eprint: arXiv:0902.2850},
	title = {Non-linear {Least}-squares {Fitting} in {IDL} with {MPFIT}},
	url = {https://ui.adsabs.harvard.edu/abs/2009ASPC..411..251M},
	doi = {10.48550/arXiv.0902.2850},
	urldate = {2025-06-05},
	publisher = {arXiv},
	author = {Markwardt, C. B.},
	month = sep,
	year = {2009},
	note = {Conference Name: Astronomical Data Analysis Software and Systems XVIII
Volume: 411
ADS Bibcode: 2009ASPC..411..251M},
}

@article{izmailov_orbits_2019,
	title = {The {Orbits} of 451 {Wide} {Visual} {Double} {Stars}},
	volume = {45},
	issn = {1063-7737},
	url = {https://ui.adsabs.harvard.edu/abs/2019AstL...45...30I},
	doi = {10.1134/S106377371901002X},
	urldate = {2025-05-25},
	journal = {Astronomy Letters},
	publisher = {Springer},
	author = {Izmailov, I. S.},
	month = jan,
	year = {2019},
	note = {ADS Bibcode: 2019AstL...45...30I},
	pages = {30--38},
}

@article{mcalister_first_2005,
	title = {First {Results} from the {CHARA} {Array}. {I}. {An} {Interferometric} and {Spectroscopic} {Study} of the {Fast} {Rotator} α {Leonis} ({Regulus})},
	volume = {628},
	issn = {0004-637X},
	url = {https://ui.adsabs.harvard.edu/abs/2005ApJ...628..439M},
	doi = {10.1086/430730},
	urldate = {2025-05-24},
	journal = {The Astrophysical Journal},
	publisher = {IOP},
	author = {McAlister, H. A. and ten Brummelaar, T. A. and Gies, D. R. and Huang, W. and Bagnuolo, Jr., W. G. and Shure, M. A. and Sturmann, J. and Sturmann, L. and Turner, N. H. and Taylor, S. F. and Berger, D. H. and Baines, E. K. and Grundstrom, E. and Ogden, C. and Ridgway, S. T. and van Belle, G.},
	month = jul,
	year = {2005},
	note = {ADS Bibcode: 2005ApJ...628..439M},
	pages = {439--452},
}

@article{anugu_mircx_2020,
	title = {{MIRC}-{X}: {A} {Highly} {Sensitive} {Six}-telescope {Interferometric} {Imager} at the {CHARA} {Array}},
	volume = {160},
	issn = {1538-3881},
	shorttitle = {{MIRC}-{X}},
	url = {https://dx.doi.org/10.3847/1538-3881/aba957},
	doi = {10.3847/1538-3881/aba957},
	language = {en},
	number = {4},
	urldate = {2025-05-24},
	journal = {The Astronomical Journal},
	publisher = {The American Astronomical Society},
	author = {Anugu, Narsireddy and Le Bouquin, Jean-Baptiste and Monnier, John D. and Kraus, Stefan and Setterholm, Benjamin R. and Labdon, Aaron and Davies, Claire L and Lanthermann, Cyprien and Gardner, Tyler and Ennis, Jacob and Johnson, Keith J. C. and Ten Brummelaar, Theo and Schaefer, Gail and Sturmann, Judit},
	month = sep,
	year = {2020},
	pages = {158},
}

@misc{giovinazzi_neid_2025,
	title = {The {NEID} {Earth} {Twin} {Survey}. {II}. {Dynamical} {Masses} in {Seven} {High}-acceleration {Star} {Systems}},
	url = {http://arxiv.org/abs/2505.12563},
	doi = {10.3847/1538-3881/add922},
	urldate = {2025-05-20},
	author = {Giovinazzi, Mark R. and Blake, Cullen H. and Robertson, Paul and Lin, Andrea S. J. and Gupta, Arvind F. and Mahadevan, Suvrath and Fernandes, Rachel B. and Wright, Jason T. and Gagliuffi, Daniella Bardalez and Dong, Jiayin and Fitzmaurice, Evan and Halverson, Samuel and Kanodia, Shubham and Logsdon, Sarah E. and Luhn, Jacob K. and McElwain, Michael W. and Monson, Andy and Ninan, Joe P. and Rajagopal, Jayadev and Roy, Arpita and Schwab, Christian and Stefánsson, Gudmundur and Terrien, Ryan and Eastman, Jason D. and Horner, Jonathan and Plavchan, Peter and Wang, Sharon X. and Wilson, Maurice L. and Wittenmyer, Robert A.},
	month = may,
	year = {2025},
	note = {arXiv:2505.12563 [astro-ph]},
}

@article{mamajek_improved_2008,
	title = {Improved {Age} {Estimation} for {Solar}-{Type} {Dwarfs} {Using} {Activity}-{Rotation} {Diagnostics}},
	volume = {687},
	issn = {0004-637X},
	url = {https://ui.adsabs.harvard.edu/abs/2008ApJ...687.1264M},
	doi = {10.1086/591785},
	urldate = {2025-04-16},
	journal = {The Astrophysical Journal},
	publisher = {IOP},
	author = {Mamajek, Eric E. and Hillenbrand, Lynne A.},
	month = nov,
	year = {2008},
	note = {ADS Bibcode: 2008ApJ...687.1264M},
	pages = {1264--1293},
}

@article{nunez_nearinfrared_2017,
	title = {A near-infrared interferometric survey of debris-disc stars. {VI}. {Extending} the exozodiacal light survey with {CHARA}/{JouFLU}},
	volume = {608},
	issn = {0004-6361},
	url = {https://ui.adsabs.harvard.edu/abs/2017A&A...608A.113N},
	doi = {10.1051/0004-6361/201730859},
	urldate = {2025-04-16},
	journal = {Astronomy and Astrophysics},
	author = {Nuñez, P. D. and Scott, N. J. and Mennesson, B. and Absil, O. and Augereau, J. -C. and Bryden, G. and ten Brummelaar, T. and Ertel, S. and Coudé du Foresto, V. and Ridgway, S. T. and Sturmann, J. and Sturmann, L. and Turner, N. J. and Turner, N. H.},
	month = dec,
	year = {2017},
	note = {ADS Bibcode: 2017A\&A...608A.113N},
	pages = {A113},
}

@article{tayar_guide_2022,
	title = {A {Guide} to {Realistic} {Uncertainties} on the {Fundamental} {Properties} of {Solar}-type {Exoplanet} {Host} {Stars}},
	volume = {927},
	issn = {0004-637X},
	url = {https://ui.adsabs.harvard.edu/abs/2022ApJ...927...31T},
	doi = {10.3847/1538-4357/ac4bbc},
	urldate = {2025-02-18},
	journal = {The Astrophysical Journal},
	publisher = {IOP},
	author = {Tayar, Jamie and Claytor, Zachary R. and Huber, Daniel and van Saders, Jennifer},
	month = mar,
	year = {2022},
	note = {ADS Bibcode: 2022ApJ...927...31T},
	pages = {31},
}

@article{harada_setting_2024,
	title = {Setting the {Stage} for the {Search} for {Life} with the {Habitable} {Worlds} {Observatory}: {Properties} of 164 {Promising} {Planet}-survey {Targets}},
	volume = {272},
	issn = {0067-0049},
	shorttitle = {Setting the {Stage} for the {Search} for {Life} with the {Habitable} {Worlds} {Observatory}},
	url = {https://ui.adsabs.harvard.edu/abs/2024ApJS..272...30H},
	doi = {10.3847/1538-4365/ad3e81},
	urldate = {2025-01-30},
	journal = {The Astrophysical Journal Supplement Series},
	publisher = {IOP},
	author = {Harada, Caleb K. and Dressing, Courtney D. and Kane, Stephen R. and Ardestani, Bahareh Adami},
	month = jun,
	year = {2024},
	note = {ADS Bibcode: 2024ApJS..272...30H},
	pages = {30},
}

@article{chen_improving_2014,
	title = {Improving {PARSEC} models for very low mass stars},
	volume = {444},
	issn = {0035-8711},
	url = {https://doi.org/10.1093/mnras/stu1605},
	doi = {10.1093/mnras/stu1605},
	number = {3},
	urldate = {2025-01-24},
	journal = {Monthly Notices of the Royal Astronomical Society},
	author = {Chen, Yang and Girardi, Léo and Bressan, Alessandro and Marigo, Paola and Barbieri, Mauro and Kong, Xu},
	month = nov,
	year = {2014},
	pages = {2525--2543},
}

@article{swayne_eblm_2024,
	title = {The {EBLM} {Project}- {XI}. {Mass}, radius, and effective temperature measurements for 23 {M}-dwarf companions to solar-type stars observed with {CHEOPS}},
	volume = {528},
	issn = {0035-8711},
	url = {https://ui.adsabs.harvard.edu/abs/2024MNRAS.528.5703S},
	doi = {10.1093/mnras/stad3866},
	urldate = {2024-11-15},
	journal = {Monthly Notices of the Royal Astronomical Society},
	publisher = {OUP},
	author = {Swayne, M. I. and Maxted, P. F. L. and Triaud, A. H. M. J. and Sousa, S. G. and Deline, A. and Ehrenreich, D. and Hoyer, S. and Olofsson, G. and Boisse, I. and Duck, A. and Gill, S. and Martin, D. and McCormac, J. and Persson, C. M. and Santerne, A. and Sebastian, D. and Standing, M. R. and Acuña, L. and Alibert, Y. and Alonso, R. and Anglada, G. and Bárczy, T. and Barrado Navascues, D. and Barros, S. C. C. and Baumjohann, W. and Baycroft, T. A. and Beck, M. and Beck, T. and Benz, W. and Billot, N. and Bonfils, X. and Borsato, L. and Bourrier, V. and Brandeker, A. and Broeg, C. and Carmona, A. and Charnoz, S. and Collier Cameron, A. and Cortés-Zuleta, P. and Csizmadia, Sz and Cubillos, P. E. and Davies, M. B. and Deleuil, M. and Delfosse, X. and Delrez, L. and Demangeon, O. D. S. and Demory, B. -O. and Dransfield, G. and Erikson, A. and Fortier, A. and Forveille, T. and Fossati, L. and Fridlund, M. and Gandolfi, D. and Gillon, M. and Güdel, M. and Günther, M. N. and Hara, N. and Hébrard, G. and Heidari, N. and Hellier, C. and Helling, Ch and Isaak, K. G. and Kerschbaum, F. and Kiefer, F. and Kiss, L. L. and Kunovac, V. and Lalitha, S. and Lam, K. W. F. and Laskar, J. and Lecavelier des Etangs, A. and Lendl, M. and Magrin, D. and Marafatto, L. and Martioli, E. and Miller, N. J. and Mordasini, C. and Moutou, C. and Nascimbeni, V. and Ottensamer, R. and Pagano, I. and Pallé, E. and Peter, G. and Piazza, D. and Piotto, G. and Pollacco, D. and Queloz, D. and Ragazzoni, R. and Rando, N. and Rauer, H. and Ribas, I. and Santos, N. C. and Scandariato, G. and Ségransan, D. and Simon, A. E. and Smith, A. M. S. and Southworth, R. and Stalport, M. and Szabó, Gy M. and Thomas, N. and Udry, S. and Ulmer, B. and Van Grootel, V. and Venturini, J. and Walton, N. A. and Willett, E. and Wilson, T. G.},
	month = mar,
	year = {2024},
	note = {ADS Bibcode: 2024MNRAS.528.5703S},
	pages = {5703--5722},
}

@article{leeuwen_validation_2007,
	title = {Validation of the new {Hipparcos} reduction},
	volume = {474},
	copyright = {© ESO, 2007},
	issn = {0004-6361, 1432-0746},
	url = {https://www.aanda.org/articles/aa/abs/2007/41/aa8357-07/aa8357-07.html},
	doi = {10.1051/0004-6361:20078357},
	language = {en},
	number = {2},
	urldate = {2024-08-09},
	journal = {Astronomy \& Astrophysics},
	publisher = {EDP Sciences},
	author = {Leeuwen, F. van},
	month = nov,
	year = {2007},
	note = {Number: 2},
	pages = {653--664},
}

@article{kervella_stellar_2019,
	title = {Stellar and substellar companions of nearby stars from {Gaia} {DR2}. {Binarity} from proper motion anomaly},
	volume = {623},
	issn = {0004-6361},
	url = {https://ui.adsabs.harvard.edu/abs/2019A&A...623A..72K},
	doi = {10.1051/0004-6361/201834371},
	urldate = {2024-07-01},
	journal = {Astronomy and Astrophysics},
	author = {Kervella, Pierre and Arenou, Frédéric and Mignard, François and Thévenin, Frédéric},
	month = mar,
	year = {2019},
	note = {ADS Bibcode: 2019A\&A...623A..72K},
	pages = {A72},
}

@misc{southworth_rediscussion_2022,
	title = {Rediscussion of eclipsing binaries. {Paper} {XI}. {ZZ} {Ursae} {Majoris}, a solar-type system showing total eclipses and a radius discrepancy},
	url = {http://arxiv.org/abs/2209.10384},
	urldate = {2023-02-21},
	publisher = {arXiv},
	author = {Southworth, John},
	month = oct,
	year = {2022},
	note = {arXiv:2209.10384 [astro-ph]},
}
\bibliographystyle{aasjournalv7}

\end{document}